\documentclass[a4paper,11pt]{article}
\pdfoutput=1
\usepackage[utf8]{inputenc}
\usepackage{jcappub}
\usepackage{natbib}
\setcitestyle{square,numbers}
\usepackage{xcolor}
\usepackage{float}
\usepackage{multirow,array,longtable}
\usepackage{arydshln} 
\usepackage[makeroom]{cancel}
\usepackage{amsfonts}
\usepackage{dcolumn}
\usepackage{bm}
\usepackage{graphicx}
\usepackage{amssymb,amsmath}
\usepackage{multirow}
\usepackage{slashed,xfrac}
\usepackage[makeroom]{cancel}
\usepackage{rotating}
\usepackage{color,url}
\usepackage[colorlinks=true
,urlcolor=blue
,anchorcolor=blue
,citecolor=blue
,filecolor=blue
,linkcolor=blue
,menucolor=blue
,linktocpage=true
,pdfproducer=medialab
,pdfa=true
]{hyperref}
\usepackage{epsfig,psfrag,rotating,soul}
\usepackage{rotfloat}
\usepackage[normalem]{ulem}
\usepackage{framed,fancybox,xcolor,xfrac,geometry}
\usepackage{hhline}
\renewcommand\eqref[1]{Eq.~(\ref{#1})}
\newcommand\eqrefs[2]{Eqs.~(\ref{#1})-(\ref{#2})}
\newcommand\figref[1]{Fig.~\ref{#1}}

\newcommand\tabrefs[2]{Tables~\ref{#1}-\ref{#2}}
\newcommand\secref[1]{Section~\ref{#1}}
\newcommand\appref[1]{Appendix~\ref{#1}}
\newcommand\apprefs[2]{Appendices~\ref{#1}-\ref{#2}}
\evensidemargin \oddsidemargin
\allowdisplaybreaks

\def\gY{g'}
\def\gYc{g^{'\,2}}
\def\cw{c_{\rm w}}
\def\sw{s_{\rm w}}

\def\cosw{c_{\rm w}}
\def\sinw{s_{\rm w}}

\def\mw{m_{\rm W}}
\def\mz{m_{\rm Z}}
\def\mh{m_{\rm H}}

\def\vev{{\it v}}

\def\Zf{Z}
\def\div{\Delta_\epsilon}
\def\deltaCT{{\delta_\epsilon}}

\def\gmunu{g^{\mu\nu}}
\def\kVnu{k_1^\nu}
\def\kVmu{k_2^\mu}
\def\kTnu{k_2^\nu}
\def\kTmu{k_1^\mu}

\newcommand{\nn}{\nonumber}
\newcommand{\be}{\begin{equation}}
\newcommand{\ee}{\end{equation}}
\newcommand{\bear}{\begin{eqnarray}}
\newcommand{\eear}{\end{eqnarray}}
\newcommand{\mL}{\mathcal{L}}
\newcommand{\mO}{\mathcal{O}}

\newcommand{\mV}{\mathcal{V}}
\def\1loop{one-loop}

\def\greenfR{\hat{\Gamma}}
\def\greenfT{\Gamma^{\rm Tree}}
\def\greenfL{\Gamma^{\rm Loop}}
\def\SMgreenfL{\overline{\Gamma}^{\rm Loop}}
\def\greenfC{\Gamma^{\rm CT}}
\def\amp{\mathcal{A}}

\def\propR{\hat{\Delta}}
\def\SER{\hat{\Sigma}}
\def\SMSigma{\overline{\Sigma}}
\def\SERSM{\hat{\overline{\Sigma}}}
\def\tadR{\hat{T}}
\def\treeLtwo{{\rm EChL}^{(2)}_{\rm Tree}}
\def\treeLfour{{\rm EChL}^{(4)}_{\rm Tree}}
\def\treeLtwoLfour{{\rm EChL}^{(2+4)}_{\rm Tree}}
\def\loopLtwo{{\rm EChL}^{(2)}_{\rm Loop}}
\def\full{{\rm EChL}_{\rm Full}}
\def\fullSM{{\rm SM}_{\rm Full}}
\def\treeSM{{\rm SM}_{\rm Tree}}
\def\loopSM{{\rm SM}_{\rm Loop}}

\usepackage{soul}
\usepackage{lineno} 

\title{One-loop renormalization of VBS with the electroweak chiral Lagrangian in covariant gauges}
\author[a]{Maria J.  Herrero,}
\author[a]{and Roberto A.  Morales}
\affiliation[a]{Departamento de F\'{\i}sica Te\'orica and Instituto de F\'{\i}sica Te\'orica, IFT-UAM/CSIC,\\
Universidad Aut\'onoma de Madrid, Cantoblanco, 28049 Madrid, Spain}
\emailAdd{maria.herrero@uam.es}
\emailAdd{robertoa.morales@uam.es}

\abstract{This work presents a first full one-loop computation of vector boson scattering (VBS) within the non-linear effective field theory  given by the bosonic sector of the usually called electroweak chiral Lagrangian (EChL). 
The computation is performed in the most general case of covariant $R_\xi$ gauges and is compared through all this work with the Standard Model case,  whose computation in these covariant gauges  is also novel and is presented also here.  The calculation of the one-loop VBS amplitude is performed using the diagrammatic method by means of the one-particle-irreducible (1PI) Green functions that are involved in these scattering processes.  The central part of this work is then devoted to the renormalization of all the $n$-legs one-loop 1PI 
Green functions involved.  This renormalization is performed in the most general off-shell case with  arbitrary external legs momenta.  We then describe in full detail the renormalization program, which within this context of the EChL, implies to derive all the counterterms for both the electroweak parameters,  like  boson masses and gauge couplings,  and those for the EChL coefficients. These later are crucial for the renormalization of the new divergences typically appearing when computing loops with the lowest chiral dimension Lagrangian.  We present here the full list of involved divergences and counterterms in the $R_\xi$ gauges and derive the complete set of renormalization group equations for the EChL coefficients.  In the last part of this work, we present the EChL numerical results for the one-loop cross section in the $WZ$  channel and compare them with the SM results. }

\begin{document}
\begin{flushright}
	IFT-UAM/CSIC-21-30 
\end{flushright}
\maketitle

\section{Introduction}
\label{intro}

The use of Effective Field Theories (EFTs) to describe the phenomenology of new physics beyond the Standard Model (SM) of elementary particle interactions is nowadays a quite generalized  tool when comparing theoretical predictions with experimental data.  The most appealing feature of an Effective Field Theory (EFT) approach  is that it can be used as a generic test of the new physics without specifying the underlying ultraviolet fundamental theory that originates such low energy  theory.   For a proper EFT,  it is sufficient to require it to preserve the same symmetries of the SM,  in particular,  the $SU(3)_C \times SU(2)_L \times  U(1)_Y$ gauge invariance,  and to be able to deal with quantum corrections,  providing a framework for renomalization.   The physical active fields in these EFTs are as in the SM,  and include fermions (quarks and leptons),  gauge bosons (both electroweak (EW) and strong) and the Higgs field.   At present,   the most popular EFTs are the Standard Model Effective Field Theory (SMEFT)  and the Higgs Effective Field Theory (HEFT) (for a review see,  for instance,~\cite{Brivio:2017vri}).  The main difference between these two EFTs is the realization of the EW gauge symmetry in the scalar sector, i.e., in the system formed by the Higgs particle, $H$,   and the three EW would be Golstone bosons,   named here $\pi^a$ ($a=1,2,3$).  It is important to recall that,  within the SM context,  this scalar system is the responsible for the mass generation of all the SM particles, via the Higgs mechanism, once the EW symmetry, $SU(2)_L\times U(1)_Y$,   is spontaneously broken down to the electromagnetism subgroup, $U(1)_{\rm em}$.  One crucial point to notice is that these three associated GBs to this breaking are indeed the same three GBs that are associated to the spontaneous breaking of the EW chiral symmetry,  $SU(2)_L \times SU(2)_R \to SU(2)_{L+R}$.  This EW chiral  symmetry  is an exact global symmetry of the SM scalar sector in the limit of vanishing EW gauge couplings, i.e., $g$ and $g'$ set to zero,  and is the main responsible for protecting the mass relation $m_W=m_Z \cos \theta_W$ from large radiative corrections.  Thus,  $SU(2)_{L+R}$ is sometimes called custodial symmetry group.

The realization of both the EW gauge symmetry and the EW chiral symmetry in the scalar sector is linear in 
the SM and the SMEFT,  whereas it is non-linear in the HEFT.  The Higgs boson and the three GBs are placed  together in a linear representation in the case of SM and SMEFT,  given by the usual doublet 
$\Phi^T=(i\pi^+ \,,\,((H+v)-i\pi^3)/\sqrt{2})$,  with $\pi^\pm =(\pi^1\mp i \pi^2)/\sqrt{2}$,  whereas they are treated differently in the HEFT.  In this HEFT,  the three GBs are put together in a non-linear representation, usually by an exponential parametrization,  $U(\pi^a)={\rm exp}(i \pi^a \tau^a/v)$ with  $v=246\,  {\rm GeV}$ and $\tau^a=1,2,3$ the Pauli matrices,  and the Higgs boson, in contrast, 
 is a singlet under all the involved symmetries.  This difference between linear and non-linear behaviour of the EW scalar sector in the two EFTs,  SMEFT and HEFT,  is not just a mere choice of parametrization but it contains important differences in their respective  phenomelogical implications.  The main reason of this relevant difference  is because of the different hierarchies in the ordering and importance of the participating effective operators in both approaches.  To understand deeply this main difference one has to place the comparison of both EFTs not only at the tree level, but also beyond the tree level, i.e., including ${\cal O}(\hbar)$ loop corrections,  and dealing with the important renormalization issue in both theories.  
 In contrast to the usual ordering based on canonical dimension counting in the SMEFT,  the ordering of effective operators in the HEFT is by construction done by increasing powers of their chiral dimension,  starting in chiral dimension 2, next chiral dimension 4, and so on.  This implies that the ordering in the importance of the HEFT effective operators is given in terms of an expansion in powers of momenta,   
$(p/(4 \pi v))^n$,  where $4 \pi v \sim 3 \,  {\rm TeV}$ sets typically the  maximum energy scale of applicability in this EFT.  

In this mentioned chiral counting it is important to notice that all EW mass scales are treated as {\it soft} mass scales,  therefore,  counting equally as momentum,  namely,  
$\partial_\mu \,,\,\mw \,,\,\mz \,,\,\mh \,,\,g\vev \,,\,\gY\vev \sim \mO(p) \,$.   This  also orders the importance of the loop corrections in this EFT.  This reasoning is done in analogy to the usual Chiral Perturbation Theory (ChPT) of low energy QCD~\cite{Gasser:1983yg,Manohar:1983md,Weinberg:1978kz},  where the most important contributions come from pion loops (the so called chiral loops of QCD or GB loops) which naturally come with the loop factors $(1/(4 \pi f_\pi))^n$,  where $4 \pi f_\pi \sim 1.2 \,  {\rm GeV}$,   and the polynomial contributions in powers of the external momenta, $p^n$,  introduced by the derivative interactions of the GBs in this non-linear approach.  The renormalization program of ChPT is well known and very successful and teaches us the way to proceed in the non-linear EFTs.  It relays on the key point that the chiral dimension four operators act as well as counterterms of the \1loop  corrections generated by the lowest chiral dimension two  operators.  Thus,  when computing an observable in ChPT to ${\cal O}(\hbar)$, like for instance the pion-pion scattering amplitude, it is sufficient to compute loops of pions with the lowest order chiral dimension two Lagrangian, and to treat  the chiral dimension four  operators  at the tree level,  acting simultaneously as counterterms to renormalize the loop divergenges generated from those loops. 
The approach followed by the HEFT is very similar to the one of ChPT.  It also organizes the ordering of the effective operators by their increasing chiral dimension as ${\cal L}_{\rm HEFT}= {\cal L}_2+{\cal L}_4+...$,  and the renormalization program  is implemented in practice by computing  loops with just ${\cal L}_2$ and treating  ${\cal L}_4$ to tree level and simultaneously using it as counterterms to renormalize the loop divergences generated by those loops.  Thus, the renormalization of the EFT coefficients in ${\cal L}_4$ is dictated by the loop divergences generated from ${\cal L}_2$.  The difference with respect to ChPT is that in the HEFT  all kind of loops participate, with gauge bosons, scalar bosons (GBs and Higgs boson),  etc, not only GB loops.  This renormalization program is totally different in the case of SMEFT where the effective operators,  which are ordered by their increasing canonical dimension,  four,  five, six and so on,  as ${\cal L}_{\rm SMEFT}= {\cal L}_{c4}+{\cal L}_{c5}+{\cal L}_{c6}+..$,  are treated all at both the tree and the loop level,  and the so-called Wilson coefficients of those effective operators  participate in the loop contributions to a given observable.  This sets important differences also on the hierarchies of the quantum corrections on both EFT approaches.

We focus here on the HEFT and more concretely on the bosonic sector,  with EW gauge bosons,  GBs and the Higgs boson,  assuming  that the fermion sector is as in the SM.  The corresponding effective Lagrangian of the HEFT in that case is usually called the electroweak chiral Lagrangian (EChL). 
The main focus of this work is on the \1loop renormalization program for this EFT when applied to a practical computation of physical scattering processes.  In particular, we focus here on the most sensitive observables to the new physics in the case of an hypothetical underlying strongly interacting UV theory,  which are the vector boson scattering (VBS) processes,  with $V=W^{\pm}, Z$ vector bosons
in the external legs (for a recent review on VBS see,  for instance,  \cite{Covarelli:2021gyz}).  At high energies compared to the gauge boson masses,  and by virtue of the Equivalence Theorem (ET),  the VBS amplitude for external  longitudinal vector bosons,  is approximately equal to the scattering amplitude with the vector bosons replaced by the corresponding GBs, $\pi^{\pm}, \pi^3$,   and,  therefore, one expects  to find in observables  like the cross section $\sigma(V_L V_L \to V_L V_L)$ the most prominent signals of the underlying UV strongly interacting theory.  This expectation  is in close analogy to the case of the pion-pion scattering predictions from  ChPT being one of the best low energy hints of the strongly interacting underlying QCD dynamics.  In that case, it is well known that a linear approach to the pions self-interactions (the so-called sigma model) does not provide as good predictions of the pion-pion scattering rates as it does the non-linear ChPT approach.

In this work we  present the computation within the EChL  of the full \1loop scattering VBS amplitudes including all ${\cal O}(\hbar)$ radiative corrections from the complete bosonic sector and the corresponding cross sections. We  use the standard  Feynman diagrammatic approach and describe the full renormalization program also in terms of \1loop Feynman diagrams.  We organize this computation in terms of the involved one-particle-irreducible (1PI) Green functions, with two, three and four external legs,  and  perform by the first time in the literature this full \1loop computation in the generic $R_\xi$ gauges.  This is a crucial point of the computation since it provides a good check of the gauge invariance of the result.  For definiteness, we will describe the details of the computation for the particular VBS channel,   $WZ \to WZ$, and leave for future  works the details of the other VBS channels.  We also present a comparison of our renormalization program for the non-linear EFT with previous related works in the literature.  The \1loop renormalization of the EChL has been studied in~\cite{Guo:2015isa,Buchalla:2017jlu,Buchalla:2020kdh}. These works are focused,  in contrast to our work, in the renormalization of the Lagrangian itself and use the path integral formalism and the background field  method to compute the \1loop divergences that have to be renormalized by the redefinition of the EChL coefficients.  They use in addition the equations of motion to reduce the number of independent effective operators.  Our computation in contrast does not make use of the equations of motion since our objective is to compute off-shell renormalized 1PI Green functions.  On the other hand,  the \1loop VBS amplitudes within the EChL were computed previously in~\cite{Espriu:2013fia, Delgado:2013hxa,Delgado:2014jda} but using the ET,  namely,  taking into account only the scalar sector,  with external GBs and with pure scalar loops of GBs and Higgs bosons, and considering  massless GBs.   Also the \1loop renormalization program within the EChL in~\cite{Gavela:2014uta} is performed in the pure scalar theory.  In this latter reference they consider the off-shell effects and therefore use the full set of effective operators of the scalar sector.  Our work here represents, therefore, the first full \1loop computation of VBS with all relevant features included: off-shell renormalized 1PI Green functions,   all effective operators up to chiral dimension four taken into account, massive  gauge bosons in the VBS amplitudes (we do not use the ET),  full bosonic loops with gauge bosons and scalar bosons and full renormalization program using the $R_\xi$ gauges.  Our \1loop VBS results within the EChL are,  therefore,  complete and include all finite radiative corrections from the bosonic sector.  The renormalization conditions are fixed here in the on-shell scheme for the EW parameters,  like masses and couplings, and in the $\overline{MS}$ scheme for the effective operator coefficients.   One of the most interesting results in this work is our finding of the $\overline{MS}$ renormalized EChL coefficients and their running with the renormalization scale,  that we compute from the involved renormalized 1PI Green functions with the generic $R_\xi$ gauges and that we find independent on the $\xi$ gauge-fixing parameter. 

The paper is organized as follows.  The main features of the EChL with $R_\xi$ gauge-fixing and the relevant operators for VBS processes are presented in \secref{sec-EChL}.  The diagrammatic computation of VBS in terms of the 1PI functions is described in \secref{diag-1pi}.  The central part of the work is in \secref{sec-renorm} where the renormalization program is presented in detail,  including the prescriptions for regularization and renormalization assumed and the summary of all the results from the renormalization process.  All the divergent counterterms and the derived renormalization group equations are also presented in this section.   A discussion on the Slavnov-Taylor identities and the peculiarities of the Higgs tadpole in the non-linear EFT and the comparison with the SM case is also included in this section.  The numerical results for the one-loop $WZ$ scattering process within the EChL and the SM are presented and discussed  in \secref{sec-plots}.  Finally, we conclude in \secref{sec-conclu}.  The appendices summarize some technical aspects of the computation.  \appref{App-FRules} collects the relevant Feynman Rules (FRs) and the chosen conventions. \appref{App-oneloopdiag} contains the generic \1loop diagrams involved. \apprefs{App-SMcompu}{App-tadpole} contain the technical details of the comparison between the SM and  the EChL computation, including the discussion on the STI and the Higgs tadpole role.


\section{The electroweak chiral Lagrangian using the $R_\xi$ gauge-fixing}
\label{sec-EChL}
In this section we shortly summarize the most relevant pieces of the EChL for the present computation, and introduce some necessary notation.   
The active fields of the EChL are the EW gauge bosons, $W^a_\mu$ ($a=1,2,3$) and $B_\mu$,  that are associated to $SU(2)_L$ and $U(1)_Y$,  respectively,  the three GBs $\pi^a$ ($a=1,2,3$),  and the Higgs boson $H$.  
The GBs are introduced in a non-linear representation, usually via the exponential parametrization,  by means of the matrix $U$:
\be 
U(\pi^a) = e^{i \pi^a \tau^a/\vev} \, \, , 
\label{expo}
\ee
where, $\tau^a$, $a=1,2,3$,  are the Pauli matrices and $v=246$ GeV.  
Under a EW chiral transformation of  $SU(2)_L \times SU(2)_R$,  given by $L \in SU(2)_L$ and $R \in SU(2)_R$, the field $U$ transforms linearly as $L U R^\dagger$, whereas the GBs $\pi^ a$ transform non-linearly.  This peculiarity implies multiple GBs interactions,  not just among themselves but also with the other fields, and it is the main feature of this non-linear EFT, which is clearly manifest in the following expansion:
\bear
U (\pi^a)&=
& I_2 +i\frac{\pi^a}{\vev}\tau^a -\frac{2\pi^+\pi^-+\pi^3\pi^3}{2\vev^2}I_2 -i\frac{(2\pi^+\pi^-+\pi^3\pi^3)\pi^a}{6\vev^3}\tau^a +\ldots
\eear
where $I_2$ is the unity matrix and the dots stand for terms with four or more GBs.

The $H$ field is, in contrast to the GBs,  a singlet of the EW chiral symmetry and the EW gauge symmetry and, consequently, there are not limitations from symmetry arguments on the implementation of this field and its interactions with itself and with the other fields.  Usually,  in the EChL, the interactions of $H$  are introduced via generic polynomials.

The EW gauge bosons are introduced in the EChL by means of the $SU(2)_L \times U(1)_Y$ gauge prescription,  namely, via the covariant derivative of the $U$ matrix,  and by the $SU(2)_L$ and $U(1)_Y$  field strength tensors,  as follows:
\bear
D_\mu U &=& \partial_\mu U + i\hat{W}_\mu U - i U\hat{B}_\mu \,, \nn\\
\hat{W}_{\mu\nu} &=& \partial_\mu \hat{W}_\nu - \partial_\nu \hat{W}_\mu + i  [\hat{W}_\mu,\hat{W}_\nu ] \,, 
\quad \hat{B}_{\mu\nu} = \partial_\mu \hat{B}_\nu -\partial_\nu \hat{B}_\mu \,,  
\eear
where $\hat{W}_\mu = g W^a_\mu \tau^a/2$, and  $\hat{B}_\mu = \gY B_\mu \tau^3/2$.  For the construction of the EChL and 
in addition to these basic building blocks,  it is also customary to use the following objects:
\bear
 \mV_\mu&=&(D_\mu U)U^\dagger   \,, 
\quad {\cal D}_\mu O=\partial_\mu O+i[\hat{W}_\mu,O]\, .
\eear
The physical gauge fields are given,  as usual,  by:
\be
W_{\mu}^\pm = \frac{1}{\sqrt{2}}(W_{\mu}^1 \mp i W_{\mu}^2) \,,\quad
Z_{\mu} = c_W W_{\mu}^3 - s_W B_{\mu} \,,\quad
A_{\mu} = s_W W_{\mu}^3 + c_W B_{\mu} \,,
\label{eq-gaugetophys}
\ee
where we use the short notation $s_W=\sin \theta_W$ and $c_W=\cos \theta_W$,  with $\theta_W$ the weak angle.

For the present computation we  include  in the EChL all the relevant effective operators,  organized as usual by their chiral dimension into two terms:  ${\cal L}_2$,  with chiral dimension two and  ${\cal L}_4$
with chiral dimension four.  As already said in the introduction,  for this chiral counting,  we consider that all involved masses count equally as momentum,  namely,  with chiral dimension one.  Consequently,  $\partial_\mu \,,\,\mw \,,\,\mz \,,\,\mh \,,\,g\vev \,,\,\gY\vev \,,\lambda v\, \sim \mO(p) \,$. Thus, 
for the present work,  the relevant $SU(2)_L \times U(1)_Y$ gauge invariant EChL  is given, in short,  by
\be
{\cal L}_{\rm EChL}=\mL_2+ \mL_4 \, .
\label{EChL}
\ee
The relevant chiral dimension two  Lagrangian is as follows,
\bear
\mL_2 &=& \frac{v^2}{4}\left(1+2a\frac{H}{v}+b\left(\frac{H}{v}\right)^2 +\ldots\right){\rm Tr}\Big[ 
 D_\mu U^\dagger D^\mu U \Big]+\frac{1}{2}\partial_\mu H\partial^\mu H-V(H)  \nn\\
&&-\frac{1}{2g^2} {\rm Tr}\Big[ \hat{W}_{\mu\nu}\hat{W}^{\mu\nu}\Big]
-\frac{1}{2\gYc}{\rm Tr}\Big[ 
\hat{B}_{\mu\nu}\hat{B}^{\mu\nu}\Big]  +\mL_{GF} +\mL_{FP} \,, 
\label{eq-L2}
\eear
where,  $ \mL_{GF}$,  and  $\mL_{FP}$,  are the gauge-fixing  Lagrangian and Fadeev-Popov  Lagrangian,  respectively,  and $V(H)$ is the Higgs potential,  which we take here as,
\be
V(H) = \frac{1}{2}\mh^2 H^2 +\kappa_3\lambda\vev H^3+\kappa_4\frac{\lambda}{4}H^4 \,,
\ee
with $m_H^2= 2 \lambda v^2$. The new physics BSM in $\mL_2$ is encoded in the chiral coefficients $a$, $b$, $\kappa_3$ and $\kappa_4$. These are all generically different from one, which is their reference SM value.  
In Eq.(\ref{eq-L2}) above,  the dots stand for terms with more than two Higgs bosons (allowed by the fact that $H$ is a singlet in this non-linear EFT) but we omit them here since they do not enter in our VBS processes of interest, neither at tree level nor at \1loop level.

The relevant  chiral dimension four Lagrangian is organized here as follows: 
\bear
{\mL}_{4}&=& {\mL}_{4}^{\rm no-Higgs}+ {\mL}_{4}^{\rm one-Higgs}+ {\mL}_{4}^{\rm two-Higgs}+\ldots,  
\eear
where `no-Higgs' means effective operators not including the Higgs field,  `one-Higgs' means effective operators including one Higgs field,  and so on,  and the dots again means operators which do not enter in the present computation.  

The list of operators in ${\mL}_{4}^{\rm no-Higgs}$ were given long ago in~\cite{Longhitano:1980iz} and summarize  the complete set of CP and $SU(2)_L \times U(1)_Y$ gauge invariant effective operators, including both custodial preserving and custodial breaking ones.  We take them from this reference,  although using a different notation:
\bear
{\mL}_{4}^{\rm no-Higgs}&=& \,a_0 \left( \mz^2-\mw^2 \right) {\rm Tr}\Big[ U \tau^3 U^\dagger {\cal V}_\mu \Big] {\rm Tr}\Big[ U \tau^3 U^\dagger {\cal V}^\mu \Big]  \nn\\
&& + a_1 {\rm Tr}\Big[ U \hat{B}_{\mu\nu} U^\dagger \hat{W}^{\mu\nu}\Big] + i a_2  {\rm Tr}\Big[ U \hat{B}_{\mu\nu} U^\dagger [{\cal V}^\mu, {\cal V}^\nu ]\Big]  - i a_3 {\rm Tr}\Big[\hat{W}_{\mu\nu}[{\cal V}^\mu, {\cal V}^\nu]\Big]  \nn\\
&& + a_4 {\rm Tr}\Big[{\cal V}_\mu {\cal V}_\nu\Big] {\rm Tr}\Big[{\cal V}^\mu {\cal V}^\nu\Big] + a_5 {\rm Tr}\Big[{\cal V}_\mu {\cal V}^\mu\Big] {\rm Tr}\Big[{\cal V}_\nu {\cal V}^\nu\Big]  \nn\\
&& + a_6 {\rm Tr}\Big[{\cal V}_\mu {\cal V}_\nu\Big]{\rm Tr}\Big[U \tau^3 U^\dagger{\cal V}^\mu\Big] {\rm Tr}\Big[U \tau^3 U^\dagger{\cal V}^\nu\Big]  
+ a_7 {\rm Tr}\Big[{\cal V}_\mu {\cal V}^\mu\Big]{\rm Tr}\Big[U \tau^3 U^\dagger{\cal V}^\nu\Big] {\rm Tr}\Big[U \tau^3 U^\dagger{\cal V}_\nu\Big]  \nn\\
&&-\frac{a_8}{4}{\rm Tr}\Big[U \tau^3 U^\dagger\hat{W}_{\mu\nu}\Big]{\rm Tr}\Big[U \tau^3 U^\dagger\hat{W}^{\mu\nu}\Big] -i\frac{a_9}{2}{\rm Tr}\Big[U \tau^3 U^\dagger\hat{W}_{\mu\nu}\Big]{\rm Tr}\Big[ U \tau^3 U^\dagger [{\cal V}^\mu, {\cal V}^\nu ]\Big]  \nn\\
&&+ a_{10} {\rm Tr}\Big[U \tau^3 U^\dagger{\cal V}^\mu\Big] {\rm Tr}\Big[U \tau^3 U^\dagger{\cal V}_\mu\Big]{\rm Tr}\Big[U \tau^3 U^\dagger{\cal V}^\nu\Big] {\rm Tr}\Big[U \tau^3 U^\dagger{\cal V}_\nu\Big] \nn\\
&&+ a_{11} {\rm Tr}\Big[{\cal D}_\mu {\cal V}^\mu {\cal D}_\nu {\cal V}^\nu\Big] +a_{12}{\rm Tr}\Big[U \tau^3 U^\dagger {\cal D}_\mu {\cal D}_\nu {\cal V}^\nu\Big] {\rm Tr}\Big[U \tau^3 U^\dagger {\cal V}^\mu\Big]  \nn\\
&&+\frac{a_{13}}{2}{\rm Tr}\Big[U \tau^3 U^\dagger {\cal D}_\mu {\cal V}_\nu\Big]{\rm Tr}\Big[U \tau^3 U^\dagger {\cal D}^\mu {\cal V}^\nu\Big] \,.
\label{eq-L4-Longhitano}
\eear
The relation with the notation in~\cite{Longhitano:1980iz} is:  $a_0= \beta_1$,  $a_1=(g/g') \alpha_1$, $a_2=(g/g') \alpha_2$,
$a_3= -\alpha_3$, $a_4= \alpha_4$, $a_5= \alpha_5$; $a_6=\alpha_6$, $a_7=\alpha_7$, $a_8=-g^2\alpha_8$, $a_9=-g\alpha_9$, $a_{10}=\alpha_{10}/2$, $a_{11}=\alpha_{11}$, $a_{12}=\alpha_{12}/2$ and $a_{13}=\alpha_{13}$.

The set of relevant chiral dimension four effective operators with one-Higgs field and two-Higgs fields are taken from~\cite{Brivio:2013pma},  but using a different notation:
\bear
{\mL}_{4}^{\rm one-Higgs}&=& - a_{HBB} \frac{H}{v}\, {\rm Tr} \Big[\hat{B}_{\mu\nu} \hat{B}^{\mu\nu} \Big]  - a_{HWW} \frac{H}{v} {\rm Tr}\Big[\hat{W}_{\mu\nu} \hat{W}^{\mu\nu}\Big] +a_{\Box\mV\mV}\frac{\Box H}{\vev} {\rm Tr}\Big[{\cal V}_\mu {\cal V}^\mu\Big]  \nn\\
&& +a_{H0} \frac{H}{v} \left( \mz^2-\mw^2 \right) {\rm Tr}\Big[ U \tau^3 U^\dagger {\cal V}_\mu \Big] {\rm Tr}\Big[ U \tau^3 U^\dagger {\cal V}^\mu \Big] + a_{H1} \frac{H}{v} {\rm Tr}\Big[ U \hat{B}_{\mu\nu} U^\dagger \hat{W}^{\mu\nu}\Big]  \nn\\
&& -\frac{a_{H8}}{4}\frac{H}{\vev} {\rm Tr}\Big[U \tau^3 U^\dagger\hat{W}_{\mu\nu}\Big]{\rm Tr}\Big[U \tau^3 U^\dagger\hat{W}^{\mu\nu}\Big] +a_{H11}\frac{H}{\vev} {\rm Tr}\Big[{\cal D}_\mu {\cal V}^\mu {\cal D}_\nu {\cal V}^\nu\Big]  \nn\\
&& +\frac{a_{H13}}{2}\frac{H}{\vev} {\rm Tr}\Big[U \tau^3 U^\dagger {\cal D}_\mu {\cal V}_\nu\Big]{\rm Tr}\Big[U \tau^3 U^\dagger {\cal D}^\mu {\cal V}^\nu\Big] + \ldots  \nn\\
&& +a_{\Box 0}\left( \frac{\mz^2-\mw^2}{\vev^2} \right)\frac{\Box H}{\vev} {\rm Tr}\Big[ U \tau^3 U^\dagger {\cal V}_\mu \Big] {\rm Tr}\Big[ U \tau^3 U^\dagger {\cal V}^\mu \Big]  \nn\\
&& +ia_{d1}\frac{\partial^\nu H}{v} {\rm Tr}\Big[ U \hat{B}_{\mu\nu} U^\dagger {\cal V}^\mu\Big] +ia_{d2}\frac{\partial^\nu H}{v} {\rm Tr}\Big[ \hat{W}_{\mu\nu} {\cal V}^\mu\Big] +a_{d3}\frac{\partial ^\nu H}{\vev} {\rm Tr}\Big[{\cal V}_\nu {\cal D}_\mu {\cal V}^\mu \Big]  \nn\\
&&+ia_{d4}\frac{\partial^\nu H}{v} {\rm Tr}\Big[U \tau^3 U^\dagger\hat{W}_{\mu\nu}\Big]{\rm Tr}\Big[ U \tau^3 U^\dagger {\cal V}^\mu \Big] +a_{d5}\frac{\partial^\nu H}{v} {\rm Tr}\Big[U \tau^3 U^\dagger {\cal D}_\mu {\cal V}^\mu\Big]{\rm Tr}\Big[ U \tau^3 U^\dagger {\cal V}_\nu \Big]  \, \,,    \,\nonumber \\
{\mL}_{4}^{\rm two-Higgs}&=& a_{\Box\Box}\frac{\Box H\,\Box H}{\vev^2}+ \ldots \, \, ,  
\label{eq-L4-withH}
\eear
where the dots again mean operators which are not relevant for the present computation.  
The relation of our EChL coefficients in the above $\mL_4$,  which are refer jointly as $a_i$ coefficients,  with those in~\cite{Brivio:2013pma} is as follows: $a_{HBB}\leftrightarrow P_B$, $a_{HWW}\leftrightarrow P_W$, $a_{\Box{\cal V}{\cal V}}\leftrightarrow P_7$, $a_{H0}\leftrightarrow P_T$, $a_{H1}\leftrightarrow P_1$, $a_{H8}\leftrightarrow P_{12}$, $a_{H11}\leftrightarrow P_9$, $a_{H13}\leftrightarrow P_{16}$, $a_{\Box 0}\leftrightarrow P_{25}$, $a_{d1}\leftrightarrow P_4$, $a_{d2}\leftrightarrow P_5$, $a_{d3}\leftrightarrow P_{10}$, $a_{d4}\leftrightarrow P_{17}$, $a_{d5}\leftrightarrow P_{19}$ and $a_{\Box\Box}\leftrightarrow P_{\Box H}$.  
Notice that we have used a specific notation for some of the $a_i$'s in ${\mL}_{4}^{\rm one-Higgs}$ that refer explicitly to  those operators that are replicas of the structures in ${\mL}_{4}^{\rm no-Higgs}$ but with an extra factor given by $(H/v)$.  For instance,  $a_{H0}$ versus $a_0$,  $a_{H1}$ versus $a_1$, etc.    Alternative sets of effective operators,  with a reduced number of operators in the list by the use of the equations of motion,  can be found in the literature (see, for instance,~\cite{Buchalla:2020kdh}).

Regarding the quantization of the EChL we choose here to use the same gauge-fixing Lagrangian, $\mL_{\rm GF}$, as in the SM for the linear covariant $R_\xi$ gauges~\cite{Fujikawa:1972fe}, in which the tree level mixing between gauge bosons with their corresponding GB are absent.  Some features of $R_\xi$ gauge-fixing and renormalization within the context of the EChL were already studied long ago in~\cite{Herrero:1993nc,Herrero:1994iu} when the Higgs particle was not included explicitly in the Lagrangian (see also~\cite{ST-Espriu}). Generically, the quantization of the EChL requires the insertion of appropriate gauge-fixing functions $F_j$ involving the EW gauge bosons and the GBs. This gauge-fixing Lagrangian can be written in terms of the physical basis as follows:
\be
\mL_{\rm GF} = -F_+F_- -\frac{1}{2}F_{Z}^2 -\frac{1}{2}F_{ A}^2 \,,
\label{GF-lag}
\ee
where the gauge-fixing functions are:
\bear
F_\pm=\frac{1}{\sqrt{\xi}}(\partial^{\mu}W_{\mu}^\pm-\xi \mw \pi^\pm) \,,\quad F_Z=\frac{1}{\sqrt{\xi}}(\partial^{\mu}Z_{\mu}-\xi \mz \pi^{3}) \,,\quad F_A=\frac{1}{\sqrt{\xi}}(\partial^{\mu}A_{\mu}) \,,
\label{gaugefixingfunctions}
\eear
and $\xi$ is the typical gauge-fixing parameter of the $R_\xi$ gauges.

From the above gauge-fixing functions, $F_\pm$, $F_Z$ and $F_A$,  one derives the corresponding Faddeev-Popov Lagrangian~\cite{Faddeev:1967fc}, by: 
\be
\mL_{\rm FP} = \sum_{i,j=+,-,Z,A} \bar{c}^{i} \frac{\delta F_i}{\delta \alpha_j} c^j \,,
\label{FP-lag}
\ee
where $c^j$ are the ghost fields and $\alpha_j$ ($j=+,-, Z,A$) are the corresponding gauge transformation parameters
under the local transformations $SU(2)_L\times U(1)_Y$ given by $L=e^{ig\vec{\tau}\cdot\vec{\alpha}_L(x)/2}$ and $R=e^{i\gY\tau^3\alpha_Y(x)/2}$.
The corresponding gauge field transformations are as in the SM.
However, the scalar transformations in this non-linear EFT are 
\bear
\delta H &=& 0 \,,  \nn\\
\delta\pi^\pm &=& \,\mw\alpha^\pm \mp\frac{ig}{2}\alpha^\pm\pi^3 \pm \frac{g(\cw^2-\sw^2)}{2\cw}\alpha_Z\pi^\pm \pm ig\sw\alpha_A\pi^\pm  \nn\\
&&+\frac{\mw}{3\vev^2}(-\alpha^\pm\pi^+\pi^-+\alpha^\mp\pi^\pm\pi^\pm-\alpha^\pm\pi^3\pi^3) +\frac{\mz}{3\vev^2}\alpha_Z\pi^3\pi^\pm +\ldots  \nn\\
\delta\pi^3 &=& \,\mz\alpha_Z -\frac{ig}{2}(\alpha^-\pi^+-\alpha^+\pi^-)  \nn\\
&&+\frac{\mw}{3\vev^2}(\alpha^+\pi^-\pi^3+\alpha^-\pi^+\pi^3) -\frac{2\mz}{3\vev^2}\alpha_Z\pi^+\pi^- +\ldots
\eear
which differ from the corresponding ones in the SM.
In particular, two main differences respect to the SM arise from this ghost Lagrangian which are worth mentioning: 1) within  the EChL  there are not interactions among the Higgs boson and ghost fields,  and 2) there are new interactions with multiple GBs and two ghost fields. 

As a general comment to the EChL above,  notice that we do not use the equations of motion to reduce the number of effective operators since, as already stated in the introduction,  we are interested on the computation of VBS in terms of the off-shell renormalized 1PI Green functions, thus we keep the track of all the involved EChL coefficients into our computation. For the same reason,   we do not apply either any Higgs field redefinition and keep explicitly the track as well of all the off-shell Higgs effects.

The summary of all the relevant FRs for the present computation is presented in \appref{App-FRules}. We have also added the corresponding FRs within the SM, for a clear comparison.

\section{Diagrammatic computation:  VBS from  1PI Green functions}
\label{diag-1pi}

For definiteness in the description of our computational procedure,   and to fix some convenient notation, we focus in this section on the VBS channel $WZ \to WZ$.  The detailed study of other VBS channels is postponed for future works.  The amplitude of our interest here is,  therefore,  
$\amp(WZ\to WZ)$.  Following the standard counting rules of the EChL, the full \1loop scattering amplitude can be splitted into two parts,  as follows:  
\be
\amp(W Z \to W Z)^{{\rm EChL}} = \amp^{(0)}(W Z \to W Z)  + \amp^{(1)}(W Z \to W Z) \equiv \amp^{\full}\,,
\label{EChLfull}
\ee
where the leading order (LO), ${\cal O}(p^2)$, and the next to leading order contributions (NLO), ${\cal O}(p^4)$, are
denoted as  $\amp^{(0)}$ and $\amp^{(1)}$ respectively, and are given by:
\bear
 \amp^{(0)}(W Z \to W Z)&\equiv &  \amp^{\treeLtwo}\,,    \\
 \amp^{(1)}(W Z \to W Z)& \equiv &  \amp^{\treeLfour}+ \amp^{\loopLtwo}\,.
\label{LOandNLOamplitudes}
\eear
This means, that to compute the LO amplitude in this EChL context, one uses the Feynman rules from ${\cal L}_2$ at the tree level, and to compute the NLO correction one adds the contribution from ${\cal L}_4$ at the tree level and the contribution from the loops using ${\cal L}_2$.  This reflects the typical double role of ${\cal L}_4$ in the chiral Lagrangian approach. On the one hand,  it adds an extra tree level contribution to the scattering amplitude,  and on the other hand it also acts as generator of new counterterms that are needed to remove the extra divergences emerging from the loops computed with  ${\cal L}_2$ and that are not removable by a simple redefinition (counterterms) of the parameters in ${\cal L}_2$.  This renormalization procedure of the extra divergences,  which is well known in the context of ChPT,  works also properly  here for the EChL and its application to the computation of finite VBS amplitudes,  as will be seen in detail in the next section.  Notice also that the previous splitting in \eqref{EChLfull} of the EChL amplitude  into 
$\amp^{(0)}$ and $\amp^{(1)}$ can be done in a different way,   attending instead to the two parts of different order in the quantum corrections expansion,  i.e., in powers of $\hbar$.  Thus, one can split the full \1loop amplitude,  alternatively,  as follows:
\be
 \amp^{\full}= \amp^{\treeLtwoLfour}+\amp^{\loopLtwo} \, , 
\label{tree-oneloop}
\ee
where, the tree level amplitude, ${\cal O} (\hbar^0)$ , is: 
\be
\amp^{\treeLtwoLfour}=\amp^{\treeLtwo}+ \amp^{\treeLfour}\,, 
\label{EChLtree}
\ee
and the \1loop correction, ${\cal O} (\hbar^1)$,  is $\amp^{\loopLtwo}$.  Since, in our posterior analysis of the full \1loop results within the EChL we plan to compare them with the corresponding SM full results,  which we have also computed here in parallel,  it is then convenient to also fix the notation here for the SM case.  The full \1loop SM amplitude is then defined as the sum of the SM leading-order (LO) contribution,  that, in this case,  is the tree level part of ${\cal O} (\hbar^0)$, and the SM next-to-leading order (NLO) contribution, that is the \1loop part of ${\cal O} (\hbar^1)$.  In short, it is given by:
\be
\amp^{\fullSM}=\amp^{\treeSM}+ \amp^{\loopSM}\,. 
\label{SMfull}
\ee
To our knowledge,  the  SM computation presented here is the first full bosonic \1loop computation in the literature of $WZ \to WZ$ scattering  using the $R_\xi$  gauges,  and,  therefore,  we believe it is interesting by itself, in addition to be the obvious reference to compare our EChL computation with.
    
As already said,  our purpose in this paper is to organize the computation of the full \1loop amplitude in terms of the basic building blocks in any QFT which are the 1PI Green functions.
Taking into account that we work in generic $R_\xi$ gauges,  this procedure by means of 1PI functions has the advantage of requiring a more demanding renormalization program,  since all the off-shell effects of external legs in the 1PI functions must be considered.  Concretely,  in the next section we will fix our renormalization program in the $R_\xi$ gauges  that makes finite all the relevant 1PI Green functions for arbitrary external legs (off-shell) momenta and no transversality condition for the EW gauge bosons is applied.  The $\xi$-dependence of all the divergences  involved in these 1PI functions will be followed explicitly and,  in this sense,  the extraction of the new counterterms given by the EChL coefficients $a_i$  will be more complicated in these $ R_\xi$ gauges,  but they will provide an excellent check of the gauge invariance of these new counterterms.  We continue next with the presentation of the amplitude in terms of the 1PI functions and postpone all these renormalization issues for the next section,  \secref{sec-renorm}.

Generically,  the full amplitude can be organized in terms of the contributing 1PI functions as follows:
\be
\amp^{\full}=\amp^{(0)}+\amp_{1-leg}+\amp_{2-legs}+\amp_{3-legs}+\amp_{4-legs}+\amp_{res} \,,
\label{amp-by1PI}
\ee
where $\amp_{n-legs}$ means the $\mO(\hbar)$ contribution from the $n$-legs 1PI function to the amplitude,  and we have separated explicitly the extra contribution from the finite residues of the external gauge bosons $\amp_{res}$.  
Regarding the 1-leg 1PI function (i.e., the tadpole $\tadR$ of the Higgs boson),  {\it a priori},  it enters in many parts of the different diagrams contributing to the amplitude,  but as it will be seen in the next section,  one can set this renormalized Green function equals to zero by a convenient renormalization condition,  and then $\amp_{1-leg}$ simply vanishes.

\begin{figure}[H]
\begin{center}
    \includegraphics[width=.9\textwidth]{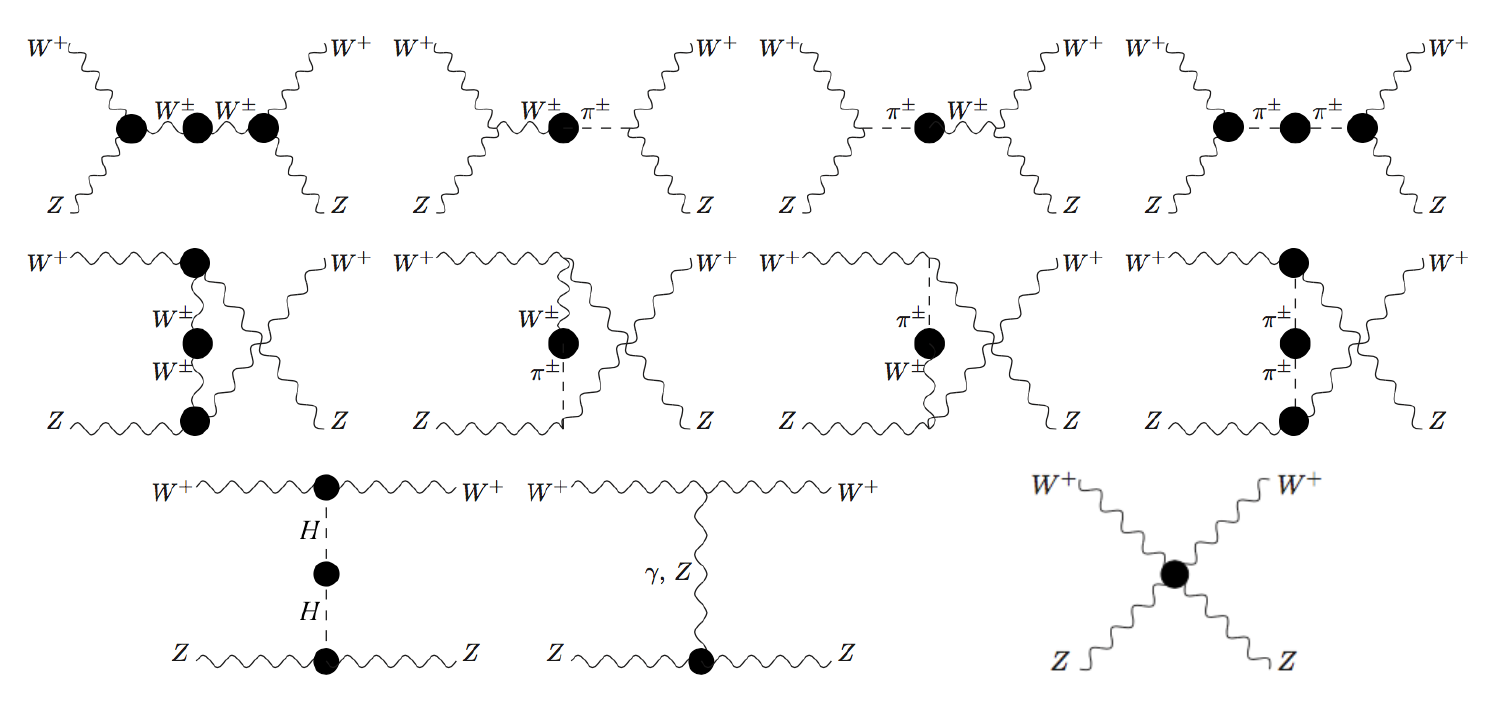} 
 \caption{Full 1PI functions (black balls) contributing to the full \1loop  amplitude $ \amp(WZ \to WZ)$ }
\label{1PIdiagsWZWZinSTUC}
\end{center}
\end{figure}

\begin{figure}[H]
\begin{center}
 \includegraphics[width=.85\textwidth]{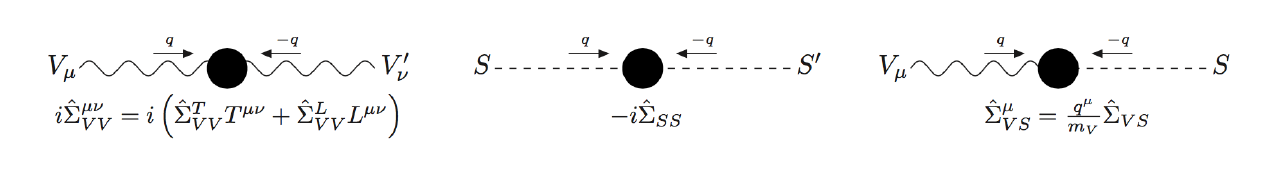}
\caption{Diagrammatic Lorentz conventions corresponding to the self-energies of two gauge bosons (left), two scalars (middle) and gauge boson with a scalar (right). Incoming momenta are understood.}
\label{our-SE-Lorentzconventions}
\end{center}
\end{figure}

We next specify the particular 1PI functions contributing to the amplitude of our interest.  For definitess,  here and in the following of this work,  we fix the notation for the momenta assignments and Lorentz indexes as follows: 
\be
W^+_\lambda(p_+)\,Z_\rho(p_0) \to W^+_\sigma(k_+)\,Z_\kappa(k_0)\,,
\label{WZscattering}
\ee
where $p_{+,0}$ and $k_{+,0}$ (with $p_+ +p_0 =k_+ +k_0$) are the incoming and outcoming momenta of the gauge bosons,  with polarization vectors $\epsilon_\lambda(p_+)\,,\epsilon_\rho(p_0)\,,\epsilon_\sigma(k_+)$ and $\epsilon_\kappa(k_0)$, respectively. 
Thus,  the amplitude $\amp$ can be written as:
\be
\amp = A^{\lambda\rho\sigma\kappa}\,\epsilon_\lambda(p_+)\epsilon_\rho(p_0)\epsilon_\sigma(k_+)^*\epsilon_\kappa(k_0)^* \,, 
\label{amp-rg4}
\ee
where the tensor amplitude with explicit Lorentz indexes is defined by $A^{\lambda\rho\sigma\kappa}$. 
The complete set of full 1PI functions and full propagators that contribute to the full \1loop amplitude 
$ \amp(WZ \to WZ)$ 
are displayed graphically in \figref{1PIdiagsWZWZinSTUC}.  These full \1loop functions are represented  in this figure  by black balls and are denoted by quantities with a hat,  in order  to be distinguished from the corresponding leading order quantities (without the hat).  These full functions include:  1) the full propagators,  $\propR^{HH}$,   $\propR^{\pi\pi}$, $\propR^{WW}$, $\propR^{W\pi}$ and $\propR^{\pi W}$; 2) the full 1PI vertex functions with three-legs,  $\greenfR_{WWZ}$, $\greenfR_{\pi WZ}$,   
$\greenfR_{HWW}$,  $\greenfR_{HZZ}$,  
$\greenfR_{ZZA}$ and $\greenfR_{ZZZ}$; and 3) the full 1PI vertex function with four-legs,  $\greenfR_{WZWZ}$. Notice, that some of these full functions contain contributions of both orders, ${\cal O} (\hbar^0)$ and ${\cal O} (\hbar^1)$,  whereas others are pure ${\cal O} (\hbar^1)$ functions.  For instance,  the functions 
$\propR^{W\pi}$, $\propR^{\pi W}$, $\greenfR_{ZZA}$ and $\greenfR_{ZZZ}$ get only NLO contributions since they vanish at LO.   Consequently, when providing the amplitude to a given order in powers of $\hbar$ one has to keep accordingly the same powers in the involved products of full functions.  Thus,  in products like  
$\greenfR_{WWZ}^{\mu\lambda\rho}\,\propR_{\mu\nu}^{WW}\,\greenfR_{WWZ}^{\nu\sigma\kappa}$ there are several contributions inside: 1)  LO contributions $\Gamma_{WWZ}^{\mu\lambda\rho}\,\Delta_{\mu\nu}^{WW}\,\Gamma_{WWZ}^{\nu\sigma\kappa}$ and 2) NLO contributions from the corresponding corrections in each of the full functions. 
On the other hand,  the propagators $\Delta^{AA}$, $\Delta^{ZZ}$ and the vertex function $\Gamma_{WWA}$,   enter just at leading order  in this computation.  
We can then simply read the various contributions by grouping them together,  for instance,  by $s$, $t$, $u$ and contact $c$ channels as follows:
\be
A^{\lambda\rho\sigma\kappa}= A_s^{\lambda\rho\sigma\kappa}+A_t^{\lambda\rho\sigma\kappa}+A_u^{\lambda\rho\sigma\kappa}+A_c^{\lambda\rho\sigma\kappa} \,, 
\label{ampbychannels}
\ee 
where the amplitudes by channels in terms of the full functions that enter at \1loop level read as follows:
\bear
i A_s^{\lambda\rho\sigma\kappa} &=& i\greenfR_{WWZ}^{\mu\lambda\rho}\,(-i)\propR_{\mu\nu}^{WW}\,i\greenfR_{WWZ}^{\nu\sigma\kappa} + i\Gamma_{WWZ}^{\mu\lambda\rho}\,\propR_{\mu}^{W\pi}\,i\Gamma_{\pi WZ}^{\sigma\kappa}  \nn\\
&&+ i\Gamma_{\pi WZ}^{\lambda\rho}\,\propR_{\nu}^{\pi W}\,i\Gamma_{WWZ}^{\nu\sigma\kappa} + i\greenfR_{\pi WZ}^{\lambda\rho}\,i\propR^{\pi\pi}\,i\greenfR_{\pi WZ}^{\sigma\kappa}  \nn\\
i A_t^{\lambda\rho\sigma\kappa} &=& i\greenfR_{HWW}^{\lambda\rho}\,i\propR^{HH}\,i\greenfR_{HZZ}^{\sigma\kappa} + i\Gamma_{WWA}^{\mu\lambda\rho}\,i\Delta_{\mu\nu}^{AA}\,i\greenfR_{ZZA}^{\nu\sigma\kappa} + i\Gamma_{WWZ}^{\mu\lambda\rho}\,i\Delta_{\mu\nu}^{ZZ}\,i\greenfR_{ZZZ}^{\nu\sigma\kappa}  \nn\\
i A_u^{\lambda\rho\sigma\kappa} &=& i\greenfR_{WWZ}^{\mu\lambda\kappa}\,(-i)\propR_{\mu\nu}^{WW}\,i\greenfR_{WWZ}^{\nu\sigma\rho} + i\Gamma_{WWZ}^{\mu\lambda\kappa}\,\propR_{\mu}^{W\pi}\,i\Gamma_{\pi WZ}^{\sigma\rho}  \nn\\
&&+ i\Gamma_{\pi WZ}^{\lambda\kappa}\,\propR_{\nu}^{\pi W}\,i\Gamma_{WWZ}^{\nu\sigma\rho} + i\greenfR_{\pi WZ}^{\lambda\kappa}\,i\propR^{\pi\pi}\,i\greenfR_{\pi WZ}^{\sigma\rho}  \nn\\
i A_c^{\lambda\rho\sigma\kappa} &=& i\greenfR_{WZWZ}^{\lambda\rho\sigma\kappa} \,.
\label{EChLfullbychannels}
\eear
For the forthcomming  section dealing with the renormalization program,  it is convenient to write the previous full propagators in terms of 1PI functions with two-legs, therefore,  in terms of the self-energies.  For the case of gauge boson propators,   we  introduce some convenient Lorentz tensors which are usefull to decompose them into their transverse ($T$) and longitudinal ($L$) parts by: 
\bear
-i\propR_{\mu\nu}^{WW}(q)&=&-i\propR_{T}^{WW}(q^2)T_{\mu\nu}-i\propR_{L}^{WW}(q^2)L_{\mu\nu} \, ,
\eear
where, 
\bear
T^{\mu\nu} &=& g^{\mu\nu}-\frac{q^\mu q^\nu}{q^2} \,\, \, ; \,\,\, L^{\mu\nu}=\frac{q^\mu q^\nu}{q^2}\,,
\label{Lorentz-structures}
\eear
which fulfil the following useful properties:
\bear
&& T_{\mu\nu} +L_{\mu\nu} = g_{\mu\nu} \,, \quad T^{\mu\nu}T_{\nu\rho} = T^{\mu}_{\rho} \,,\quad L^{\mu\nu}L_{\nu\rho} = L^{\mu}_{\rho}\,, \nn\\
&& \quad T^{\mu\nu}L_{\nu\rho} = 0 \,, \quad  T_{\mu\nu}q^\nu =0 \quad \,,  \quad 
L_{\mu\nu}q^\nu =q_\mu . 
\label{projectors}
\eear
The full self-energies, involving gauge bosons can similarly  be decomposed as:
\be
\SER^{\mu\nu}_{WW}(q) = \SER^T_{WW}(q^2)T^{\mu\nu} + \SER^L_{WW}(q^2)L^{\mu\nu} \,.
\ee
For the mixed self-energies connecting a $W$ with a $\pi$, we use the following decomposition:
\be
\SER^{\mu}_{W\pi}(q) = \frac{q^\mu}{m_{\rm W}}\SER_{W\pi}(q^2)\,,
\ee
and we have the corresponding one to the full propagator:
\be
\propR_{\mu}^{W\pi}(q) = \frac{q_\mu}{m_{\rm W}}\propR^{W\pi}(q^2)\,.
\ee
Then, the simple expressions of the full propagators in terms of the full self-energies,  $\SER$,  to be used at the \1loop level,  are summarized by:  
\bear
i\hat{\Delta}^{HH}(q^2) &=& i\Delta^{HH} + i\Delta^{HH}\,(-i)\SER_{HH}\,i\Delta^{HH} \,,  \nn\\
i\propR^{\pi\pi}(q^2) &=& i\Delta^{\pi\pi} + i\Delta^{\pi\pi}\,(-i)\SER_{\pi\pi}\,i\Delta^{\pi\pi} \,,  \nn\\
-i\propR^{WW}_T(q^2) &=& -i\Delta^{WW}_T -i\Delta^{WW}_T\,i\SER_{WW}^T\,(-i)\Delta^{WW}_T \,,  \nn\\
-i\propR^{WW}_L(q^2) &=& -i\Delta^{WW}_L -i\Delta^{WW}_L\,i\SER_{WW}^L\,(-i)\Delta^{WW}_L \,,  \nn\\
\propR^{W\pi}(q^2) &=& \Delta^{W\pi} -i\Delta^{WW}_L\,\SER_{W\pi}\,i\Delta^{\pi\pi} \,,
\label{oneloop-prop}
\eear
where all functions on the right hand side are functions of $q^2$ and we have used the conventions for signs, `$i$' factors, and Lorentz decompositions as shown in \figref{our-SE-Lorentzconventions} for the self-energies $\SER$.
The involved $R_\xi$ gauges propagators at  LO are summarized by:
\bear
i\Delta^{HH} &=& \frac{i}{q^2 -\mh^2} \,,\qquad -i\Delta^{WW}_T = \frac{-i}{q^2 -\mw^2} \,,  \nn\\
-i\Delta^{WW}_L &=& \frac{-i\xi}{q^2 -\xi\mw^2}  \,,\quad i\Delta^{\pi\pi} = \frac{i}{q^2 -\xi\mw^2} \,,\quad \Delta^{W\pi} = 0 \,.  \nn\\
-i\Delta^{AA}_T &=& \frac{-i}{q^2} \,,\quad -i\Delta^{AA}_L = \frac{-i\xi}{q^2}  \,,\quad -i\Delta^{ZZ}_T = \frac{-i}{q^2 -\mz^2} \,,\quad -i\Delta^{ZZ}_L = \frac{-i\xi}{q^2 -\xi\mz^2}  \,.
\label{propsLO}
\eear
As commented previously, only $\Delta^{HH}$, $\Delta^{WW}_T$, $\Delta^{WW}_L$ and $\Delta^{\pi\pi}$ are involved in the LO contribution to the amplitude in \eqref{EChLfull}.
On the other hand, the relevant vertex functions at LO  are: 
\bear
i\Gamma_{WWZ}^{\mu\nu\rho} &=& -i g \cw V^{\mu\nu\rho}  \,,\quad i\Gamma_{\pi WZ}^{\sigma\kappa} = \gY\sw\mw g^{\sigma\kappa} \,,  \nn\\
i\Gamma_{HWW}^{\lambda\rho} &=& i a g\mw g^{\lambda\rho}  \,,\quad i\Gamma_{HZZ}^{\sigma\rho} = i a g \frac{\mz}{\cw} g^{\sigma\rho} \,,  \nn\\
i\Gamma_{WZWZ}^{\lambda\rho\sigma\kappa} & =& -i g^2 \cw^2 S^{\lambda\sigma,\rho\kappa} \,,  \nn\\
i\Gamma_{WWA}^{\mu\nu\rho} &=& -i g \sw V^{\mu\nu\rho}  \,,\quad i \Gamma_{ZZA}^{\nu\sigma\kappa}=
i\Gamma_{ZZZ}^{\nu\sigma\kappa}=0 \,,
\label{vertexLO}
\eear
where the Lorentz structures $V^{\mu\nu\rho}$ and $S^{\lambda\sigma,\rho\kappa}$ are summarized in 
\eqref{self-gauge-vertex}.

Finally, for completeness, we end this section presenting  the computation of the LO amplitude using the $R_\xi$ gauges.  This can be easily done by plugging the corresponding LO functions of \eqref{propsLO} and \eqref{vertexLO} in \eqref{EChLfullbychannels}.  Namely, using $\Gamma$ instead of $\hat \Gamma$, and $\Delta$ instead of $\hat \Delta$. 
\begin{figure}[H]
\begin{center}
    \includegraphics[width=.85\textwidth]{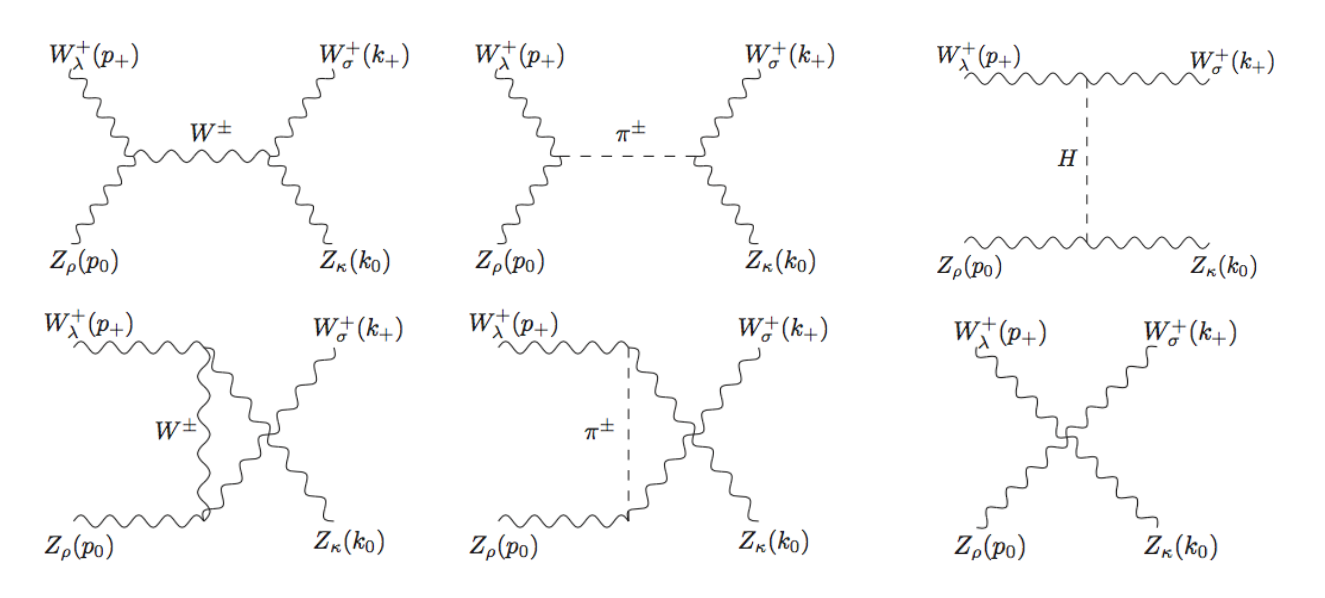}
\caption{Diagrams at LO contributing to $W^+Z\to W^+Z$ in the covariant $R_\xi$ gauges.}
\label{diagsWZWZatLO}
\end{center}
\end{figure}
Then, we get the result for the LO amplitude in the $R_\xi$ gauge,  which corresponds to the tree level diagrams in \figref{diagsWZWZatLO}.  It is given by:
\bear
\amp^{(0)} &=& \left( A^{(0)\,\lambda\rho\sigma\kappa}_s
+A^{(0)\,\lambda\rho\sigma\kappa}_t
+A^{(0)\,\lambda\rho\sigma\kappa}_u 
+A^{(0)\,\lambda\rho\sigma\kappa}_c
 \right)\,\epsilon_\lambda(p_+)\epsilon_\rho(p_0)\epsilon_\sigma(k_+)^*\epsilon_\kappa(k_0)^* \quad 
\label{amp-LO}
\eear
with:
\bear
A^{(0)\,\lambda\rho\sigma\kappa}_s &=& g^2\cw^2\frac{1}{s-\mw^2}V_{WWZ}^{\mu\lambda\rho}T_{\mu\nu}V_{WWZ}^{\nu\sigma\kappa}  \nn\\
&&+g^2\cw^2\frac{\xi}{s-\xi\mw^2}V_{WWZ}^{\mu\lambda\rho}L_{\mu\nu}V_{WWZ}^{\nu\sigma\kappa} -\gY^2\sw^2\mw^2\frac{1}{s-\xi\mw^2}g^{\lambda\rho}g^{\sigma\kappa} \,,  \nn\\
A^{(0)\,\lambda\rho\sigma\kappa}_t &=& -a^2g^2\frac{\mw\mz}{\cw} \frac{1}{t-\mh^2} \,g^{\lambda\sigma}g^{\rho\kappa} \,,  \nn\\
A^{(0)\,\lambda\rho\sigma\kappa}_u &=& g^2\cw^2\frac{1}{u-\mw^2}V_{WWZ}^{\mu\lambda\kappa}T_{\mu\nu}V_{WWZ}^{\nu\sigma\rho}  \nn\\
&&+g^2\cw^2\frac{\xi}{u-\xi\mw^2}V_{WWZ}^{\mu\lambda\kappa}L_{\mu\nu}V_{WWZ}^{\nu\sigma\rho} -\gY^2\sw^2\mw^2\frac{1}{u-\xi\mw^2}g^{\lambda\kappa}g^{\sigma\rho} \,,  \nn\\
A^{(0)\,\lambda\rho\sigma\kappa}_c&=& -g^2\cw^2 S^{\lambda\sigma,\rho\kappa} \,.
\label{amp-LO-bychannels}
\eear
Some more comments on this LO result are in order.  First, notice  that the corresponding result of the SM amplitude at LO is simply obtained from this same formula by setting $a=1$.
Second, we have checked explicitly the gauge invariance of the LO result,  namely,  that the dependence on $\xi$ disappears in the final amplitude as expected.  The cancellation of the $\xi$-dependent terms is achieved once the external gauge bosons are on-shell, i.e., by contracting the tensorial amplitude with their corresponding polarization vectors in $\amp^{(0)}$ of \eqref{amp-LO}.
Concretely,  the cancellation of the $\xi$-dependent terms occurs separately in the two channels $s$ and $u$,  and it happens between the two terms in the second lines (coming from the $W$  and GB propagators) of the corresponding  $s$ and $u$ amplitudes in \eqref{amp-LO-bychannels}.  More explicitly, using the transversality and on-shell conditions of the external gauge bosons one gets in the $s$ channel:
\bear
\hspace{-5mm}V_{WWZ}^{\mu\lambda\rho}L_{\mu\nu}V_{WWZ}^{\nu\sigma\kappa}\epsilon_\lambda(p_+)\epsilon_\rho(p_0)\epsilon_\sigma(k_+)^*\epsilon_\kappa(k_0)^* &=& \frac{(\mz^2-\mw^2)^2}{s}\epsilon(p_+)\cdot\epsilon(p_0)\,\epsilon(k_+)^*\cdot\epsilon(k_0)^*  
\eear
and the $\xi$-dependence from the two terms in the second line of the $s$ channel amplitude in \eqref{amp-LO-bychannels} then cancels.  The contribution from these two terms to ${\cal A}_s^{(0)}$  then finally reduce to:
\be
-g^2\frac{\sw^4}{\cw^2}\frac{\mw^2}{s}\epsilon(p_+)\cdot\epsilon(p_0)\,\epsilon(k_+)^*\cdot\epsilon(k_0)^* \, ,
\ee
consequently,  leading to a gauge invariant amplitude, as expected. 
The cancellation in the $u$ channel contribution,  ${\cal A}_u^{(0)}$ proceeds in a similar way.

\section{Renormalization program}
\label{sec-renorm}
\subsection{Generalities}
In this section we present our renormalization program to compute the renormalized 1PI functions within the EChL in covariant $R_\xi$ gauges which are the basic pieces in our computation of the VBS amplitudes.  Within a diagrammatic approach,  these renormalized 1PI functions,  denoted here generically by $\greenfR_{n-legs}$, can be decomposed into three pieces as follows: 
\be
\greenfR_{n-legs} =\greenfT_{n-legs}+\greenfL_{n-legs}+\greenfC_{n-legs} \,.
\label{fgreen}
\ee
where,  within the context of the EChL we are working with,  $\greenfT$ means contributions from tree level Lagrangian, $\mL_2 +\mL_4$,  $\greenfL$ are the contributions from the unrenormalized \1loop diagrams using the interaction vertices of $\mL_2$ only,  and $\greenfC$ summarizes the contributions from all the counterterms.  For the analytical computation of these three pieces we use the codes and tools specified next.  We have implemented our model with FeynRules~\cite{FeynRules}, generated and drawn the Feynman diagrams by FeynArts~\cite{FeynArts}, performed the main calculations with FormCalc and LoopTools~\cite{FormCalc-LT} and added some extra checks of the involved \1loop divergences  using FeynCalc~\cite{FeynCalc} and the code Package-X~\cite{packX}.
Also, a comparison with respect to the SM has been perfomed in parallel and a summary of the corresponding SM results are collected in \appref{App-SMcompu},  for completeness.

In the following of this section we describe the details of the renormalization program.  First we set the regularization and multiplicative renormalization prescriptions,  and fix the renormalization conditions.  Second,  we discuss the Slavnov-Taylor identity in these $R_\xi$ gauges.  Third,  we present the various contributions to the 1PI functions as well as the solutions for the counterterms involved.  Finally, we  derive the values of the renormalized EChL coefficients and set the corresponding renormalization group equations (RGEs) describing the running of 
these coefficients.  
\subsection{Regularization and Renormalization prescriptions}
\label{RR}
The regularization procedure of the loop contributions is performed with dimensional regularization~\cite{tHooft:1972tcz},   in $D=4-\epsilon$ dimensions,  as usual.  This method has the advantage that preserves all the relevant symmetries in the bosonic sector of the theory, including chiral invariance (remember that we do not consider the fermionic contributions and the Dirac $\gamma_5$ is not involved in this work).  Consequently,  the scale of dimensional regularization is set to $\mu$ and all the \1loop divergencies are expressed in terms of :
 \be
\div=\frac{2}{\epsilon}-\gamma_E+\log(4\pi) \,.
\label{div-definition}
\ee
Regarding the renormalization procedure, the counterterms of all the parameters and fields appearing in the tree level Lagrangian, $\mL_2+\mL_4$,  are generated by the usual multiplicative renormalization prescription that relates the bare quantities (here denoted by a specific sub- or super-script with a label $0$) and the renormalized ones (here with no specific sub- or super-script labels).  In our present case,  the relevant relations are summarized as follows:
\bear
H_0 &= & \sqrt{\Zf_H}H \,, \quad B_{0\,\mu} = \sqrt{\Zf_B}B_{\mu}\,, \quad W_{0\,\mu}^{1,2,3} = \sqrt{\Zf_W}W_{\mu}^{1,2,3} \,, \quad \pi_0^{1,2,3} = \sqrt{\Zf_\pi}\pi^{1,2,3} \,,  \nn\\
\vev_0 &= & \sqrt{\Zf_\pi} (\vev +\delta\vev)\,,
\quad \lambda_0 = \Zf_H^{-2}(\lambda +\delta \lambda)\,, \nn\\
\gY_0 &= & \Zf_B^{-1/2}(g' +\delta \gY)\,,\quad
g_0 = \Zf_W^{-1/2}(g +\delta g)\,, \quad \xi_{1,2}^0 = \xi (1+\delta\xi_{1,2})\,,  \nn\\
a^0&=&a+\delta a\,,\quad b^0=b+\delta b\,,\quad \kappa_{3,4}^0=\kappa_{3,4}+\delta\kappa_{3,4}\,,\quad a_i^0=a_i+\delta a_i\,,
\label{EChL-renorm-factors}
\eear
where the $Z_i$ renormalization constants are set as usual to $\Zf_i=1+\delta\Zf_i$. 
Some comments on this \eqref{EChL-renorm-factors} are in order.  Our final results for both the renormalized 1PI functions and the VBS scattering amplitudes are expressed in terms of the renormalized quantities,  $m_W$,  $m_Z$, $m_H$, 
 $g$,  $g'$, $v$,  $a$, $a_i$'s, etc.  
It should be noticed that for the $WZ\to WZ$ case,  the renormalization of the $b$,   $\kappa_{3,4}$ and $\lambda$ parameters do not enter and,  therefore,  the associated counterterms will not be present in the renormalized 1PI functions.  Besides,  the ghost counterterms do not enter either, and we omit to show them for shortness.  In contrast,  the renormalization of the covariant gauge parameters do enter and are relevant for the present paper.  
Notice that in \eqref{EChL-renorm-factors} we have set a common renormalized $\xi$ parameter for all the involved EW gauge bosons.  The associated counterterms $\delta \xi_{1,2}$, introduced in \eqref{charged-LGF-bare},  play an important role in the renormalization of the charged unphysical propagators.  We refer here to the charged unphysical propagators as in reference~\cite{Bohm:1986rj}, namely, formed by the charged Goldstone bosons $\pi^\pm$ and the gauge bosons $W^\pm$.  The discussion on the renormalization of the involved functions for this charged unphysical sector, $\SER_{WW}^L$, $\SER_{W\pi}$ and $\SER_{\pi\pi}$, and their relations will be presented in the subsection \ref{sec-STidentity}.  
Finally, one can easily relate the corresponding counterterms for the EW parameters in the physical basis with the previous counterterms in \eqref{EChL-renorm-factors}.  In particular,
\bear
\delta\Zf_A &=&\cw^2\delta\Zf_B +\sw^2\delta\Zf_W\,,  \nn\\
\delta\Zf_Z &=&\sw^2\delta\Zf_B +\cw^2\delta\Zf_W\,,  \nn\\
\delta\Zf_{ZA} &=&\sw\cw(\delta\Zf_W -\delta\Zf_B)=\frac{\sw\cw}{\cw^2-\sw^2}(\delta\Zf_Z -\delta\Zf_A)\,,  \nn\\
\delta\mw^2 &=& \mw^2\left( -\delta\Zf_W +\delta\Zf_\pi +2\frac{\delta g}{g} +\frac{2\delta\vev}{\vev} \right) \,,  \nn\\
\delta\mz^2 &=& \mz^2\left( -\delta\Zf_Z +\delta\Zf_\pi +2\cw^2\frac{\delta g}{g} +2\sw^2\frac{\delta \gY}{\gY} +\frac{2\delta\vev}{\vev} \right) \,,  \nn\\
\delta \mh^2 &=&\mh^2\left(-2\delta\Zf_H+\frac{\delta\Zf_\pi}{2}+\frac{\delta\lambda}{\lambda}+\frac{2\delta\vev}{\vev}\right) \, ,
\label{CTphys-relations}
\eear
where,  our conventions for the mass counterterms are:
\bear
 \mh^{0\,2} &=& \mh^2 +\delta \mh^2 \, ,  \,  \mw^{0\,2} = \mw^2 +\delta \mw^2\, ,  \,  \mz^{0\,2} = \mz^2 +\delta \mz^2 \,.
 \label{CTmass}
\eear
Finally,  regarding the renormalization conditions we adopt here a hybrid prescription 
in which we choose the on-shell (OS) scheme for the EW parameters in the lowest order Lagrangian $\mL_2$ and the $\overline{MS}$ scheme for all the EChL coefficients $a$ and $a_i$.  These particular conditions will provide the specific values of all the counterterms involved in the present computation.  Concretely,  our renormalization conditions read as follows:

\begin{itemize}
    \item Vanishing (Higgs) tadpole:
        \be
        \hat{T}=0 \,.
        \label{condOS1}
        \ee
    \item The pole of the renormalized propagator of the Higgs boson lies at $\mh^2$ and the corresponding residue is equal to 1:
        \be
        {\rm Re}\left[ \SER_{HH}(\mh^{2}) \right] =0 \,,\quad {\rm Re}\left[ \frac{d\SER_{HH}}{dq^2}(\mh^{2}) \right] =0 \,.
        \label{condOS2}
        \ee
    \item Properties of the photon: residue equal one; no $A-Z$ mixing propagators; and the electric charge defined like in QED, since there is a remnant $U(1)_{\rm em}$ electromagnetic gauge symmetry:
        \be
        {\rm Re}\left[ \frac{d\SER^T_{AA}}{dq^2}(0) \right] =0 \,,\quad \SER^T_{ZA}(0) =0  \,,\quad \greenfR^\mu_{\gamma e e}\vert_{\rm OS}=i e \gamma^\mu \,. 
        \label{condOS3}
        \ee    
    \item The poles of the transverse renormalized propagators of the $W$ and $Z$ bosons lie at $q^2=\mw^2$ and $q^2=\mz^2$,  respectively:
        \be
        {\rm Re}\left[ \SER_{WW}^T(\mw^{2}) \right] =0 \,,\quad {\rm Re}\left[ \SER_{ZZ}^T(\mz^{2}) \right] =0 \,.
        \label{condOS4}
        \ee
 \item The poles of the renormalized propagators in the unphysical charged sector $\{W^\pm,\pi^\pm\}$ lie at $q^2=\xi\mw^2$. Therefore:
        \be
        {\rm Re}\left[ \SER_{WW}^L(\xi\mw^{2}) \right] =0 \,,\quad {\rm Re}\left[ \SER_{\pi\pi}(\xi\mw^{2}) \right] =0 \,.  
        \label{condOS5}
        \ee
    \item $\overline{MS}$ scheme for all the involved EChL coefficients. \\
    In particular for $a$,  $b$,  $\kappa_{3,4}$  in \eqref{eq-L2} and the $a_i$'s  in \eqrefs{eq-L4-Longhitano}{eq-L4-withH}.  
\end{itemize}

It is important to stress that the previous renormalization conditions on all the EChL parameters determine both the divergent and finite parts of all  the counterterms. In particular,  the physical masses $m_W$, $m_Z$ and $m_H$  coincide with the corresponding renormalized ones in the OS scheme.  On the other hand, notice that the residue for the Higgs and photon fields are set to one in the previous conditions,  but the residues ${\cal Z}_{W,Z}$, for the gauge bosons $W$ and $Z$, are not set to one.  These finite residues  are given,  respectively,  by:
\be
{\cal Z}_W=1+{\rm Re}\left[ \frac{d\SER^T_{WW}}{dq^2}(\mw^2) \right] \quad  ; \quad    {\cal Z}_Z=1+{\rm Re}\left[ \frac{d\SER^T_{ZZ}}{dq^2}(\mz^2) \right] \,.
\label{WZresidues}
\ee
Therefore,  they contribute {\it via}  $\amp_{res}$ in \eqref{amp-by1PI} to the $WZ$ scattering as follows:
\be
\amp_{res}=\left({\rm Re}\left[ \frac{d\SER^T_{WW}}{dq^2}(\mw^2) \right]+{\rm Re}\left[ \frac{d\SER^T_{ZZ}}{dq^2}(\mz^2) \right]\right)\amp^{(0)} \,,
\label{ampRes}
\ee
since each external $W(Z)$ provides a factor ${\cal Z}^{1/2}_{W(Z)}$ to the observable $S$ matrix.
 

\subsection{Summary of contributions to the renormalized 1PI functions}
\label{1PIR}
\begin{figure}[H]
\begin{center}
    \includegraphics[width=.9\textwidth]{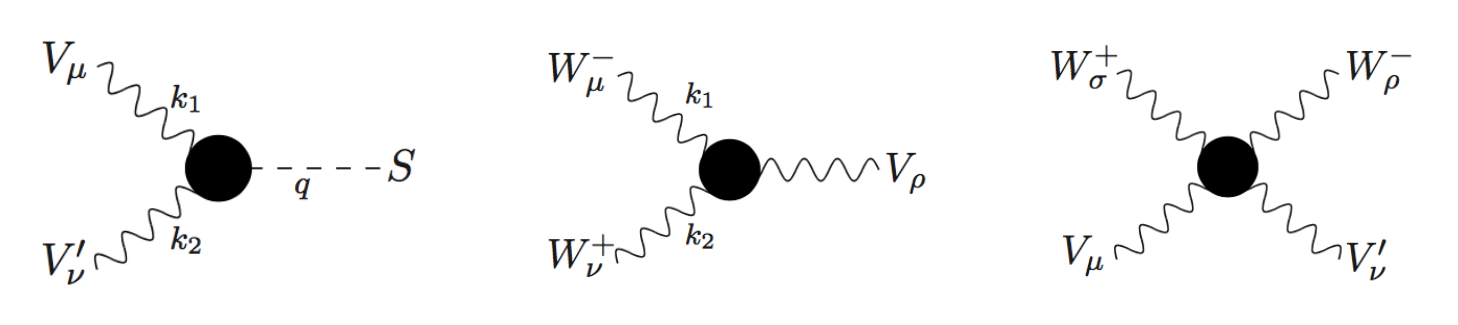}
\caption{Diagrammatic Lorentz conventions corresponding to the 3-legs ($i\greenfR_{SVV'}^{\mu\nu}$, $i\greenfR_{WWV}^{\mu\nu\rho}$) and 4-legs ($i\greenfR_{VV'WW}^{\mu\nu\rho\sigma}$) Green functions, relevant for our observables of interest.  The generic scalar here, $S$,  can be a Goldstone boson or a Higgs boson. The  vector bosons $V$ here,  refer generically to the corresponding  EW gauge boson,  $W$, $Z$, $\gamma$.  All momenta are incoming.}
\label{our-VertexBox-Lorentzconventions}
\end{center}
\end{figure}

We collect here the various contributions to the  renormalized 1PI functions, including the specific counterterms according to our previous prescriptions especified in subsection \ref{RR}.  For the 2-legs functions and the corresponding self-energies 
we use the conventions defined in  \figref{our-SE-Lorentzconventions}.  For the 3-legs and 4-legs functions we use the conventions defined in \figref{our-VertexBox-Lorentzconventions}.  All these functions depend on the momenta of the external legs which are generically off-shell.   By using momentum conservation one can reduce the independent momenta to 2 and 3 for the 3-legs and 4-legs functions respectively.  
For the generic 3-legs function
with one external scalar $S$ (with momentum $q$) and two EW gauge bosons $V_\mu$, $V'_\nu$ (with momenta $k_1$, $k_2$) we choose the gauge bosons momenta as independent variables and the generic function can be decomposed into Lorentz structures as $g^{\mu\nu}$, $k_2^\mu k_1^\nu$, $k_1^\mu k_1^\nu$, $k_1^\mu k_2^\nu$ and $k_2^\mu k_2^\nu$.  Notice that the last three structures can be omitted if we impose the transversality conditions to the gauge bosons, i.e. $k_i\cdot\epsilon(k_i)=0$.  However we keep them here since we are interested on the full vertex functions off-shell.
Similarly, we choose the momenta of the $W$ bosons as independent variables of the $\greenfR_{WWV}^{\mu\nu\rho}$ (middle panel of \figref{our-VertexBox-Lorentzconventions}) resulting in the Lorentz structures $g^{\mu\nu}k_1^\rho$, $g^{\mu\nu}k_2^\rho$, $g^{\nu\rho}k_1^\mu$, $g^{\nu\rho}k_2^\mu$, $g^{\rho\mu}k_1^\nu$ and $g^{\rho\mu}k_2^\nu$.  For the case of $\greenfR_{ZZA}$ and $\greenfR_{ZZZ}$ we choose the $Z$ bosons momenta.
The 4-legs vertex functions of the right panel have Lorentz structures $g^{\mu\nu}g^{\rho\sigma}$, $g^{\mu\rho}g^{\nu\sigma}$ and $g^{\mu\sigma}g^{\nu\rho}$ even at $\mO(\hbar)$ as we found explicitly.

We list in the following the various contributions to the renormalized  $n$-legs functions.  We include not only the 1PI functions that are needed for the present computation of $WZ \to WZ$ scattering,  already mentioned in \secref{diag-1pi},  but also some additional ones since we wish to determine separately the renormalization of all the involved EW parameters and  EChL coefficients in \eqrefs{eq-L4-Longhitano}{eq-L4-withH}. 

\subsubsection{1-leg function (Higgs tadpole)}
\bear
i\hat{T} &=& iT^{\rm Loop} -i\delta T\,,  \quad
\delta T = (-\delta \mh^2/2 +2\lambda\vev^2 \left(\frac{\delta\vev}{\vev}+\frac{\delta\Zf_\pi}{2}\right) +\delta\lambda \vev^2 -2\lambda \vev^2\delta\Zf_H)\vev
\label{Htadpole}
\eear 

\subsubsection{2-legs functions}

\bear
-i\hat{\Sigma}_{HH}(q^2) &=& -i\Sigma_{HH}^{\rm Loop}(q^2) +i\left(\delta\Zf_H (q^2-\mh^2) -\delta\mh^2\right) +i\frac{2a_{\Box\Box}}{\vev^2}q^4 \,,
\label{HH}
\eear
\bear
i\hat{\Sigma}_{WW}^T (q^2) &=& i\Sigma_{WW}^{T\,{\rm Loop}} (q^2) -i\left( \delta\Zf_W\left( q^2 -\mw^2 \right) -\delta\mw^2 \right)\,,  \nn\\
i\SER_{AA}^T (q^2) &=& i\Sigma_{AA}^{T\,{\rm Loop}} (q^2) -i\delta\Zf_A q^2 -ig^2\sw^2 (a_8-2a_1)q^2 \,,  \nn\\
i\SER_{ZZ}^T (q^2) &=& i\Sigma_{ZZ}^{T\,{\rm Loop}} (q^2) -i\left( \delta\Zf_Z\left( q^2 -\mz^2 \right) -\delta\mz^2\right)  \nn\\ 
&&-i\left(2\gY^2\mz^2a_0+(2g^2\sw^2 a_1+g^2\cw^2a_8+(g^2+\gY^2)a_{13}) q^2 \right)\,,  \nn\\
i\SER_{ZA}^T (q^2) &=& i\Sigma_{ZA}^{T\,{\rm Loop}} (q^2) -i\left( \delta\Zf_{ZA}q^2 -\mz^2\sw\cw\left(\delta g/g -\delta\gY/\gY \right) \right)  \nn\\
&&-i \left(g^2\sw\cw a_8-g\gY(\cw^2-\sw^2) a_1 \right)q^2\,.
\label{SE-phys-renorm}
\eear
\bear
i\SER_{WW}^L (q^2) &=& i\Sigma_{WW}^{L\,{\rm Loop}} (q^2) +i\left( -\frac{1}{\xi}\left( q^2 -\xi\mw^2 \right)\delta\Zf_W +\delta\mw^2 +q^2\frac{\delta\xi_1}{\xi}\right) -iq^2g^2a_{11} \,, \nn\\
\SER_{W\pi} (q^2) &=& \Sigma_{W\pi}^{\rm Loop} (q^2) +\frac{\delta\xi_2-\delta\xi_1}{2}\mw^2 +q^2g^2a_{11} \,,  \nn\\
-i\SER_{\pi\pi} (q^2) &=& -i\Sigma_{\pi\pi}^{\rm Loop} (q^2) +i\left( \left( q^2 -\xi\mw^2 \right)\delta\Zf_\pi -\xi\delta\mw^2 -\xi\mw^2\delta\xi_2 \right) -i\frac{g^2}{\mw^2}q^4a_{11}\,.
\label{charged-unphys-SE-renorm}
\eear

\subsubsection{3-legs functions}

\bear
i\greenfR_{HAA}^{\mu\nu} &=& i\greenfL_{HAA} +i\frac{g^2\sw^2}{\vev} (a_{HBB}+a_{HWW}-a_{H1}+a_{H8}/2)((q^2-k_1^2-k_2^2)\gmunu-2 \kVmu \kVnu) \,,  \nn\\
i\greenfR_{HAZ}^{\mu\nu} &=& i\greenfL_{HAZ} +i\frac{a g^2\sw \vev}{2 \cw} \left(\frac{\delta g}{g}-\frac{\delta \gY}{\gY}\right)\gmunu  \nn\\
&&\hspace{-15mm} +i\frac{g^2\sw }{\vev\cw} (-\sw^2a_{HBB}+\cw^2a_{HWW}+(1/2-\cw^2)a_{H1}+\cw^2a_{H8}/2)((q^2-k_1^2-k_2^2)\gmunu-2 \kVmu \kVnu)  \nn\\
&&+i\frac{g^2\sw }{4\vev\cw}(a_{d1}+a_{d2}+2a_{d4})((q^2+k_1^2-k_2^2)\gmunu-2 (\kTmu+\kVmu) \kVnu) \,,  \nn\\
i\greenfR_{HZZ}^{\mu\nu} &=& ia\frac{g^2\vev}{2\cw^2}\gmunu +i\greenfL_{HZZ} +\frac{ a g^2 \vev}{2 \cw^2}\left(\frac{\delta a}{a}+\frac{2 \delta g}{g}\cw^2+\frac{2 \delta \gY}{\gY}\sw^2+\frac{\delta\vev}{\vev}+\frac{\delta\Zf_H}{2}+\frac{\delta\Zf_\pi}{2}\right)\gmunu  \nn\\
&& -i\frac{g^2}{2\vev\cw^2}\left(\left(4\sw^2\mz^2a_{H0} +\left(-2\cw^4a_{HWW} -2\sw^4a_{HBB} -2\sw^2\cw^2a_{H1} -\cw^4a_{H8} -a_{H13} \right.\right.\right.  \nn\\
&&\left.\left.\left.\hspace{15mm} +\sw^2a_{d1} -\cw^2a_{d2} - 2\cw^2a_{d4} - 4\sw^2\frac{\mz^2}{\vev^2}a_{\Box 0}-2a_{\Box{\cal V}{\cal V}}\right)q^2 \right.\right.  \nn\\
&&\left.\left.\hspace{15mm}+(2\cw^4a_{HWW}+2\sw^4a_{HBB}+2\sw^2\cw^2a_{H1}+\cw^4a_{H8}+a_{H13})(k_1^2+k_2^2)\right) \gmunu \right.  \nn\\
&&\left.\hspace{10mm}+\left(-\sw^2a_{d1}+\cw^2a_{d2}+a_{d3}+2\cw^2a_{d4}+2a_{d5}\right)\left(\kTmu\kVnu+\kVmu\kTnu\right) \right.  \nn\\
&&\left.\hspace{10mm}+2\left(a_{d3}+2a_{d5}-a_{H11}\right)\kTmu\kTnu \right.  \nn\\
&&\left.\hspace{-3mm}+2\left(2\cw^4a_{HWW}+2\sw^4a_{HBB}+2\sw^2\cw^2a_{H1}+\cw^4a_{H8}-\sw^2a_{d1}+\cw^2a_{d2}+2\cw^2a_{d4}\right)\kVmu\kVnu \right) \,,  \nn\\
i\greenfR_{HW^-W^+}^{\mu\nu} &=& ia\frac{g^2\vev}{2}\gmunu +i\greenfL_{HWW} +i\frac{ a g^2 \vev}{2}\left(\frac{\delta a}{a}+\frac{2 \delta g}{g}+\frac{\delta\vev}{\vev}+\frac{\delta\Zf_H}{2}+\frac{\delta\Zf_\pi}{2}\right)\gmunu  \nn\\
&& -i\frac{g^2}{2\vev}\left(\left(-(2a_{HWW}+a_{d2}+2a_{\Box{\cal V}{\cal V}})q^2+2a_{HWW}(k_1^2+k_2^2)\right) \gmunu \right.  \nn\\
&&\left.\hspace{12mm}+\left(a_{d2}+a_{d3}\right)\left(\kTmu\kVnu+\kVmu\kTnu\right)+2\left(a_{d3}-a_{H11}\right)\kTmu\kTnu+2\left(2a_{HWW}+a_{d2}\right)\kVmu\kVnu \right) \,,  \nn\\
i\greenfR_{\pi^+ W^-A}^{\mu\nu} &=& -\frac{g\gY \vev\cw}{2}\gmunu +i\greenfL_{\pi WA} -\frac{g\gY \vev\cw}{2}\left(\frac{\delta g}{g}+\frac{\delta\gY}{\gY}+\frac{\delta\vev}{\vev}+\delta\Zf_\pi\right)\gmunu  \nn\\
&& +\frac{g^2\sw}{\vev} \left(\left(-(a_1-a_2+a_3-a_8+a_9-2a_{11})q^2\right.\right.  \nn\\
&& \left.\left.\hspace{20mm}+(a_1-a_2+a_3-a_8+a_9)k_1^2+(a_1+a_2-a_3-a_8-a_9)k_2^2\right) \gmunu \right)  \nn\\
&& \left.\hspace{-1mm}+4a_{11}\kTmu\kVnu+2a_{11}\kTmu\kTnu+2\left(a_1-a_2+a_3-a_8+a_9\right)\kVmu\kVnu-2(a_2-a_3-a_9)\kVmu\kTnu\right) \,,  \nn\\
i\greenfR_{\pi^+ W^-Z}^{\mu\nu} &=& \frac{g\gY \vev\sw}{2}\gmunu +i\greenfL_{\pi WZ} +\frac{g\gY \vev\sw}{2}\left(\frac{\delta g}{g}+\frac{\delta\gY}{\gY}+\frac{\delta\vev}{\vev}+\delta\Zf_\pi\right)\gmunu  \nn\\
&& -\frac{ g^2}{\vev\cw} \left(\left(4\sw^2\mz^2a_0-(\sw^2a_1-\sw^2a_2+\sw^2a_3+\cw^2a_8-\cw^2a_9-2\sw^2a_{11}+2a_{12}+a_{13})q^2 \right.\right.  \nn\\
&&\left.\left.\hspace{18mm}+(\sw^2a_1-\sw^2a_2-(1+\cw^2)a_3+\cw^2a_8-\cw^2a_9+a_{13})k_1^2  \right.\right.  \nn\\
&&\left.\left.\hspace{18mm}+(\sw^2a_1+\sw^2a_2+(1+\cw^2)a_3+\cw^2a_8+\cw^2a_9+a_{13})k_2^2  \right) \gmunu\right.  \nn\\
&&\left.\hspace{12mm} +2\left(a_3+(\sw^2-\cw^2)a_{11}-a_{12}\right)\kTmu\kVnu +2(\sw^2a_{11}-a_{12})\kTmu\kTnu\right.  \nn\\
&&\left.\hspace{12mm} +2\left(-\sw^2a_2-\cw^2a_3-\cw^2a_9+a_{11}-2a_{12}+a_{13})\right)\kVmu\kVnu\right.  \nn\\
&&\left.\hspace{12mm} +2\left(\sw^2a_1-\sw^2a_2+\sw^2a_3+\cw^2a_8-\cw^2a_9+a_{13})\right)\kVmu\kVnu\right) \,,
\label{SVV-renorm}
\eear
%
\bear
i\greenfR_{W^-W^+A}^{\mu\nu\rho} &=& -ig\sw V^{\mu\nu\rho} +i\greenfL_{WWA} -ig \sw V^{\mu\nu\rho}  \nn\\
&& +ig^3\sw \left(\gmunu(a_{11}k_1^\rho-(a_1-a_2+a_3-a_8+a_9-a_{11})k_2^\rho) \right.  \nn\\
&&\left.\hspace{10mm}+g^{\rho\nu}(a_{11}k_1^\mu+(a_1-a_2+a_3-a_8+a_9)k_2^\mu)\right) \,,  \nn\\
%
i\greenfR_{W^-W^+Z}^{\mu\nu\rho} &=& -ig\cw V^{\mu\nu\rho} +i\greenfL_{WWZ} -ig\cw V^{\mu\nu\rho}  \nn\\
&& \hspace{-18mm}+i\frac{g^3}{\cw} \left(\gmunu((a_3-\sw^2a_{11}+a_{12})k_1^\rho+(\sw^2a_1-\sw^2a_2-\cw^2a_3+\cw^2a_8-\cw^2a_9-\sw^2a_{11}+a_{12}+a_{13})k_2^\rho) \right.  \nn\\
&&\left.\hspace{-5mm}+g^{\nu\rho}((a_3-\sw^2a_{11}+a_{12})k_1^\mu-(\sw^2a_1-\sw^2a_2-(1+\cw^2)a_3+\cw^2a_8-\cw^2a_9+a_{13})k_2^\mu) \right.  \nn\\
&&\left.\hspace{-5mm}+g^{\rho\mu}(-2a_3k_1^\nu-a_3k_2^\nu)\right) \,,  \nn\\
i\greenfR_{ZZA}^{\mu\nu\rho} &=& i\greenfL_{ZZA}  \,,  \nn\\
i\greenfR_{ZZZ}^{\mu\nu\rho} &=& i\greenfL_{ZZZ} \,.
\label{WWV-renorm}
\eear
\subsubsection{4-legs functions}

\bear
i\greenfR_{AAW^-W^+}^{\mu\nu\rho\sigma} &=& -ig^2\sw^2 S^{\mu\nu,\rho\sigma} +i\greenfL_{AAWW} -ig^2\sw^2 S^{\mu\nu,\rho\sigma}  \nn\\
&& -ig^4\sw^2a_{11}(g^{\mu\rho}g^{\nu\sigma}+g^{\mu\sigma}g^{\nu\rho}) \,,  \nn\\
i\greenfR_{AZW^-W^+}^{\mu\nu\rho\sigma} &=& -ig^2\sw\cw S^{\mu\nu,\rho\sigma} +i\greenfL_{AZWW} -ig^2\sw\cw S^{\mu\nu,\rho\sigma}  \nn\\
&& +i\frac{g^4\sw}{\cw}\left(2a_3\gmunu g^{\rho\sigma}-(-a_3+\sw^2 a_{11}-a_{12})(g^{\mu\rho}g^{\nu\sigma}+g^{\mu\sigma}g^{\nu\rho})\right) \,,  \nn\\
i\greenfR_{ZZW^-W^+}^{\mu\nu\rho\sigma} &=& -ig^2\cw^2 S^{\mu\nu,\rho\sigma} +i\greenfL_{ZZWW} -ig^2\cw^2 S^{\mu\nu,\rho\sigma}  \nn\\
&& +i\frac{g^4}{\cw^2}\left(2(2\cw^2a_3+a_5+a_7)\gmunu g^{\rho\sigma}\right.  \nn\\
&&\left. \hspace{12mm}-(2\cw^2a_3-a_4-a_6+\sw^4 a_{11}-2\sw^2a_{12})(g^{\mu\rho}g^{\nu\sigma}+g^{\mu\sigma}g^{\nu\rho})\right) \,,  \nn\\
i\greenfR_{W^-W^+W^-W^+}^{\mu\nu\rho\sigma} &=& ig^2 S^{\mu\rho,\nu\sigma} +i\greenfL_{WWWW} +ig^2 S^{\mu\rho,\nu\sigma}  \nn\\
&& +i\frac{g^4}{\cw^2}\left(2(-2a_3+a_4+a_8-2a_9+2a_{13})g^{\mu\rho}g^{\nu\sigma}\right.  \nn\\
&&\left. \hspace{12mm}-(-2a_3-a_4-2a_5+a_8-2a_9+2a_{13})(\gmunu g^{\rho\sigma}+g^{\mu\sigma}g^{\nu\rho})\right) \,.
\label{VVWW-renorm}
\eear

Notice that,  to simplify the notation,  in all the previous equations,  \eqref{Htadpole} through \eqref{VVWW-renorm},  we have dropped the superindex $0$ in all the bare $a_i^0$ coefficients which are the ones entering in these equations.  Remember that these bare coefficients $a_i^0$ are the sum of the renormalized coefficients plus the counterterms,  according to our prescription defined in \eqref{EChL-renorm-factors}.  Therefore,  the $a_i$ coefficients  appearing  in the previous equations,  \eqref{Htadpole} through \eqref{VVWW-renorm},  really mean  $a_i+\delta a_i$,  where $a_i$ is the rermormalized EChL coefficient and $\delta a_i$ is the corresponding counterterm.

Regarding the contributions from the \1loop diagrams to the Green functions (generated by $\mL_2$),  we classify them generically into four categories accordingly to the particles in the loops: i) loops with just scalar bosons, i.e., with $H$ and/or $\pi$ (called `chiral' loops in short),  ii) loops with both gauge and scalar bosons (called `mix' in short reference to mixed loops), iii) loops with only gauge bosons (called `gauge' loops in short), and iv) loops with only ghosts (called `ghost' loops in short). Then we compute the loops contributions in the $R_\xi$ gauges,  generically called here by $\greenfL_{n-legs}$,  as the sum of all these loops:
\be
\greenfL_{n-legs}=\Gamma^{\rm{chiral}}_{n-legs}+\Gamma^{\rm{mix}}_{n-legs}+\Gamma^{\rm{gauge}}_{n-legs}+\Gamma^{\rm{ghost}}_{n-legs} \,.
\label{amp-classification}
\ee
For this purpose, we have calculated analytically the many Feynman loop diagrams (there are more than 500 in the computation of the $WZ \to WZ$ scattering amplitude) by using Mathematica~\cite{WMathematica}. More precisely, we have implemented our model with FeynRules~\cite{FeynRules}, generated and drawn all the Feynman diagrams by FeynArts~\cite{FeynArts}, performed the main calculations with FormCalc and LoopTools~\cite{FormCalc-LT} and incorporate some additional checks using FeynCalc~\cite{FeynCalc} and Package-X~\cite{packX}.  The generic loop diagrams entering in the specific 1PI functions that participate in the present computation of  the $WZ \to WZ$ scattering amplitude are summarized in Appendix \appref{App-oneloopdiag}. 

On the other hand,  the loop diagrams contributing to the 1-leg funtion (EChL Higgs tadpole), not explicitly drawn, are the loops with just scalars ($H$ and $\pi$), and the loops with just gauge bosons ($W$ and $Z$).  
The analytical result for these loop contributions to the Higgs tadpole in the $R_\xi$ gauges is:
\be
iT^{\rm Loop} = i\frac{1}{32\pi^2\vev}(3\kappa_3\mh^2A_0(\mh^2)+2a(6\mw^2A_0(\mw^2)-4\mw^4+3\mz^2A_0(\mz^2)-2\mz^4)) \,,
\label{TadLoop}
\ee
where $A_0(m^2)=\left(\Delta_\epsilon+\log(\mu^2/m^2)+1\right)m^2$.

At this point, it is important to remark that the Higgs tadpole in the EChL is $\xi$-independent, in contrast to the SM case. Therefore, the tadpole is a gauge invariant quantity within the EChL. This interesting feature is a consequence of the Higgs boson being a singlet field in this non-linear EFT approach,  in contrast to the SM (and SMEFT) case where the Higgs is a component of a doublet field.  Furthermore,  the Higgs field does not couple to the ghost fields in the EChL (in contrast to the linear realization of SM and SMEFT where the Higgs does couple to the ghosts), as we have seen in the presentation of the EChL Feynman rules in \secref{sec-EChL}.  Therefore,  the ghosts loops do not participate in this 1-leg function and also decouple from other 1PI functions with external $H$ fields.  The comparison with the SM case is summarized in the \apprefs{App-SMcompu}{App-tadpole}.

\subsection{Divergences of the \1loop contributions}
Due to the lengthy full analytical results of all the involved 1PI functions to \1loop in the $R_\xi$ gauge,   we do not include the explicit formulas for these  $\greenfL_{n-legs}$  here,  and instead we present just their corresponding analytical divergent (singular) contributions.  The list with these ${\cal O}(\Delta_\epsilon)$ results is  shown in the following.

The loops divergencies in 1- and 2-legs 1PI functions are: 
\bear
iT^{\rm Loop}\vert_{div} &=& i\frac{\Delta_\epsilon}{16\pi^2} \frac{3} {2\vev}
\left(\kappa_3 \mh^4 +2a\left(2 \mw^4+\mz^4\right) \right)  \nn\\
\hspace{-3mm}-i\Sigma_{HH}^{\rm Loop}(q^2)\vert_{div} &=& i\frac{\Delta_\epsilon}{16\pi^2}\frac{3}{2\vev^2} 
\left(a^2q^4 -2a^2(2\mw^2+\mz^2)q^2 +(3\kappa_3^2+\kappa_4)\mh^4+(4a^2+2b)(2\mw^4+\mz^4)\right)  \nn\\
i\Sigma_{WW}^{T\,{\rm Loop}} (q^2)\vert_{div} &=& i\frac{\Delta_\epsilon}{16\pi^2}\frac{g^2}{12}\left((51-a^2-12\xi)q^2 +3(a^2-b)\mh^2 +3(9-3a^2+4\xi)\mw^2 -9\mz^2\right)  \nn\\
i\Sigma_{AA}^{T\,{\rm Loop}} (q^2)\vert_{div} &=& i\frac{\Delta_\epsilon}{16\pi^2}e^2(4-\xi)q^2  \nn\\
i\Sigma_{ZZ}^{T\,{\rm Loop}} (q^2)\vert_{div} &=& i\frac{\Delta_\epsilon}{16\pi^2}\frac{g^2}{12}\left((4-\frac{1+a^2}{\cw^2}+12\cw^2(4-\xi))q^2 \right.  \nn\\
&&\left.\hspace{16mm}+3(a^2-b)\frac{\mh^2}{\cw^2}+12\mw^2(3+\xi)-18\mz^2-9a^2\frac{\mz^2}{\cw^2} \right)  \nn\\ 
i\Sigma_{ZA}^{T\,{\rm Loop}} (q^2)\vert_{div} &=& i\frac{\Delta_\epsilon}{16\pi^2}\frac{e g}{2\cw}\left((1/3+2\cw^2(4-\xi))q^2 +\mw^2(3+\xi)\right)  \nn\\
i\Sigma_{WW}^{L\,{\rm Loop}} (q^2)\vert_{div} &=& i\frac{\Delta_\epsilon}{16\pi^2}\frac{g^2}{4}\left(a^2q^2 +(a^2-b)\mh^2+(9-3a^2+4\xi)\mw^2-3\mz^2\right) \nn\\
i\Sigma_{W\pi}^{\rm Loop} (q^2)\vert_{div} &=& i\frac{\Delta_\epsilon}{16\pi^2}\frac{g^2}{4}\left(-a^2q^2 -(a^2-b)\mh^2-(3-3a^2+8\xi/3)\mw^2+(3-2\xi/3)\mz^2\right)  \nn\\
-i\Sigma_{\pi\pi}^{\rm Loop} (q^2)\vert_{div} &=& i\frac{\Delta_\epsilon}{16\pi^2}\left(\frac{a^2}{\vev^2}q^4 +\frac{q^2}{\vev^2}((a^2-b)\mh^2-(3+3a^2-4\xi/3)\mw^2-(3-4\xi/3)\mz^2)\right)
\label{Loop-SE}
\eear

The loops divergencies in the $S(q)V_\mu(k_1)V'_\nu(k_2)$ 1PI functions  are:
\bear
\greenfL_{HAA}\vert_{div} &=& 0  \nn\\
i\greenfL_{HAZ}\vert_{div} &=& i\frac{\Delta_\epsilon}{16\pi^2}\frac{g^2\sw}{\vev}a\mw\mz(\xi +3) \gmunu \nn\\
i\greenfL_{HZZ}\vert_{div} &=& i\frac{\Delta_\epsilon}{16\pi^2} \frac{g^2}{12\vev\cw^2}  \left(\left(3a(2+a^2)q^2 +a(a^2-b)(k_1^2+k_2^2)  \right.\right. \nn\\
&&\hspace{14mm}\left.\left.-3(a^2-b)(2a-3\kappa_3)\mh^2 -18ab\mz^2+24a \mw^2\cw^2 (\xi +3)-36a \mw^2 \right)\gmunu \right.  \nn\\
&&\left.\hspace{20mm} +2a(a^2+2b)(\kTmu\kVnu+\kVmu\kTnu) +12a^3\kTmu\kTnu \right)  \nn\\
i\greenfL_{HWW}\vert_{div} &=& i\frac{\Delta_\epsilon}{16\pi^2} \frac{g^2}{12\vev} \left(\left(3a(2+a^2)q^2 +a(a^2-b)(k_1^2+k_2^2)  \right.\right. \nn\\
&&\left.\left.\hspace{20mm}-3(a^2-b)(2a-3\kappa_3)\mh^2 -18ab\mw^2+6 a\mw^2 (4 \xi +9)-18a \mz^2\right)\gmunu \right.  \nn\\
&&\left.\hspace{18mm} +2a(a^2+2b)(\kTmu\kVnu+\kVmu\kTnu) +12a^3\kTmu\kTnu \right)  \nn\\
i\greenfL_{\pi WA}\vert_{div} &=& -\frac{\Delta_\epsilon}{16\pi^2}\frac{ g^2\sw}{6\vev}\left( \gmunu\left( 3a^2q^2 +(1-a^2)k_2^2 \right.\right. \nn\\
&& \left.\left. \hspace{28mm}+3(a^2-b)\mh^2 +9(1-a^2)\mw^2 -9 \mz^2 +2(4\mw^2+\mz^2)\xi \right) \right.  \nn\\
&&\left.\hspace{20mm}+6a^2\kTmu\kVnu+3a^2\kTmu\kTnu-(1-a^2)\kVmu\kTnu  \right)  \nn\\
i\greenfL_{\pi WZ}\vert_{div} &=& \frac{\Delta_\epsilon}{16\pi^2}\frac{g^2}{12\vev\cw}\left( \gmunu\left( 6\sw^2a^2q^2+(1-a^2)k_1^2-(1-a^2)(\cw^2-\sw^2)k_2^2 \right.\right.  \nn\\
&& \left.\left.\hspace{28mm}+2\sw^2\left( 3(a^2-b)\mh^2 +9(1-a^2)\mw^2 -9a^2\mz^2 +2(4\mw^2+\mz^2)\xi \right)\right)\right.  \nn\\
&& \left. +(-1+7a^2-12\cw^2a^2)\kTmu\kVnu+6\sw^2a^2\kTmu\kTnu+(-1+2\cw^2+7a^2-2\cw^2a^2)\kVmu\kTnu\right)
\label{Loop-SVV}
\eear

The loops divergencies in the $W^-_\mu(k_1)W^+_\nu(k_2)V_\rho$ 1PI functions  are:
\bear
i\greenfL_{WWA}\vert_{div} &=& i\frac{\Delta_\epsilon}{16\pi^2}\frac{g^3\sw}{12} \left(\gmunu \left(k_1^\rho \left(33-4 a^2-18 \xi\right)-k_2^\rho \left(33+2 a^2-18 \xi\right)\right) \right.  \nn\\
&&\left.\hspace{20mm}+g^{\nu\rho}(k_1^\mu(33-4 a^2-18 \xi)-2k_2^\mu(-33+a^2+18\xi))\right.  \nn\\
&&\left.\hspace{20mm}+(-33+a^2+18\xi)g^{\rho\mu}(2k_1^\nu+k_2^\nu)\right)  \nn\\
i\greenfL_{WWZ}\vert_{div} &=& i\frac{\Delta_\epsilon}{16\pi^2}\frac{g^3}{24\cw} \left(\gmunu \left(k_1^\rho \left(1+5a^2+\cw^2(66-8a^2-36\xi)\right) \right.\right. \nn\\
&&\left.\left.\hspace{29mm}-k_2^\rho\left(1-7a^2+\cw^2(66+4a^2-36\xi)\right)\right) \right.  \nn\\
&&\left.\hspace{2mm}+g^{\nu\rho}(k_1^\mu(1+5a^2+\cw^2(66-8a^2-36\xi))-2k_2^\mu(-1+a^2+\cw^2(-66+2a^2+36\xi)))\right.  \nn\\
&&\left.\hspace{2mm}+(-1+a^2+\cw^2(-66+2a^2+36\xi))g^{\rho\mu}(2k_1^\nu+k_2^\nu)\right) 
\label{Loop-WWV}
\eear

The $Z_\mu(k_1)Z_\nu(k_2)V_\rho$ 1PI functions are finite.

The loops divergencies in the $V_\mu V'_\nu W^-_\rho W^+_\sigma$ 1PI  functions  are:
\bear
i\greenfL_{AAWW}\vert_{div} &=& -i\frac{\Delta_\epsilon}{16\pi^2}\frac{g^4\sw^2}{12}\left((24\xi+a^2-15)S^{\mu\nu,\rho\sigma}-3a^2(g^{\mu\rho}g^{\nu\sigma}+g^{\mu\sigma}g^{\nu\rho})\right)  \nn\\
i\greenfL_{AZWW}\vert_{div} &=& -i\frac{\Delta_\epsilon}{16\pi^2}\frac{g^4\sw\cw}{24}\left((48\xi-33+2\cw^2(a^2+1)+a^2)S^{\mu\nu,\rho\sigma} \right.  \nn\\
&&\left.\hspace{31mm}+8(2-3\cw^2)a^2(g^{\mu\rho}g^{\nu\sigma}+g^{\mu\sigma}g^{\nu\rho})\right)  \nn\\
i\greenfL_{ZZWW}\vert_{div} &=& -i\frac{\Delta_\epsilon}{16\pi^2}\frac{g^4}{24\cw^2}\left((48\cw^4\xi-32\cw^4-a^4+3a^2b-2a^2\sw^2(2+\cw^2) \right. \nn\\
&&\left.\hspace{26mm}-3b^2/2-1+4\sw^2\cw^2)S^{\mu\nu,\rho\sigma}\right.  \nn\\
&&\left.\hspace{5mm}-(a^4+3a^2b-3a^2\sw^2(1+2\cw^2)-3b^2/2+3\sw^2\cw^2)(g^{\mu\rho}g^{\nu\sigma}+g^{\mu\sigma}g^{\nu\rho})\right)  \nn\\
i\greenfL_{WWWW}\vert_{div} &=& i\frac{\Delta_\epsilon}{16\pi^2}\frac{g^4}{24}\left((48\xi+2a^4-30)S^{\mu\rho,\nu\sigma}-3(-2a^4+2a^2b-b^2-2)(g^{\mu\rho}g^{\nu\sigma}+g^{\mu\sigma}g^{\nu\rho})\right) \nn\\
\label{Loop-VVWW}
\eear

All these divergent contributions will set the values of the ${\cal O}(\Delta_\epsilon)$ counterterms,  both for the EW parameters and the $a_i$ coefficients,  that are relevant for our computation of the VBS amplitudes.  These will be  presented in  \secref{sec-main-div-results}.

\subsection{Slavnov-Taylor identity in the unphysical charged sector}
\label{sec-STidentity}

Before presenting our main results on the renormalization of the $\mL_2$ parameters and $a_i$ coefficients in \secref{sec-main-div-results},  it is illustrative to discuss first an important aspect related to the gauge invariance of the EChL: the Slavnov-Taylor Identity (STI) corresponding to the unphysical charged sector, in particular  the equation relating the two-legs functions of the $W$ and $\pi$ fields. 
The validity of this relation among the self-energies in this unphysical sector  at the renormalized level and for arbitrary external momentum is one of the key points for the $\xi$-invariance of the NLO EChL predictions.  Setting properly this equation at the renormalized level also guarantees
the same poles structure at the NLO as in the LO for all the unphysical propagators as reflected in \eqref{condOS5}, which is  very convenient from the practical computational aspects.
The corresponding Slavnov-Taylor relations among the self-energies in the unphysical neutral sector, i.e., for $\{A,Z,\pi^3\}$,  can also be set in a similar way.  However,  the details in this sector are more involved and we do not present them here since they are not needed for the present computation of $WZ$ scattering.  We then focus next in the charged unphyical sector and summarize the basic points of the STI.

After the introduction of $\mL_{GF}$ and $\mL_{FP}$ in \eqref{eq-L2}, the original $SU(2)_L \times U(1)_Y$ gauge invariance of the classical Lagrangian is lost.  However,  the EChL is invariant under the BRS transformations~\cite{BRSpaper} that involve the ghost fields.
The BRS symmetry induces relations among the various Green functions involved,  called the Slavnov-Taylor identities (STIs)~\cite{Slavnov:1972fg,Taylor:1971ff}.
A special case of interest here is the relation between the propagators of the unphysical charged sector, $\{W^\pm,\pi^\pm\}$ which then leads to the corresponding relations among self-energies.  

In oder to get these relations and to understand how are they written in terms of renormalized quantities let us first consider the corresponding gauge-fixing Lagrangian of the charged sector in terms of the bare quantities. This can be written as follows:
\be
\mL_{GF}^{0} = -\left( \frac{1}{\sqrt{\xi_1^0}}\,\partial^\mu W_{0\,\mu}^+ -\sqrt{\xi_2^0}\,\frac{g_0 \vev_0}{2}\,\pi_0^+ \right) \left( \frac{1}{\sqrt{\xi_1^0}}\,\partial^\mu W_{0\,\mu}^- -\sqrt{\xi_2^0}\,\frac{g_0 \vev_0}{2}\,\pi_0^- \right)\,,
\label{charged-LGF-bare}
\ee
where two independent gauge-fixing bare parameters $\xi_1^0$ and $\xi_2^0$ have been used (for this presentation we follow closely~\cite{Bohm:1986rj} and~\cite{ST-Espriu}).  Then we use the prescription in \eqref{EChL-renorm-factors}, to write this Lagragian in terms of renormalized quantities. In particular,  remember that we have chosen just one common renormalized gauge parameter $\xi$ with  $\xi_{1,2}^0 = \xi (1+\delta\xi_{1,2})$. Then, 
the  leading order GF Lagrangian in terms of the renormalized quantities corresponds to the first term on the l.h.s. of \eqref{GF-lag}, i.e., to:
\be
\mL_{GF} = -\left( \frac{1}{\sqrt{\xi}}\,\partial^\mu W_{\mu}^+ -\sqrt{\xi}\,\frac{g \vev}{2}\,\pi^+ \right) \left( \frac{1}{\sqrt{\xi}}\,\partial^\mu W_{\mu}^- -\sqrt{\xi}\,\frac{g \vev}{2}\,\pi^- \right)\,,
\label{charged-LGF}
\ee
One can use \eqref{charged-LGF-bare} to get the relevant STI 
in terms of the undressed propagators in momentum-space.  This reads as follows:
\be
q^2\Delta_{L}^{WW}(q^2) +2\sqrt{\xi_1^0\xi_2^0}\, q^2\Delta^{W\pi}(q^2) -\xi_1^0\xi_2^0\mw^{0\,2}\Delta^{\pi\pi}(q^2) =\xi_1^0 \,.
\label{STIundressed}
\ee
We then write the above propagators in terms of the undressed 2-leg functions at \1loop precision and arrive to the following relation among undressed self-energies:
\be
q^2\Sigma_{WW}^{L\,{\rm Loop}}(q^2) +2q^2\Sigma_{W\pi}^{\rm Loop}(q^2) -\mw^2\Sigma_{\pi\pi}^{\rm Loop}(q^2) = 0 \,.
\label{charged-ST-unrenorm}
\ee
Finally,  by adding the corresponding counterterms, given explicitly  in \eqref{charged-unphys-SE-renorm},  one arrives to  the wanted STI in terms of the renormalized quantities at the \1loop level.  This reads as follows:
\be
q^2\SER_{WW}^{L}(q^2) +2q^2\SER_{W\pi}(q^2) -\mw^2\SER_{\pi\pi}(q^2) = \frac{q^2 -\xi\mw^2}{\xi} f_{ST}(q^2) \,,
\label{charged-ST-renorm}
\ee
where the $q^2$ dependent function  $f_{ST}(q^2)$ on the r.h.s of this STI,  which is finite,  has been defined as:
\be
 f_{ST}(q^2)=-\left( \delta\Zf_W -\delta\xi_1 \right)q^2 +\xi\delta\mw^2 +\xi\mw^2\left( \delta\Zf_\pi +\delta\xi_2 \right).
\ee
Some comments about the previous  \eqref{charged-ST-renorm} are in order.
First,  this STI and the renormalization conditions of \eqref{condOS5} ensure that $\SER_{W\pi}(\xi\mw^2)=0$, or equivalently, the poles of the renormalized charged propagators are located at $q^2=\xi\mw^2$ at the \1loop level.
Second,  it is important to notice that the new physics effects encoded in the $a_{11}$ coefficient,  which enters separately  in  each self-energy,  cancel in this particular combination of renormalized self-energies. Then, interestingly,  this STI in \eqref{charged-ST-renorm}, using the $R_\xi$ gauges,  has the same formal expression in the EChL as in the SM~\cite{Bohm:1986rj}. However, it is also important to notice that  the role played by the tadpole in the SM  is different than in the EChL, as it is shown in \appref{App-tadpole}.  Again, the fact that the $H$ field is a singlet in the EChL case introduces some 
peculiarities respect to the linear realizations like the SM and the SMEFT.  For a verification and discussion 
of the STIs in the SM and SMEFT using the background field method see also~\cite{Corbett:2020ymv}.

Finally, due to the relevance of the STIs in the consistency of the theory, we will impose the validity of \eqref{charged-ST-unrenorm} at the renormalized level.  Namely, we set in \eqref{charged-ST-renorm}:
\be
f_{ST}(q^2)=0 \quad \forall q^2
\label{fST}
\ee 
 This is equivalent to set an additional renormalization condition in the unphysical sector valid for all external momenta (i.e., not only at $q^2=\xi m_W^2$),  and 
implies that each coefficient of this linear function of $q^2$ vanishes, resulting in: 
\bear
&&\delta\xi_1=\delta\Zf_W \quad ; \quad \delta\xi_2=-\delta\Zf_\pi-\frac{\delta\mw^2}{\mw^2}. 
\label{xi1xi2}
\eear
These conditions in \eqref{fST} (or,  equivalently,  in \eqref{xi1xi2}) and \eqref{condOS5} fix then the four counterterms of the unphysical charged sector ($\delta\xi_1$, $\delta\xi_2$, $\delta\Zf_\pi$ and $\delta a_{11}$).
It is important to stress that the particular gauge-fixing and renormalization conditions in the unphysical sector do not affect the physical $S$-matrix elements. However, the simplifications induced by the use of the STIs and the consequent specific cancellations among the $\xi$-dependent terms of the various contributions in the different channel diagrams do depend on the concrete formulation of the unphysical sector.
In this sense, among the many arbitrary prescriptions for this unphysical sector, we choose a particular one via the  renormalization conditions given by \eqref{condOS5} and \eqref{xi1xi2}.  This is mainly motivated by the interplay between the unphysical self-energies invloved in the $s$- and $u$-channels of the $WZ$ scattering process of our interest  in \figref{1PIdiagsWZWZinSTUC} and the cancellations of the $\xi$-dependent terms that we have seen occur in both channels of the LO amplitude.  Our prescription then replicates these cancellations also at the NLO. 

\subsection{Renormalization of the EFT parameters}
\label{sec-main-div-results}

In this section we present the results for the renormalization of the EFT parameters.  These include the EW parameters entering in $\mL_2$,  like $g,$  $g'$,  etc., and the EChL coefficients,  namely,  the ones entering in $ \mL_2$,   like $a$,  etc,  and the ones entering in $ \mL_4$,   i.e., the $a_i$ coefficients. 

With our previously described diagrammatic procedure of renormalization of all the 1PI functions we can derive the divergent parts,  i.e., of ${\cal O}(\Delta_\epsilon)$,  for all the involved counterterms.  These are called here in short $\delta_\epsilon$.  Alternatively,  these divergent parts can also be derived by using the renormalization conditions introduced in \secref{RR} which allow to write the parameter counterterms in terms of the undressed 1PI functions.  In fact, we have used this second procedure as a check of our results that we obtain  with the first procedure. Regarding the finite contributions to all these counterterms,  they are also determined by these renormalization conditions and, indeed,  we use them in the final numerical computation of the $WZ$ scattering in the next section.  Therefore,  we postpone the introduction of the finite contributions for the next section and focus here in the derivation of the divergent parts of the EChL counterterms. 

The  determination of the divergent parts of all the counterterms for the EChL parameters follows immediately from the requirement of getting finite all the renormalized 1PI functions at arbitrary values of the external leg momenta, i.e., off-shell 1PI functions. This means that the cancellation of the ${\cal O}(\Delta_\epsilon)$ contributions must occur at all external momenta, and it  proceeds between the loop contributions in all these 1PI functions  and the involved counterterms. One then requires that these cancellations occur in each  involved Lorentz structure and in each term in the momentum powers expansion and all these lead to a system of equations with a number of unknowns, the $\delta_\epsilon$'s,  than one solves globally.  In practice,  one can conveniently solve this system in a sequential form, first solving the 1-leg function, then the two-legs functions,  with the previous solutions then solving the three-legs and finally  the four-legs functions.  This provides the solution for all $\delta_\epsilon$ of both the EW parameters and the EChL coefficients jointly. 
For instance,  the case of the  Higgs  self-energy  illustrates clearly this procedure.  Setting the cancellation of the ${\cal O}(\Delta_\epsilon)$ contributions in all the terms of $\hat{\Sigma}_{HH}(q^2)$ of \eqref{HH}, namely, in the terms of order $q^4$,  $q^2$ and $q^0$,  and using our result for $ \Sigma_{HH}^{\rm Loop}(q^2)\vert_{div}$ in \eqref{Loop-SE} one derives easily $ \deltaCT a_{\Box\Box}$, $ \deltaCT\Zf_H $ and $\deltaCT\mh^2$.  One can proceed similarly with the other self-energies of  the gauge bosons and the $\pi$ bosons in \eqref{SE-phys-renorm} and the other 1PI functions.  At this point, it is worth remarking that the divergencies of the CTs for the EW parameters are  determined by just the 1- and 2-legs 1PI functions whereas the divergencies of the CTs for the EChL coefficients are determined by all the 1PI functions. These new CTs in the EChL coefficients cancel the extra divergencies arising in the loop diagramas (generated by $\mL_2$) of all the 1PI functions and for arbitrary (off-shell) external momenta.

The summary of the sequential settings regarding the EW parameters can be read as follows: 1) $\hat{\Sigma}_{HH}$ sets $ \deltaCT\Zf_H $ and $\deltaCT\mh^2$,  2) $\hat{\Sigma}_{WW}^T$ sets $ \deltaCT\Zf_W$ and $\deltaCT\mw^2$,  3) $\hat{\Sigma}_{\pi\pi}$ sets $ \deltaCT\Zf_\pi$ (and $\deltaCT \xi_2$),  4)  $\hat{\Sigma}_{W\pi}$ and $\hat{\Sigma}_{WW}^L$ set $\deltaCT \xi_1$ and $\deltaCT \xi_2$,  5) $U(1)_{\rm em}$ gauge invariance sets $\delta g'/g' =0$,  6) $\hat{\Sigma}_{ZA}^T$  then sets $\deltaCT g /g$,  7) with the previous $ \deltaCT\Zf_\pi$, $ \deltaCT\Zf_W$,  $\deltaCT\mw^2$ and $\deltaCT g /g$,  then one derives $\deltaCT v/v$,  8) the tadpole $\hat T$ then sets $\deltaCT \lambda /\lambda$, or it can also be derived alternatively from  $ \deltaCT\Zf_H$,  $\deltaCT\mh^2$,  $\deltaCT g /g$ and $\deltaCT v/v$,  9) finally, the CTs  in the neutral gauge sector $\deltaCT \Zf_Z$, $\deltaCT\mz^2$ and $\deltaCT \Zf_A$ are derived from $\hat{\Sigma}_{ZZ}^T$,  $\hat{\Sigma}_{AA}^T$ and $\hat{\Sigma}_{AZ}^T$ jointly.

Regarding the EChL coefficients,  the $\deltaCT$'s  are fixed, in summary,  as follows: 1) $\hat{\Sigma}_{HH}$ sets $ \deltaCT a_{\Box\Box}$,  2)  $\hat{\Sigma}_{\pi\pi}$ sets $\deltaCT a_{11}$ (it is also set by $\hat{\Sigma}_{WW}^L$ and $\hat{\Sigma}_{W\pi}$),  3) 
 $\hat{\Sigma}_{ZZ}^T$,  $\hat{\Sigma}_{AZ}^T$ and  $\hat{\Sigma}_{AA}^T$ jointly 
 set $\deltaCT a_0$,  $\deltaCT (a_1+ a_{13})$ and $\deltaCT (a_8+ a_{13})$,  4) 
 $\hat{\Gamma}_{HWW}$ sets  $\deltaCT a /a$,  5)  $\hat{\Gamma}_{HWW}$, 
  $\hat{\Gamma}_{HZZ}$,  $\hat{\Gamma}_{HAZ}$ and 
  $\hat{\Gamma}_{HAA}$ jointly set 
 all the $\deltaCT a_i$'s in ${\mL}_{4}^{\rm one-Higgs}$,  6) $ \hat{\Gamma}_{WWZ}$, $ \hat{\Gamma}_{WWA}$,  
 $\hat{\Gamma}_{\pi WZ}$ and $\hat{\Gamma}_{\pi WA}$ set jointly $\deltaCT a_2$,  $\deltaCT a_3$, $\deltaCT a_8$,  $\deltaCT a_9$, 
$\deltaCT a_{12}$ and $\deltaCT a_{13}$ (and using this latter one then solves $a_1$ separately from 3)) ,  
6) $\hat{\Gamma}_{WWWW}$ sets $\deltaCT a_4$ and $\deltaCT a_5$ and 
$\hat{\Gamma}_{ZZWW}$ sets $\deltaCT (a_4+a_6)$ and $\deltaCT (a_5+a_7)$.  Therefore,  the two together also set $\deltaCT a_6$ and $\deltaCT a_7$.  7) Finally,  one can use extra functions and relations among them as a check of the previous findings.  For instance,  one can use $\hat{\Gamma}_{AAWW}$ and $\hat{\Gamma}_{AZWW}$ regarding the checks in the 4-legs functions,  and the relations in \eqref{xi1xi2} regarding the two-legs functions in the unphysical sector.

We list in the following the results for all these divergent counterterms for the EFT parameters. We collect together the CTs  for the EW parameters,  and separately those for  the EChL coefficients. 

\subsubsection{Divergent CTs for the EW parameters}
\label{divEWparams}
The results are the following:
\bear
&&\deltaCT\Zf_H=\frac{\div}{16\pi^2}\frac{3a^2}{\vev^2}(2\mw^2+\mz^2)\,,\qquad \deltaCT T=\frac{\div}{16\pi^2} \frac{3}{2\vev}\left(\kappa_3\mh^4 +2a\left(2 \mw^4+\mz^4\right) \right)\,,  \nn\\
&&\deltaCT\mh^2=\frac{\div}{16\pi^2}\frac{3}{2\vev^2}((3\kappa_3^2+\kappa_4)\mh^4-2a^2\mh^2(2\mw^2+\mz^2)+(4a^2+2b)(2\mw^4+\mz^4))\,,  \nn\\
\vspace{1mm}
&&\deltaCT\Zf_B=-\frac{\div}{16\pi^2}\frac{\gY^2}{12}(1+a^2)\,, \qquad \deltaCT\Zf_W=\frac{\div}{16\pi^2}\frac{g^2}{12}(51-a^2-12\xi)\,,  \nn\\
&&\deltaCT\mw^2=-\frac{\div}{16\pi^2}\frac{g^2}{12}\left(3(a^2-b)\mh^2 +(78-10a^2)\mw^2 -9\mz^2\right)\,,  \nn\\
    %
&&\deltaCT\mz^2=\frac{\div}{16\pi^2}\frac{g^2}{12\cw^2}\left(-3(a^2-b)\mh^2 +(7(1+a^2)+2(-43+a^2)\cw^2)\mw^2 +(10+a^2)\mz^2\right)\,,  \nn\\
&&\deltaCT\gY/\gY=0\,,  \qquad\deltaCT g/g=-\frac{\div}{16\pi^2}\frac{g^2}{2}(3+\xi)\,,  \nn\\
\vspace{1mm}
&&\deltaCT\xi_1=\frac{\div}{16\pi^2}\frac{g^2}{12}(51-a^2-12\xi)\,,  \nn\\
&&\deltaCT\xi_2=\frac{\div}{16\pi^2}\frac{1}{3\vev^2}(6(a^2-b)\mh^2+(69-19a^2+4\xi)\mw^2-(18-4\xi)\mz^2)\,,  \nn\\
&&\deltaCT\Zf_\pi=-\frac{\div}{16\pi^2}\frac{1}{\vev^2}((a^2-b)\mh^2-(3+3a^2-4\xi/3)\mw^2-(3-4\xi/3)\mz^2)\,,  \nn\\
&&\deltaCT\vev/\vev=\frac{\div}{16\pi^2}\frac{2(\mw^2+\mz^2)}{3\vev^2}\xi\,.
\label{L2param-div}
\eear
Some comments about these results are in order.  First of all,  we notice that the $\xi$ parameter enters in these counterterms, as it is expected in renormalization with $R_\xi$ gauges.  This also happens in the SM case,  which we have also analyzed for completeness and comparison and whose results are collected in \eqref{SMparam-div}.  Concretely, we find in both the EChL and the SM, that the  $\xi$ parameter enters in the divergencies of the CTs corresponding to $g$, $\Zf_W$, $\xi_{1,2}$, $\Zf_\pi$ and $\vev$.   As we can see by an explicit comparison,  only $\deltaCT g'/g'$,  $\deltaCT g/g$ coincide in both theories.  $\deltaCT\Zf_B$, $\deltaCT\Zf_W$ and $\deltaCT \xi_1$ coincide for $a=1$. In general all the other EW CTs are different in both theories.
One of the new features of this non-linear EFT is that $\deltaCT T$, $\deltaCT\mh^2$, $\deltaCT\mw^2$ and $\deltaCT\mz^2$ are $\xi$-independent separately.  In contrast,  in the SM case,  only the proper contributions  of both tadpole and mass counterterms to the pole of the propagators  ($\widetilde{\delta m}^2$)  are $\xi$-independent,  as we show in \eqref{deltam2bar}.
Finally,  notice also that the divergent part of the combination $(\delta\vev/\vev+\delta\Zf_\pi/2)$ is $\xi$-independent in the EChL.

\subsubsection{Divergent CTs for the EChL coefficients}

The results are the following
\bear
&&\deltaCT a = \frac{\Delta_\epsilon}{16\pi^2}\frac{3}{2\vev^2}\left((a^2-b)(a-\kappa_3)\mh^2+a\left((1-3a^2+2b)\mw^2+(1-a^2)\mz^2\right)\right) \,,  \nn\\
&&\deltaCT a_0=\frac{\Delta_\epsilon}{16\pi^2}\frac{3}{8}(1-a^2) \,,\qquad\deltaCT a_1=\frac{\Delta_\epsilon}{16\pi^2}\frac{1}{12}(1-a^2) \,,\qquad \deltaCT a_2=-\deltaCT a_3=\frac{\Delta_\epsilon}{16\pi^2}\frac{1-a^2}{24}  \nn\\
&& \deltaCT a_4 =-\frac{\Delta_\epsilon}{16\pi^2}\frac{(1-a^2)^2}{12} \,,\quad \deltaCT a_5 =-\frac{\Delta_\epsilon}{16\pi^2}\frac{1}{24}(a^4+4a^2-3a^2b+\frac{3}{2}b^2+1) \,,  \nn\\
&&\deltaCT a_{6} = \deltaCT a_{7} =\deltaCT a_8=\deltaCT a_{9} = \deltaCT a_{10} = \deltaCT a_{12}  = \deltaCT a_{13} = 0 \,,\qquad \deltaCT a_{11}=\frac{\Delta_\epsilon}{16\pi^2}\frac{a^2}{4} \,,  \nn\\
&&\deltaCT a_{HBB} = \deltaCT a_{HWW} =\frac{\Delta_\epsilon}{16\pi^2}\frac{a(a^2-b)}{12} \,,\qquad \deltaCT a_{\Box{\cal V}{\cal V}} = -\frac{\Delta_\epsilon}{16\pi^2}\frac{a(2+a^2)}{4} \,,  \nn\\
&&\deltaCT a_{H0} = \frac{\Delta_\epsilon}{16\pi^2}\frac{3a (1-b)}{4} \,,\qquad \deltaCT a_{H1} = \frac{\Delta_\epsilon}{16\pi^2}\frac{a(a^2-b)}{6} \,,\qquad \deltaCT a_{H11} = -\frac{\Delta_\epsilon}{16\pi^2}\frac{a(a^2-b)}{2} \,,  \nn\\
&&\deltaCT a_{d1} = -\deltaCT a_{d2} = \frac{\Delta_\epsilon}{16\pi^2}\frac{a(a^2-b)}{6} \,,  \qquad \deltaCT a_{d3} = \frac{\Delta_\epsilon}{16\pi^2}\frac{a(a^2+b)}{2} \,,  \nn\\
&&\deltaCT a_{H8} = \deltaCT a_{H13} = \deltaCT a_{\Box 0} = \deltaCT a_{d4} = \deltaCT a_{d5} = 0 \,,  \nn\\
&&\deltaCT a_{\Box\Box}=-\frac{\Delta_\epsilon}{16\pi^2}\frac{3a^2}{4} \,.
\label{L4param-div}
\eear
Some comments about these results are in order.  First of all,  we wish to remark that we have found  no $\xi$-dependent piece in any of all these results for the CTs in the EChL coefficients,  in contrast to the previous results for the CTs of the EW parameters.  This $\xi$-independence in the results for the $\deltaCT a_i$'s is not trivial at all,  since its derivation involves $\xi$-dependent terms from the loop diagrams everywhere.  Obviously,  this result is welcome, and on the other hand, it is also expected since by construction all the operators in 
${\cal L}_4$  are separately $SU(2)_L\times U(1)_Y$ gauge invariant.  Therefore, they do not mix under gauge transformations.  Secondly, we see that just a subset of these coefficients get divergent CTs,  concretely: $a_0$,  $a_1$,  $a_2$, $a_3$, $a_4$, $a_5$, $a_{11}$,  $a$ (remember that the renormalization of $b$,  $\kappa_3$ and $\kappa_4$ do not enter here),   $a_{HWW}$,  $a_{HBB}$,   $a_{\Box{\cal V}{\cal V}}$,  $a_{H0}$,  $a_{H1}$,
$a_{H0}$, $a_{H11}$,  $a_{d1}$,  $a_{d2}$,  $a_{d3}$ and  $a_{\Box\Box}$.  The remaining  $a_i$'s do not get divergent counterterms and,  therefore,  the corresponding operators in ${\cal L}_4$  are not needed to cancel the extra divergencies generated by ${\cal L}_2$.  We also see in these results that 
some of these CTs vanish for the choice $a=b=\kappa_3=\kappa_4=1$,  and some others do not,  like $a_5$, $a_{11}$,  $a_{\Box{\cal V}{\cal V}}$ and $a_{\Box\Box}$.  The particular choice $b=a^2$ also produces some simplifications and fewer divergent CTs are found.  Concretely,  we find vanishing $\deltaCT$'s for $a_{HWW}$,  $a_{HBB}$, $a_{H1}$,  $a_{H11}$, $a_{d1}$ and $a_{d2}$ in that case. 

At this point,  we believe it is worth comparing our results in \eqref{L4param-div} with some previous results of the EChL \1loop divergencies and counterterms in the literature.  This comparison is a partial one in any case,  since our results are the only ones that apply to the most general and complete case of  off-shell \1loop 1PI functions, including 1-, 2-, 3- and 4-legs functions, and including all types of loop diagrams in the $R_\xi$ gauges. 

The closest comparison of our results in the present work is with our own  previous results  in~\cite{paperHdecays}.  In that paper we also worked with the EChL in the $R_\xi$ gauges but focused exclusively on the \1loop decay amplitudes of the Higgs boson  into $\gamma \gamma$ and $\gamma Z$.   There we found that some particular combinations of EChL coefficients enter in those amplitudes  and they turned out to be finite,  RGE invariant and, therefore,  not needed for the cancellation of the \1loop divergences involved in these Higgs decays.  Using the notation in the present paper,  these combinations are:  
\bear
c_{H\gamma \gamma} &=& a_{HBB}+a_{HWW}-a_{H1} \,,  \\
c_{H\gamma Z} &=& \frac{1}{\cosw^2}(-a_{HBB} \sinw^2+a_{HWW}\cosw^2-\frac{1}{2} a_{H1} (\cosw^2-\sinw^2)) \,.
\label{HM1}
\eear
for $H \to \gamma \gamma$ and $H \to \gamma Z$ respectively. 
 This result is in agreement with \eqref{L4param-div} since we also get here the same cancellation in those combinations of the corresponding  $\delta_\epsilon a_i$'s.
 
Another example of observable where there appear cancellations of divergences in the involved combination of EChL coefficients is in $\gamma \gamma \to WW$ and $\gamma \gamma \to ZZ$ scattering.  This was studied in~\cite{Delgado:2014jda} where
the \1loop amplitude for the scattering processes with final longitudinal gauge bosons $V_LV_L$ (with $VV=WW,ZZ$)  was computed using the equivalence theorem, i.e., using  
${\cal A}(\gamma \gamma \to V_L V_L) \simeq {\cal A}(\gamma \gamma \to \pi \pi)$ and within the approximation of considering only chiral loops, i.e., loops with Higgs and GBs.  The masses of GBs were set to cero (as in the Landau gauge, i.e., for $\xi=0$).  The combinations of EChL coefficients found in~\cite{Delgado:2014jda} for ${\cal A}(\gamma \gamma \to \pi \pi)$,  which are finite and RGE invariant, using the notation of the present paper,  are:
\bear
c_{\gamma} = a_{HBB}+a_{HWW}-a_{H1}  \,\,\, &\rm{in}& \,\,\, \pi^+\pi^-  \,\,\, \rm{and}  \,\,\,\pi^0\pi^0\,, \\
\text{and also }\qquad(a_1-a_2+a_3) \,\,\, &\rm{in}&  \,\,\,\pi^+\pi^- \,.
\label{gammagamma}
\eear
Again this result is in agreement with \eqref{L4param-div}, as can be checked by  the cancellation of the corresponding combinations of $\delta_\epsilon a_i$'s.  Also in~\cite{Delgado:2014jda} the EW precision `oblique' $S$ parameter was related with the coefficient $a_1$.  The divergent CT  $\deltaCT a_1$ in \eqref{L4param-div}  is also in agreement with the corresponding divergence involved in the computation of $S$ in~\cite{Delgado:2014jda}.

The VBS processes with $W_L$ and $Z_L$ in the external legs were studied to \1loop within the EChL previously in~\cite{Espriu:2013fia, Delgado:2013hxa}.  It was done by means of the ET,  i.e., replacing the external $V_L$'s by the corresponding $\pi$'s,  considering just chiral loops and assuming massless GBs (as in Landau gauge again, i.e., for $\xi=0$).  In those works only interactions among scalars were involved,  and only $a$, $b$, $a_4$ and $a_5$ appeared in the VBS amplitudes,  since they assumed the so-called isospin limit where no custodial breaking operators appear.  Comparing the divergences for $a$, $a_4$ and $a_5$ (the renormalization of $b$ does not enter in VBS) found in~\cite{Espriu:2013fia, Delgado:2013hxa} with our results in \eqref{L4param-div}, we find agreement for $a$ and $a_4$.  The case of $a_5$ is more tricky.  We have checked that when taking the isospin limit in the VBS amplitudes with physical on-shell external gauge bosons (this limit implies:  $m_Z \to m_W$,  $g' \to 0$,  $c_W \to 1$ and no custodial breaking operators) and considering the full set of operators in ${\cal L}_4$,   it turns out that $a_5$ allways appears combined with others.  Concretely,  it appears in the following combination of EChL coefficients:
\bear
\tilde{a}_5 & =& a_5 -\frac{a}{2} a_{\Box{\cal V}{\cal V}} + \frac{a^2}{4} a_{\Box\Box}.
\label{a5tilde}
\eear
From our results in \eqref{L4param-div},  our prediction for the divergence in this $\tilde{a}_5$ is:
\bear
\deltaCT \tilde{a}_5 & =& 
- \frac{\Delta_\epsilon}{32\pi^2}\left(\frac{1}{8}(b-a^2)^2+\frac{1}{12}(1-a^2)^2\right) \, , 
\label{a5tildeCT}
\eear
which is in agreement with the result for $a_5$ in~\cite{Espriu:2013fia, Delgado:2013hxa},  where these two coefficients, $a_{\Box{\cal V}{\cal V}}$ and $a_{\Box\Box}$,  were not considered.  At this point, it is interesting to remark that
this combination in \eqref{a5tildeCT} indeed vanishes for $a=b=1$. 

The renormalization of the EChL coefficients was also studied in~\cite{Gavela:2014uta}.  In this case,  they considered the pure scalar theory, i.e., only the Higgs and GBs sector of the EChL and worked with massless GBs (as in Landau gauge, with $\xi=0.$).  No gauge or ghost fields were included and, therefore,  no gauge-fixing.  The considered operators were custodial preserving. The renormalization of \1loop 1PI functions,  for 1-leg,  2-legs, 3-legs and 4-legs,  was performed for off-shell external legs.  We find agreement in the divergences found for the subset of $a_i$'s involved in the scalar sector (the coefficients in the notation of~\cite{Gavela:2014uta} are specified inside the parentheses). Concretely,  we agree in: $a$ ($a_C$),  $a_4$ ($c_{11}$), $a_5$ ($c_6$), $a_{11}$ ($c_9$),  $a_{\Box{\cal V}{\cal V}}$ ($c_7$),  $a_{d3}$ ($c_{10}$),  and $a_{\Box\Box}$ ($c_{\Box H}$). 

The renormalization of the EChL was studied in the path integral formalism,  using the background field method,  in~\cite{Guo:2015isa, Buchalla:2017jlu, Buchalla:2020kdh}.  The most complete comparison of our results should be done with the bosonic loop results of~\cite{Buchalla:2017jlu,Buchalla:2020kdh} since they also included all loops of scalar and gauge particles.  However, the comparison with the path integral results is tricky since they use the equations of motion to reduce the number of operators in the Lagrangian.  Therefore,  some off-shell divergences do not appear in their results and some others are redefined by the use of the equations of motion.  They also use redefinitions of the fields (in particular the Higgs field) to reach the canonical kinetic term in the Lagrangian.  On the other hand, the parametrization used in~\cite{Buchalla:2017jlu,Buchalla:2020kdh} is also very different than here and not straightforward to compare with. In any case,  solving these differences of parametrizations  and doing  some algebra,  we have found agreement in the divergences of the following subset of coefficients (the coefficients in the notation of~\cite{Buchalla:2017jlu,Buchalla:2020kdh} are specified inside the parentheses): $a_0$ ($\beta_1$), $a_1$ ($XU1$), $a_2$ ($XU7$),  $a_3$ ($XU8$),   $a_4$ ($D_2$) and  ${\tilde a}_5$ ($D_1$).  

 \subsubsection{RGEs and running EChL  coefficients}
 Finally,  to close  the renormalization section we include here the results for the RGEs  and running   EChL coefficients.  These are easily derived from the previous results in \eqref{L4param-div} and taking into account the relation between the renormalized and  bare coefficients given in \eqref{EChL-renorm-factors} by $a^0_i=a_i+\delta a_i$.   In the $\overline{MS}$ scheme (recall that $\mu$ is the scale of dimensional regularization in $D=4 -\epsilon$ dimensions), the running  $a_i(\mu)$ can  be written as follows:
 \bear
a_i(\mu)& =& a_i^0 -\delta a_i(\mu) \,,\,\,\, \delta a_i(\mu)=\deltaCT a_i - \frac{\gamma_{a_i}}{16\pi^2}\log \mu^2 \,,\,\,\,
\deltaCT a_i =\frac{\Delta_\epsilon}{16\pi^2}\gamma_{a_i} \,, 
\label{airun}
\eear
where we have written the divergent $\deltaCT a_i$ in terms of the anomalous dimension $\gamma_{a_i}$  of the corresponding effective operator. The running and renormalized $a_i$'s can then be related, in practice,  by:
 \bear
 a_i(\mu)& =&a_i+\frac{\gamma_{a_i}}{16\pi^2}\log \mu^2
 \label{airun-air}
 \eear
The set of RGEs for all the $a_i$'s then  immediately follow:
\bear
a_i(\mu)&=&a_i(\mu')+\frac{1}{16\pi^2}\gamma_{a_i}\log\left(\frac{\mu^2}{\mu'^2}\right) \,,
\label{RGE1}
\eear
where the specific value of $\gamma_{a_i}$ for each coefficient can be read from \eqref{L4param-div}.
The RGEs can also be written,  alternatively,  as:
\bear
\frac{d a_i(\mu)}{d \log \mu} &=& \frac{\gamma_{a_i}}{8\pi^2}
\label{RGE2}
\eear
Notice that the coefficients (or combination of coefficients) with vanishing $\deltaCT$ do not run,  therefore,  they are RGE invariants.  Finally,  notice also that,  in the scattering amplitude of our interest in this work,  
${\cal A}(WZ \to WZ)$ only the following reduced list of running coefficients enter: $a(\mu)$, $a_0(\mu)$,  $a_1(\mu)$,  $a_2(\mu)$,  $a_3(\mu)$,  $a_4(\mu)$,   $a_5(\mu)$,  $a_{11}(\mu)$,  $a_{HWW}(\mu)$,  $a_{HBB}(\mu)$, $a_{H1}(\mu)$, $a_{d1}(\mu)$,  $a_{d2}(\mu)$,  $a_{\Box{\cal V}{\cal V}}(\mu)$ and $a_{\Box \Box}(\mu)$.

\section{Numerical results for $WZ$ scattering}
\label{sec-plots}

In this section we present the numerical results of the \1loop  radiative corrections to the VBS cross section for the particular channel  $WZ \to WZ$ and  postpone the other VBS channels for future works.  This $WZ$  channel illustrates the main features of the complex NLO cross section computation within the EChL and also serves as a good reference case to compare with the SM case.   For the numerical estimation of the cross section we use LoopTools~\cite{FormCalc-LT}.  Due to the lengthy and time consuming computation with many \1loop diagrams involved (more than 500),  we have chosen in this numerical computation the Feynman 't Hooft gauge ($\xi=1$), and we have restricted ourselves to the custodial preserving operators that are needed as counterterms.  Furthermore,  we have assumed the relation $b=a^2$ and taken the values $\kappa_3=\kappa_4=1$ in the EChL coefficients of ${\cal L}_2$.  With these simplifying assumptions,  the EChL coefficients  participating in our computation of $\sigma(WZ \to WZ)$ are reduced to the following ones:
$a$, $a_1$, $a_2$, $a_3$, $a_4$, $a_5$,  $a_{11}$, $a_{\Box {\cal V}{\cal V}}$ and $a_{\Box \Box}$.  
Regarding the EW input parameters, for the numerical estimates we have chosen $\mw$, $\mz$ and $G_F$.

Before presenting the numerical results,  we comment briefly on three preliminar checks that we have done on the computing procedure.  

Firstly,  in addition to the previously commented  analytical checks of cancellations of the divergent ${\cal O}(\Delta_\epsilon)$ contributions,  we have also checked  numerically  that with this subset of EChL coefficients  the full \1loop cross section $\sigma(WZ \to WZ)$ is finite.   For this check, we use an indirect method: Concretely,  we have checked numerically the independence of the full \1loop cross section with the value of the $\mu$ parameter of dimensional regularization,  which is a high precision test since this $\mu$ parameter enters in the very many loops participating  in the radiative corrections. 

Secondly,  we have also checked analytically that when using the so-called isospin limit,  which implies taking $\mw=\mz$,   the tree level amplitude from ${\cal L}_4$ depends on an even more reduced subset of coefficients: 
$a$,  $a_3$, $a_4$, $a_5$,  $a_{\Box {\cal V}{\cal V}}$ and $a_{\Box \Box}$.  Then,  taking  the high energy limit in this amplitude,  $\sqrt{s} \gg \mw,\mh$,  the involved parameters are finally reduced to:  $a$,  $a_3$, $a_4$, and ${\tilde a}_5$,  with $ \tilde{a}_5 = a_5 - (a/2) a_{\Box{\cal V}{\cal V}} + (a^2/4) a_{\Box\Box}$, already mentioned in \eqref{a5tilde}.  Here,  we do not use those approximations of taking the isosplin limit nor the high energy limit,  but it is worth knowing these analytical results for the interpretation of the final complete numerical results presented in this section. 

Thirdly,  we have also checked the already known results for $\sigma(WZ \to WZ)$ at the tree level in \cite{paperUnitarity} that showed that under the hypothesis of similar size for all the $a_i$ coefficients,  the ones that contribute the most to the cross section are $a_4$ and $a_5$ (and $a_3$ to a milder level),  specially at the high energies of ${\cal O}(1\,  {\rm TeV})$.  In this reference (and others,  see for instance \cite{Delgado:2017cls})  it was also shown the dominance, at large energies, of the polarized cross section with longitudinal external gauge bosons.   

Considering all these features summarized above,  we have restricted our numerical analysis of the NLO cross-section to the longitudinal polarization state $W_L Z_L \to W_L Z_L$ channel,   studied the effects from the most relevant EChL renormalized coefficients,  $a$, $a_4$ and $a_5$,  and fixed $b=a^2$ and $a_i=0$ for the rest of renormalized coefficients in all our plots in this section.  In that case, notice that $a_5$ and $\tilde{a}_5$ coincide. 
Then,  we analyze in the following the results for the six chosen benchmark cases:
\begin{itemize}
\item [1)] $a=1$,  $a_4 \neq 0$,  $a_5=0$
\item [2)] $a=1$,  $a_4=0$, $a_5 \neq 0$
\item [3)] $a=0.9$,  $a_4 \neq 0$,  $a_5=0$
\item [4)] $a=0.9$,  $a_4= 0$, $a_5 \neq 0$
\item [5)] $a=1$,  $a_4 \neq 0$,  $a_5 \neq 0$ 
\item [6)] $a=0.9$,  $a_4 \neq 0$,  $a_5 \neq 0$
\end{itemize}
In these benchmark cases we have considered two different values of the $a$ parameter: the case of $a=1$ that connects with the SM results,  and the  BSM case of $a=0.9$ (compatible with present data, see for instance \cite{ATLAS:2019nkf}).  The choice of non-vanishing $a_4$ and $a_5$ have been set, to both positive and negative values, with modulus within the interval  $(10^{-3}$, $10^{-4}$) which are compatible with present data (see, for instance \cite{CMS:2019qfk}).  However, it should be noticed that the concrete constraints on these parameters depend, in fact,  on the unitarization method used (for a recent study see,  for instance,  \cite{paperUnitarity}).  On the other hand, notice also that the special choice $a=1$ reduces considerably the number of involved divergences and therefore the number of divergent counterterms in the EChL coefficients.  Concretely,  for $a=1$,   there only remain non vanishing $\deltaCT$'s for  $a_5$,  $a_{11}$,  $a_{\Box {\cal V}{\cal V}}$ and $a_{\Box \Box}$,  whereas $a_4$ and  the combination defining ${\tilde a}_5$  give vanishing $\deltaCT$ therefore leading to RGE invariant coefficients. 

\begin{figure}[h!]
\begin{center}
    \includegraphics[width=1.\textwidth]{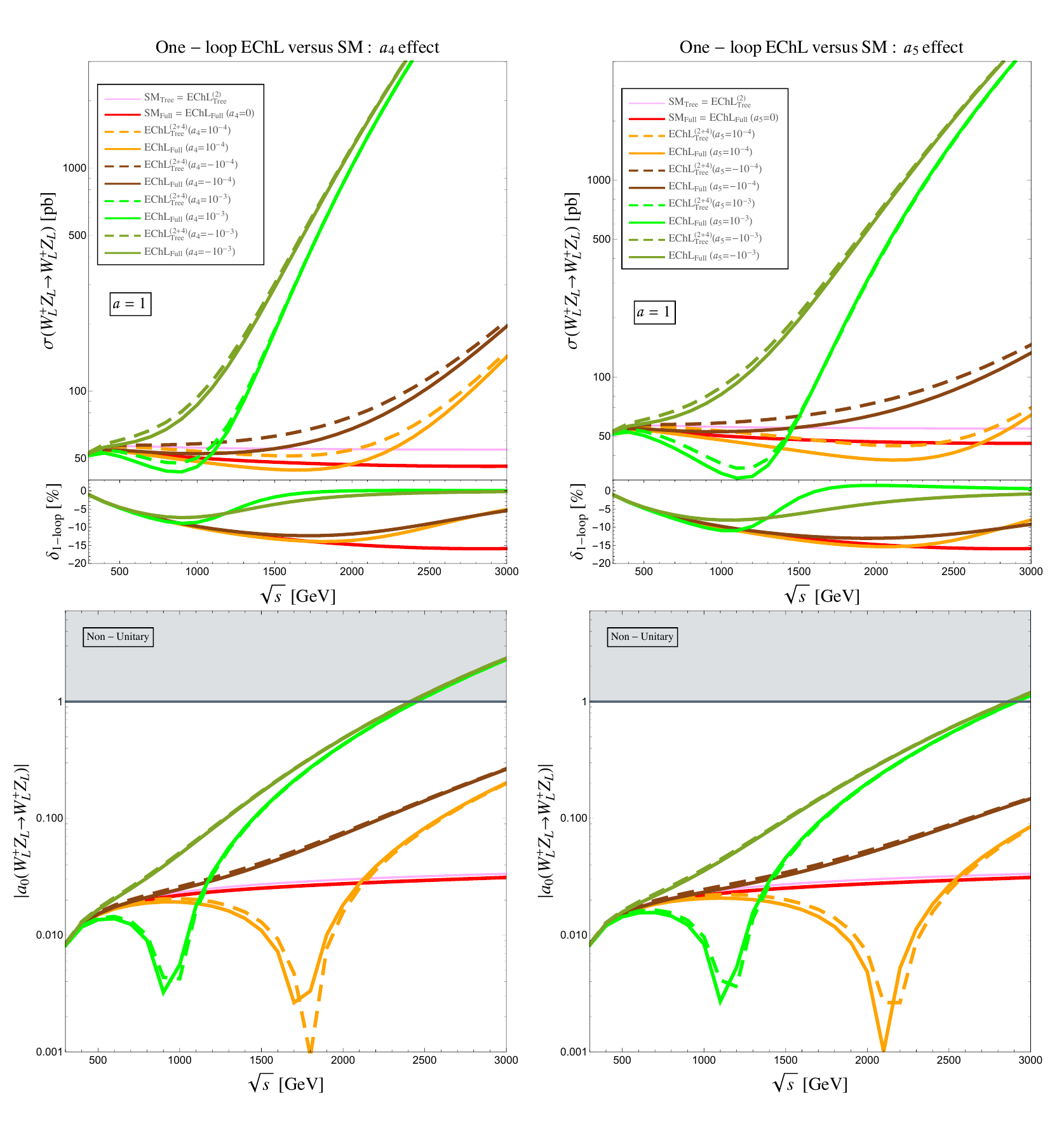}
\caption{Cross-section predictions for $W_L^+ Z_L\to W_L^+ Z_L$ within the EChL corresponding to $a=1$, $b=a^2$, $\kappa_3=\kappa_4=1$ and varying $a_4$ (left) and $a_5$ (right).  The rest of $a_i$'s  are set to zero.  The $ \mu$ scale is set to  $\mu=1$TeV.  Dashed lines are the tree level predictions.  Solid lines are the full one-loop predictions. The solid red (pink) lines correspond to the full (tree) SM prediction.  The relative size of the one-loop prediction respect to the tree level one defined by means of $\delta_{\rm{\rm{1-loop}}}$ is also shown (lower plots in the upper panels).  The predictions for the lowest partial wave $a_J$ with $J=0$ are also displayed (lower panels).  The unitarity violating regions in these plots are indicated by the shadowed areas. }
\label{plots-aSM-a4ya5}
\end{center}
\end{figure}

\begin{figure}[h!]
\begin{center}
    \includegraphics[width=1.\textwidth]{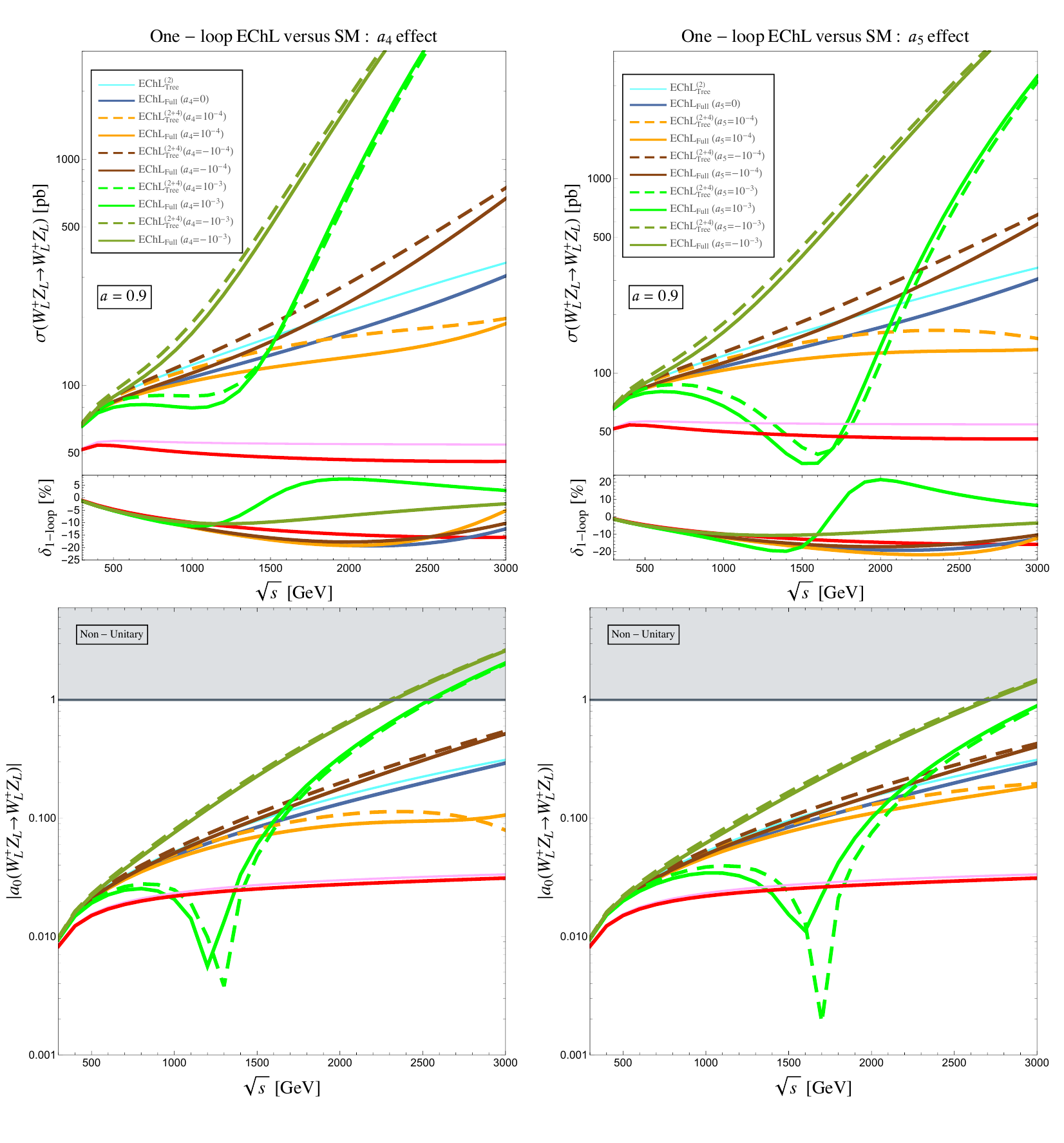}
\caption{Cross-section predictions for $W_L^+ Z_L\to W_L^+ Z_L$ within the EChL corresponding to $a=0.9$, $b=a^2$, $\kappa_3=\kappa_4=1$ and varying $a_4$ (left) and $a_5$ (right).  The rest of $a_i$'s  are set to zero.  The $ \mu$ scale is set to  $\mu=1$TeV.  Dashed lines are the tree level predictions.  Solid lines are the full one-loop predictions. The solid red (pink) lines correspond to the full (tree) SM prediction.  The relative size of the one-loop prediction respect to the tree level one defined by means of $\delta_{\rm{1-loop}}$ is also shown (lower plots in the upper pannels).  The predictions for the lowest partial wave $a_J$ with $J=0$ are also displayed (lower pannels).  The unitarity violating regions in these plots are indicated by the shadowed areas. }
\label{plots-apoint9-a4ya5}
\end{center}
\end{figure}

The results for the benchmark cases 1) and 2) are shown in \figref{plots-aSM-a4ya5},  left and right panels,  respectively. 
Those for 3) and 4) are shown in \figref{plots-apoint9-a4ya5},  left and right panels,  respectively. 
Finally,  the cases 5) and 6) are shown in \figref{plots-botha4a5},  left and right panels,  respectively. 
In all cases we have computed: a)  the total cross section (upper panels),  b)  the lowest partial wave amplitude $|a_J|$ with $J=0$  (lower panels) and c) the quantity $\delta_{\rm{1-loop}}$ that provides  the relative size of the \1loop correction in the full \1loop results respect to the tree level prediction. It is defined by:
\be
\delta_{\rm{1-loop}}=(\sigma_{\rm Full}-\sigma_{\rm Tree})/\sigma_{\rm Tree}
\ee
All these evaluated quantities are displayed  as functions of the center of mass energy $\sqrt{s}$ of the $WZ \to WZ$ scattering process, in the interval $300\, {\rm GeV} < \sqrt{s} < 3000\, {\rm GeV}$.  The disallowed areas by the partial wave unitarity bound $|a_0|>1$ are also displayed (in grey)  in the b) plots.  The corresponding predictions for the SM case are also shown in all the plots, for comparison.  Specifically,  using our notation introduced in \secref{diag-1pi},  we include the predictions for: 
${\rm SM}_{\rm Full}$,  ${\rm SM}_{\rm Tree}$,  ${\rm EChL}_{\rm Full}$,  ${\rm EChL}_{\rm Tree}^{(2+4)}$ and the corresponding  $\delta_{\rm{1-loop}}$. 

Let us first discuss the results in \figref{plots-aSM-a4ya5}.  Looking at these plots we learn the following features.   Since $a=1$ in these plots,  there is a connection between the EChL predictions and the SM ones.  Indeed the matching of the EChL and SM predictions occur for $a_i=0$,  as expected, both at the tree and \1loop level.  Similarly,  all the lines in this figure converge to the SM line at low energies,  both at the tree level and at the \1loop level.  

The (EW) \1loop prediction within the SM is by itself interesting. 
To our knowledge,  the analytical EW one-loop SM result in the $R_\xi$ gauge has not been provided in the literature for this particular $WZ  \to WZ$ channel.  The previous full one-loop evaluations of the EW radiative corrections affecting the VBS processes are numerical and have been done in the context of the LHC. These numerical estimates therefore involve many more diagrams than those considered here, since the relevant subprocesses for the production of two  EW gauge bosons at the LHC include the initial quarks in the protons and not all of them have the VBS configuration.  In addition,   final state fermions from the final EW gauge boson decays are also usually considered.  In particular,  the EW one-loop corrections at the LHC for the  doubly charged channel,  $W^+W^+$,  were computed  in  \cite{Biedermann:2016yds} and for the $WZ$ channel  in \cite{Denner:2019tmn}.  The EW one-loop corrections in the polarized case for $WZ$ have been studied  in \cite{Baglio:2019nmc}.
Although the comparison of our SM numerical results with those commented LHC works is not immediate nor can be done with high precision for the reasons  explained above,  it is however possible to make a rough comparison of the relative size of the correction,  $\delta_{\rm 1-loop}$.  For $W^+W^+$,   it was found in \cite{Biedermann:2016yds}  a large and negative \1loop radiative correction of up to $\delta_{\rm 1-loop} \sim -16 \%$ in the VBS energy range of ${\cal O}(1\, {\rm TeV})$ typically reached at the LHC.  For $WZ$,  it was found  in  \cite{Denner:2019tmn} a similar size of the EW one-loop correction of up to $\delta_{\rm 1-loop} \sim -16 \%$ at the relevant LHC energies.  Our SM numerical results for the EW one-loop radiative corrections in $WZ \to WZ$ scattering,  shown in  \figref{plots-aSM-a4ya5},  are clearly compatible with those LHC results. 
First, \figref{plots-aSM-a4ya5} shows that the SM tree level cross-section  for the $WZ$ channel (pink line) is nearly flat with $\sqrt{s}$ (reaching about $\sim$55 pb),  whereas the SM full \1loop cross-section (red line) is slightly decreasing with energy and it lays below the tree level one in the whole studied energy interval.  Therefore the interference effect from the  \1loop contribution is destructive.  The size of the \1loop correction relative to the tree level prediction (red line in the $\delta_{\rm{1-loop}}$ plot) reaches the maximum negative value of about $\delta_{\rm 1-loop} \sim -16 \%$ at the upper part of the studied interval $\sqrt{s} \sim 3000\, {\rm GeV}$ which is similar to the  previous \1loop results,  in  \cite{Biedermann:2016yds} and \cite{Denner:2019tmn},  commented above.

Next we comment on the EChL results in this \figref{plots-aSM-a4ya5}.  Regarding the tree level estimates,  first we see that the EChL predictions at the LO, i.e., from ${\cal L}_2$,  coincide with  the SM predictions at LO (pink line),   i.e., ${\rm EChL}_{\rm Tree}^{(2)}|_{a=1}={\rm SM}_{\rm Tree}$,  as expected.   Second,  we see that the tree level EChL predictions including both 
${\cal L}_2$ and ${\cal L}_4$,  given by $ {\rm EChL}_{\rm Tree}^{(2+4)}$ (dashed lines),  display a different behaviour with $\sqrt{s}$ than $ {\rm EChL}_{\rm Tree}^{(2)}$.  This is due to the effect from the involved $a_i$ which can be summarized, at the amplitude level,  as $(f_1+f_2 \, a_i)$,  where at high energies $f_1$ and for $a=1$ tends to a constant value (like in the SM),  whereas $f_2$ goes as $s^2$,   giving the  polynomial ${\cal O}(p^4)$ dependence in powers of the external momentum,  typical in  the chiral expansion.  These chiral Lagrangian features can also be seen in the predictions of the partial wave amplitudes in the lower plots.  When squaring the amplitude,  the tree level cross section receives  a linear contribution in the EChL coefficient,  of ${\cal O} (a_i)$,  and a quadratic contribution  of ${\cal O} (a_i^2)$.  The effect beyond the SM from the $a_i$ coefficient then comes from the competition of these two contributions and their comparison with the LO one.  The sign of $a_i$ also matters.  It turns out that the contribution from the linear term is destructive for positive $a_i$ and constructive for negative $a_i$.  The quadratic contribution is always constructive.  And this is true for both coefficients,  $a_4$ (left panel) and $a_5$ (right panel).  For the lowest $|a_4|$ ($|a_5|$) values of $10^{-4}$ the linear term dominates up to about 2000 GeV (2500 GeV)  and explains the behaviour with energy from threshold up around this energy, giving a tree level prediction below (above) the LO one for positive (negative) coefficient,  as the dashed orange (brown) lines indicate compared with the pink lines.   Above that energies 
(i.e., above about 2000 GeV for $a_4$ and 2500 GeV for $a_5$)  the quadratic term starts dominating the $a_i$ effect and all the tree level predictions $ {\rm EChL}_{\rm Tree}^{(2+4)}$ grow faster with energy, crossing above  the LO prediction,  $ {\rm EChL}_{\rm Tree}^{(2)}$.   For larger $|a_{4,5}|$ values of $10^{-3}$ this crossing above the LO prediction occurs at lower energies and produces the pattern shown in the dashed light green lines in these plots, corresponding to the positive coefficients. The tree level prediction first decreases with energy,  producing a minimum (at about 900 GeV in the $a_4$ case and at about 1200 GeV in the $a_5$ case) and then increases with energy.  The tree level predictions reach the highest values at the highest studied energies,  leading to very large cross sections (generically larger  for $a_4$ than for $a_5$)  which indeed enter into the unitarity violating region (see the corresponding plot for the partial wave below) for the largest $10^{-3}$ values  at around 2400 GeV for $a_4$ and  2800 GeV for $a_5$.  The predictions for the lower values of $10^{-4}$ lay all in the unitarity preserving region.

Regarding the NLO predictions within the EChL in \figref{plots-aSM-a4ya5},  they can be generically and jointly summarized by noticing that all the $ {\rm EChL}_{\rm Full}$ lines (solid lines other than the red and pink ones) follow approximately the same pattern with energy as the corresponding tree level ones,  $ {\rm EChL}_{\rm Tree}^{(2+4)}$ (dashed lines).  This is in clear concordance with the typical behaviour in ChPT where both the $a_i$ contributions from ${\cal L}_4$ and the loop contributions from ${\cal L}_2$ count equally in the chiral counting,  and provide together the NLO term in the chiral expansion (i.e., in the expansion in powers of the external momentum).  Furthermore,  we also see in these plots that all these EChL \1loop predictions for the cross-section lay below the EChL tree level ones and,  therefore,  they provide a negative \1loop correction $\delta_{\rm{1-loop}}$,  as it happens in the SM case.  For $|a_4|$ (or $|a_5|$)  of $10^{-4}$ it reaches a maximum of about  $\delta_{\rm{1-loop}}\sim -15\%$,  and for $|a_4|$ ($|a_5|$) of $10^{-3}$ the maximum decreases slightly to around $\delta_{\rm{1-loop}}\sim -10\%$ ($-12\%$).  For the high energy region  the relative size of the \1loop correction in general decreases due to the dominance of the quadratic terms already mentioned above.

Let us now move on to the results in  \figref{plots-apoint9-a4ya5} that summarize our findings for $a \neq 1$.   First of all, we see that for $a=0.9$ the  equivalence between the LO prediction within the EChL and the SM is lost.  Specifically,  the $ {\rm EChL}_{\rm Tree}^{(2)}|_{a=0.9}$ prediction (light blue lines) clearly separates from the SM one (pink lines),  growing with $\sqrt{s}$ and reaching values of up to $\sim$350 pb at the highest studied energies of  3000 GeV.  Then all the other EChL predictions shown in these plots converge at the lowest energies to this LO-EChL result and not to the LO-SM one.  When looking at the pattern with growing energy,  the qualitative behaviour of all the lines basically replicate those of the previous figure, but now the departures and interference effects occur respect to this LO-EChL prediction (light blue lines).  Again, the effects from the $a_{4,5}$ coefficients show a similar pattern as before with constructive (destructive) contributions for the negative (positive) cases.  The main difference respect to previous figure for $a=1$, is in the size of the full \1loop cross section which gets  larger for $a=0.9$ and also the relative \1loop contribution respect the tree level one gets larger and it to happens to change the sign for the largest studied values of the $a_i$ coefficients at the high energies.  For   $|a_{4,5}|$  of $10^{-4}$ we get a maximum negative value of about $\delta_{\rm{1-loop}}\sim -20\%$,  and for $|a_4|$ ($|a_5|$) of $10^{-3}$ we get a maximum negative value of around $\delta_{\rm{1-loop}}\sim -12\%$ ($-20\%$) and a maximum positive value of around $\delta_{\rm{1-loop}}\sim 8\%$ ($20\%$). 

We next move to \figref{plots-botha4a5}, where the effects from both  $a_4$ and $a_5$ are considered together for the two cases $a=1$ (left panels) and $a=0.9$ (right panels).  In general,  these two coefficients lead to combined effects that provide larger cross sections when they are of same sign than when they are of opposite sign. The pattern of the EChL tree level cross section predictions as a function of these two parameters  for fixed $\sqrt{s}$,  provides the typical  contours with elliptical shape in the $(a_4,a_5)$ plane.  For high energies in the ${\cal O}(\rm TeV)$ range,  and for relative large values of these parameters  of ${\cal O}(0.01)$ the ellipsis has the longer axis within the quadrants with opposite sign $a_{4,5}$ and the shorter axis within the quadrants with same sign $a_{4,5}$, leading to stronger constraints from experimental data in this later case (see for instance,  \cite{paperUnitarity} and {\cite{ATLAS:2016nmw}).  These same features of the EChL tree level prediction for the two cases with same/opposite signs for $a_4$ and $a_5$ can be seen in \figref{plots-botha4a5},  but now as a function of $\sqrt{s}$.  As we can see in  this figure, the dominance of the cross section for the same sign case over the opposite sign case occurs for the $a=1$ case (left panel) and for very large energies,  above 2400 GeV,  due to the small values $|a_i \sim 10^{-4}|$ assumed here. 
The most important results in this figure are the EChL \1loop predictions that again lay below the corresponding tree level ones for all the $a_{4,5}$ values considered in the case $a=1$ (left plots) and they are above the tree level  predictions only at the highest energies for $a=0.9$ (purple lines in the right plots).  This implies that the \1loop corrections get negative and the maximum reached values for  $\delta_{\rm{1-loop}}$ are around $\sim -15\%$ for $a=1$ and around $\sim -20\%$ for $a=0.9$.  It should be noticed that in this figure,  in addition to $\pm 10^{-4}$ we have chosen another  different  value for $\vert a_{4,5}\vert$.  Then the predictions in grey in this \figref{plots-botha4a5}  correspond to  those specific values taken in \cite{Delgado:2017cls} to which we wish to compare with.  In this reference an approximation to compute the EChL \1loop corrections in $W_LZ_L \to W_LZ_L$ was done, consisting in taking just the loop-corrections from scalars (i.e., the chiral loops) to evaluate the real part of the amplitude, whereas the imaginary part was not computed but instead was replaced by the LO squared (by means of the Optical Theorem). 
Our full results for $a=0.9$ (solid grey lines in the right panel),  including all loops give interestingly very similar numerical predictions for high energies,  say above 1 TeV,  than those in figure 2 of \cite{Delgado:2017cls},  
indicating that these chiral loops are indeed the dominant ones at these high energies.  This was indeed expected from the general features of ChPT but getting this result from the complex \1loop computation within the $R_\xi$ gauges is not a trivial task at all. 
\begin{figure}[h!]
\begin{center}
    \includegraphics[width=1.\textwidth]{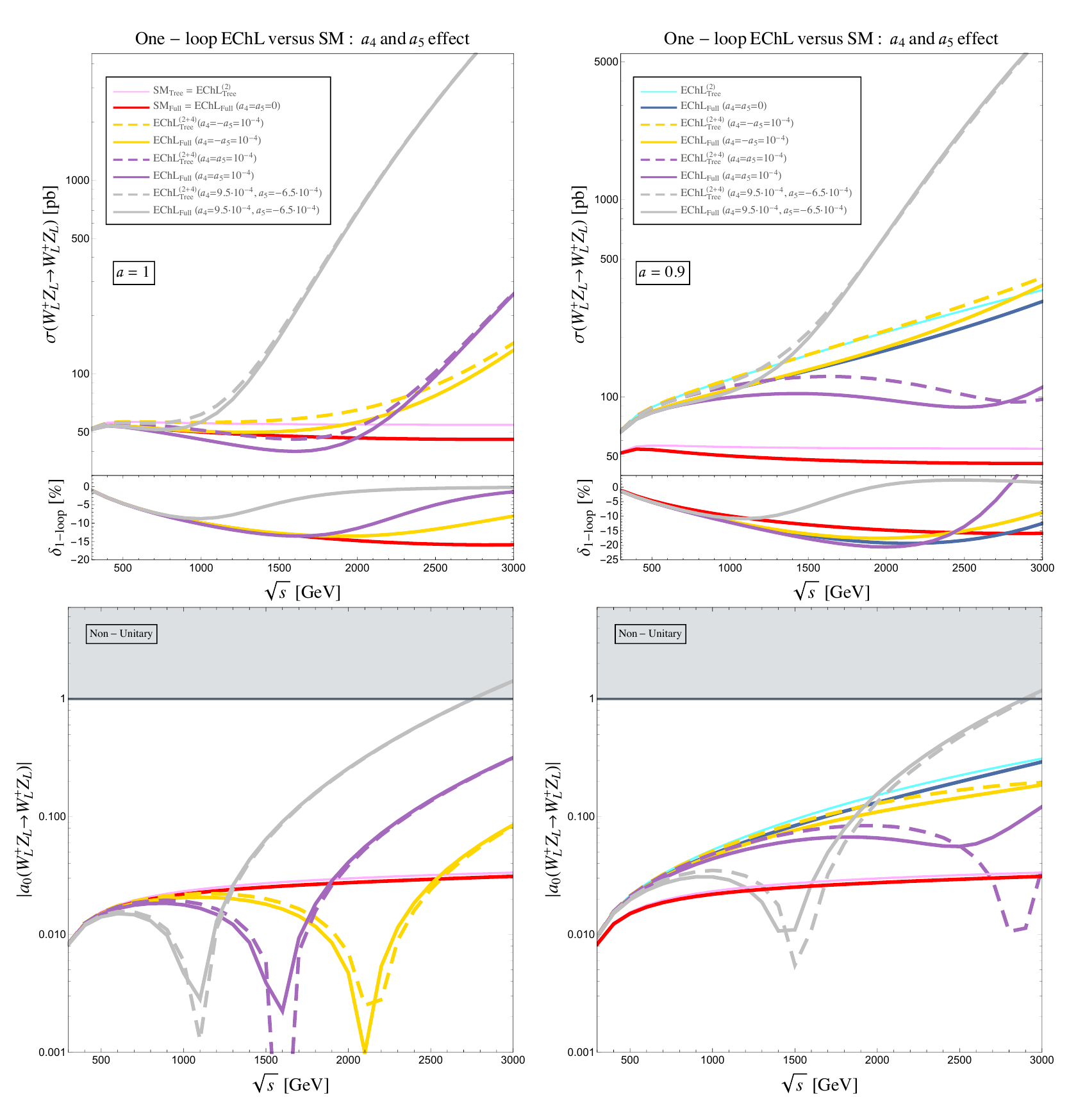}
\caption{Cross-section predictions for $W_L^+ Z_L\to W_L^+ Z_L$ within the EChL corresponding to $a=1$ (upper-left), $a=0.9$ (upper-right), $b=a^2$, $\kappa_3=\kappa_4=1$ and varying $a_4$ and $a_5$ simultaneously. The rest of $a_i$'s  are set to zero. The $ \mu$ scale is set to  $\mu=1$TeV.  Dashed lines are the tree level predictions.  Solid lines are the full one-loop predictions. The solid red (pink) lines correspond to the full (tree) SM prediction. The relative size of the one-loop prediction respect to the tree level one defined by means of $\delta_{\rm{1-loop}}$ is also shown (lower plots in the upper panels).  The predictions for the lowest partial wave $a_J$ with $J=0$ are also displayed (lower panels).  The unitarity violating regions in these plots are indicated by the shadowed areas.}
\label{plots-botha4a5}
\end{center}
\end{figure}
\newpage


Finally, to end this section,  we have analyzed further the issue of perturbative unitarity  within the EChL and compare with the SM case.  To perform such analysis one must use the full \1loop results for the partial wave amplitudes and check the validity of the unitarity condition involving the real and imaginary parts of the partial wave, namely, checking the validity of the equation Im$[a_{J}]=|a_{J}|^2$ that is derived from the elastic unitarity condition of the scattering matrix,  according to the Optical Theorem.  We have checked in \figref{plot-Uperturbative} that this condition for $J=0$ is fullfiled perturbatively at all the studied energies in both the full \1loop SM (dots in red) and full \1loop  EChL predictions (dots in blue).  Obviously the tree level predictions fail this relation since they are real for both the SM and EChL.  The interesting finding is that, in the EChL case,  this unitarity condition is fulfilled  perturbatively in the sense of ChPT, namely,   
Im$[a_{0}^{(1)}]= |a_{0}^{(0)}|^2$.  Or in other words, the NLO correction unitarizes the LO prediction. In our opinion, this a valuable result given the complexity of the full \1loop computation involving many loop diagrams of all types (more than 500),  that generate finally this imaginary part.  The unitarity condition is then fullfiled perturbatively but it is not accomplished, however,  with the full quantities in both the imaginary and real part.  Namely,   the values of Im$[a_{0}^{\rm Full}]$ and 
$|a_{0}^{\rm Full}|^2$ differ in the EChL,  mainly for large $a_{4,5}$ EChL coefficients and at high energies,  and this leads to the crossing over the unitary bound,  $|a_0|>1$,  as  can be seen for instance in some the lines in the previous figures of this section.
\begin{figure}[h!]
\begin{center}
    \includegraphics[width=0.62\textwidth]{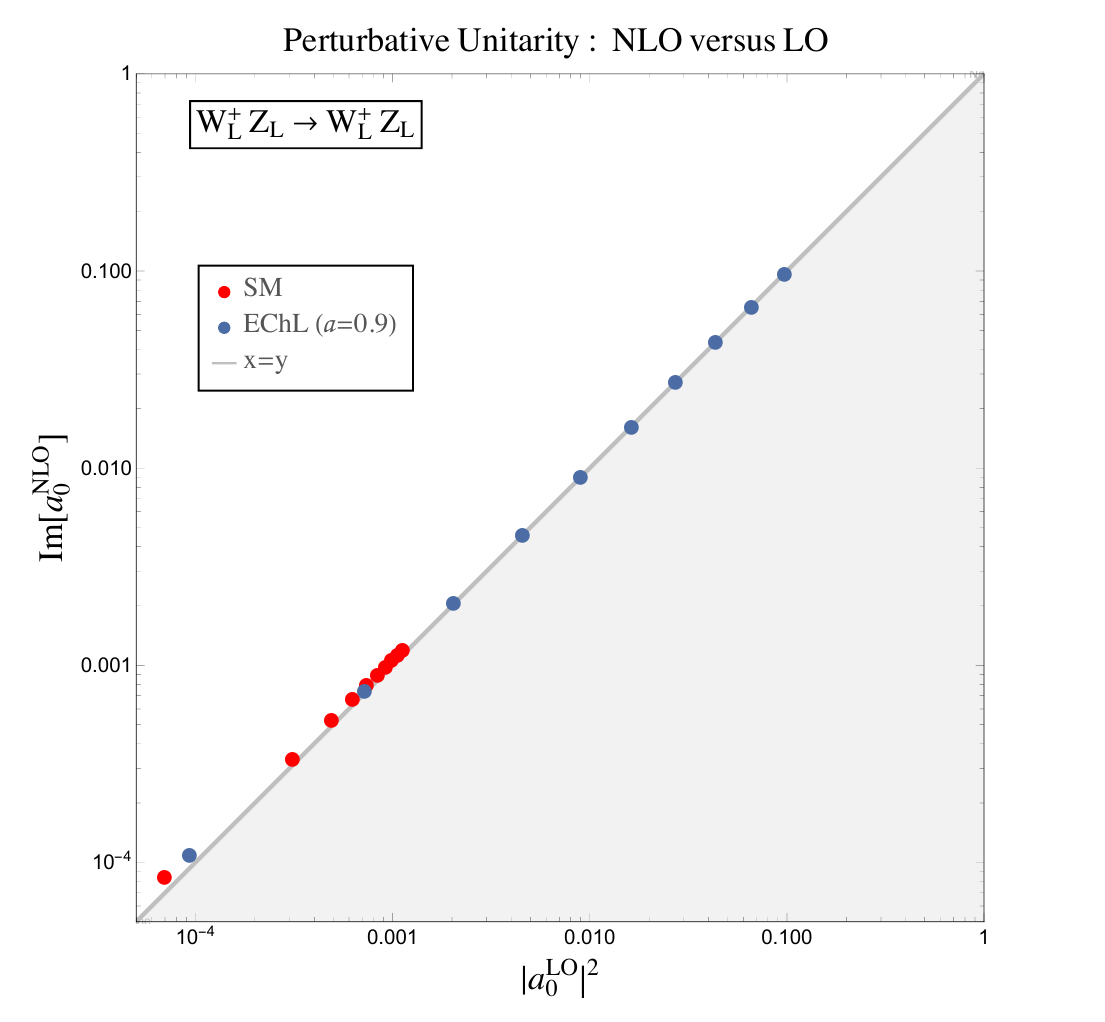}
\caption{Check of perturbative unitarity in the one-loop predictions for the lowest partial wave $a_0$ in both the EChL with $a=0.9$ and the  SM.  The 10  blue (red) points are the EChL (SM) predictions for the ten chosen energies,  $\sqrt{s} \, [{\rm GeV}]= 300,\,600,\,900\,,1200,\,1500,\,1800,\,2100,\,2400,\,2700,\,3000$,  from left to right along the diagonal line in this plot. }
\label{plot-Uperturbative}
\end{center}
\end{figure}

\section{Conclusions}
\label{sec-conclu}
In this paper we have computed the one-loop corrections to $WZ \to WZ$ scattering within the context of the non-linear EFT determined by the bosonic sector of the EChL.  This is a novel and full (EW) one-loop computation of the VBS amplitude within the covariant $R_\xi$ gauges that accounts for  all kind of bosonic loop diagrams including: EW gauge bosons,  Goldstone bosons,  the Higgs boson and the ghost fields.  For this computation we have used a diagrammatic method where the renormalization of the full one-loop amplitude is performed by first computing the  involved renormalized 1PI Green functions which is a more demanding computation since it requires the renormalization to be implemented at the off-shell  Green functions level, i.e., for arbitrary external legs momenta.   We have then established the systematics for the renormalization program of all the off-shell 1PI one-loop functions.  The regularization method that we have used through this work is dimensional regularization in $D=4-\epsilon$ dimensions where all the divergences are given in terms of $\Delta_\epsilon=2/\epsilon-\gamma_E+\log(4\pi)$  and the associated energy scale $\mu$  appears typically in logarithms.   As renormalization prescription we use multiplicative renormalization and  fix the renormalization parameters and associated counterterms in a hybrid way:  for the EW parameters involved (EW wave functions, masses and couplings) we use  the OS scheme, whereas for all the EChL coefficients denoted generically by $a_i$ we chose the $\overline{MS}$ scheme.  

After computing the very many loop diagrams involved in the renormalization of the 1-leg,  2-legs, 3-legs and 4-legs 1PI functions  we extract all the divergences in the $R_\xi$ gauges and present here all the needed counterterms  to renormalize these divergences.  We  summarize in \eqrefs{L2param-div}{L4param-div} the analytical results that we have found for all of  the divergent counterterms, both for the EW parameters and the EChL coefficients,  respectively.  Whereas the $\xi$ appears in the analytical results of the EW parameters counterterms, we find in contrast,  that all the $\delta a_i$' s counterterms are $\xi$-independent.  This implies that the EChL  coefficients are gauge invariant. Within this non-linear EFT,  the EChL coefficients of the effective operators in the Lagrangian with chiral dimension four act as countertems of the extra divergencies generated by the Lagrangian with chiral dimension two.  Thus  our  results  for the $\delta a_i$'s, which are $\xi$-independent,  together with the ones for the EW parameters,  summarize all the needed counterterms to find finite off-shell 1PI and, therefore, also to arrive to a finite prediction for the VBS amplitude at one-loop.  We have also presented the derived results for the running coefficients $a_i(\mu)$ and their corresponding RGEs.  Out of the full set of $a_i$'s only, a subset of them have non-vanishing divergent counterterms $\delta_\epsilon$, and therefore, only the corresponding  subset of running coefficients $a_i(\mu)$ do run indeed with the scale.  Saying this in different manner, the remaining coefficients or combination of coefficients with $\delta_\epsilon =0$ are RGE invariants.  
On the other hand, the fact that the RGEs found are written separately  for each  coefficient $a_i(\mu)$ and not as a coupled system of differential equations implies that these coefficients do not mix under RGEs.
In addition, we have also discussed in this work, the relevance of the STI and the role played by the Higgs tadpole which we  also find different in the EChL than in the SM.

We have also presented the results in parallel for the one-loop computation for the SM VBS  amplitude and the corresponding 1PI Green functions involved within the $R_\xi$ gauges which we believe are also novel results.  Our SM results are interesting by themselves and deserve some discussion to them, but we have focused  on the  SM comparison with our 
EChL results. 

In the last part of this work we have presented the numerical results for the full one-loop cross section of the longitudinal polarized gauge bosons  $\sigma(W_LZ_L \to W_L Z_L)$.  For this numerical computation we have set the Feynman 't Hooft gauge and considered only the subset of EChL coefficients of the custodial preserving effective operators that are needed as counterterms.  We have further assumed the simplifying relation $b=a^2$ among the most relevant coefficients, $a$ and $b$  of the lowest order Lagrangian and set $\kappa_3=\kappa_4=1$ in the Higgs boson potential,  as in the SM.  The  full one-loop cross section  $\sigma(W_LZ_L \to W_L Z_L)$ is then analyzed as a function of the most relevant renormalized EChL coefficients, $a$, $a_4$ and $a_5$ and as a function of the energy of the VBS process.  To cover different scenarios for these coefficients we have explored six benchmark cases,  which    address the study of the effects from $a_4$ and $a_5$ varying their modulus in the interval $(10^{-3}-10^{-4})$ and considering the two cases of positive and negative coefficients. The $a$ parameter is fixed to either 1, the SM reference value, or  to the BSM value of 0.9.  From our  systematic study of the size of the one-loop corrections in the cross section  respect the tree level one by means of the $\delta_{\rm 1-loop}$ quantity we conclude that the one-loop corrections within the EChL are of  comparable size to the SM ones which has also been computed and presented here.  For the SM one-loop corrections in the explored energy interval of (300 GeV,  3000 GeV) we find negative values reaching a maximum size of $\delta_{\rm 1-loop}\sim -16\%$.  For the EChL case and varying  the EChL coefficients in the benchmark cases mentioned above,  we find in this same energy  interval negative values reaching a maximum size of  $\delta_{\rm 1-loop}\sim -20\%$,  and we also find positive values reaching a maximum size of $\delta_{\rm 1-loop}\sim +20\%$.  These numerical  results are just for the $WZ \to WZ$ channel but we expect similar sizeable corrections  in other VBS and VBF processes.  These other processes are left for future works.  

\section*{Acknowledgments}

The present work has received financial support from the ``Spanish Agencia Estatal de Investigaci\'on'' (AEI) and the EU ``Fondo Europeo de Desarrollo Regional'' (FEDER) through the projects FPA2016-78022-P,   PID2019-108892RB-I00/AEI/10.13039/501100011033 and from the grant IFT Centro de Excelencia Severo Ochoa SEV-2016-0597.  We also acknowledge financial support  from the European Union's Horizon 2020 research and innovation program under the Marie Sklodowska-Curie grant agreement No 674896 and No 860881-HIDDeN and the RISE INVISIBLESPLUS H2020-MSCA-RISE-2015//690575. 
The work of RM is also supported by the ``Atracci\'on de Talento'' program (Modalidad 1) of the Comunidad de Madrid (Spain) under the grant number 2019-T1/TIC-14019.
\newpage

\section*{Appendices}
\appendix

\section{Relevant Feynman rules}
\label{App-FRules}

In this appendix we summarize all the relevant Feynman rules participating in the computation using the $R_\xi$ gauges.
In \tabrefs{relevant-FR1}{ghost-FR} we collect those relevant Feynman rules for the EChL interaction vertices coming from $\mL_2$ of \eqref{eq-L2} and the corresponding SM Feynman rules for a clear comparison between them. We use the following short notation for the standard Lorentz tensors of the gauge boson self couplings:
\bear
V^{\mu\nu\rho}(p_-,p_+,p_0) &=& g^{\mu\nu}(p_--p_+)^\rho+g^{\nu\rho}(p_+-p_0)^\mu+g^{\rho\mu}(p_0-p_-)^\nu \,,  \nn\\
S^{\mu\nu,\rho\sigma} &=& 2g^{\mu\nu}g^{\rho\sigma} -g^{\mu\rho}g^{\nu\sigma} -g^{\mu\sigma}g^{\nu\rho} \,,
\label{self-gauge-vertex}
\eear
and, to shorten the tables, we present together the photon and $Z$ boson interaction vertices in most of the Feynman rules. In that cases, the first coupling corresponds to the photon and the second one to the $Z$.
\begin{table}[H]
\includegraphics[width=160mm]{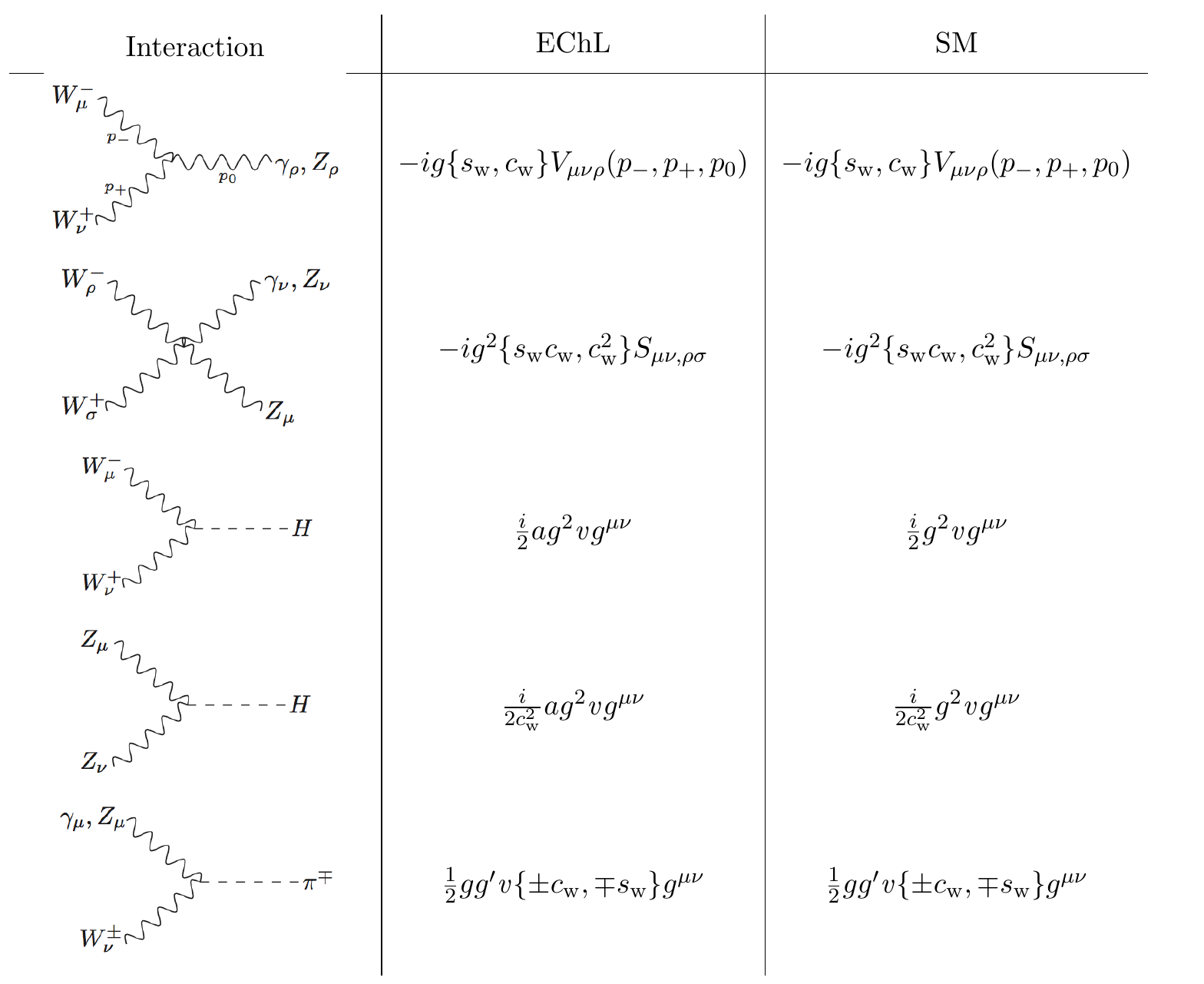} 
\caption{Relevant EChL Feynman rules from $\mL_2$ entering in VBS processes at LO.  All momenta are incoming. In the rules with `$\{\,,\}$', the first and second arguments correspond to the photon and $Z$ boson, respectively. We also include the corresponding SM Feynman rules for comparison.}
\label{relevant-FR1}
\end{table}
\begin{table}[H]
\includegraphics[width=170mm]{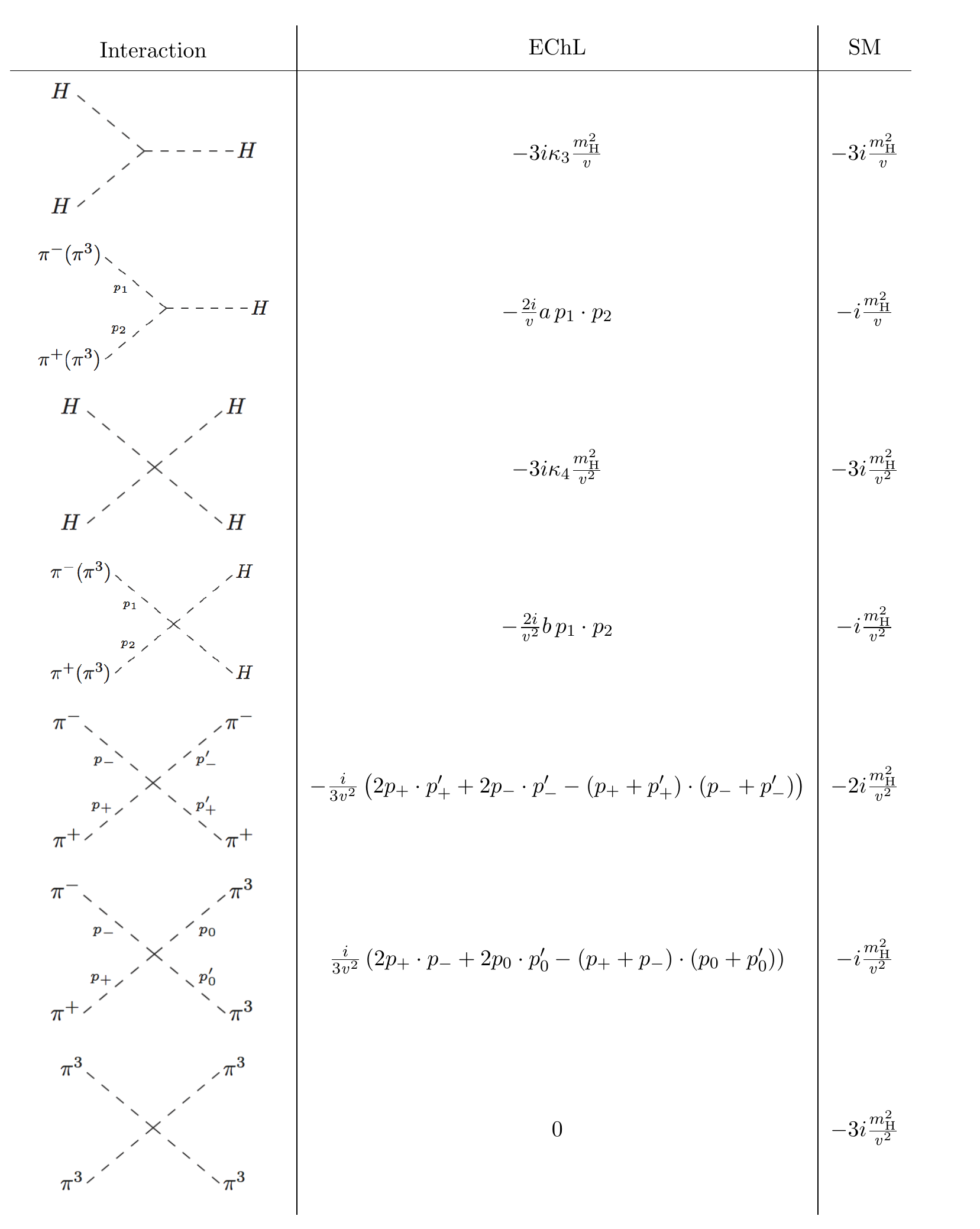} 
\caption{Additional relevant EChL Feynman rules from $\mL_2$, in the scalar sector, participating in the loops contributions to VBS processes. All momenta are incoming. The corresponding SM rules are included for comparison.}
\label{purescalars_FR}
\end{table}
\begin{table}[H]
\includegraphics[width=150mm]{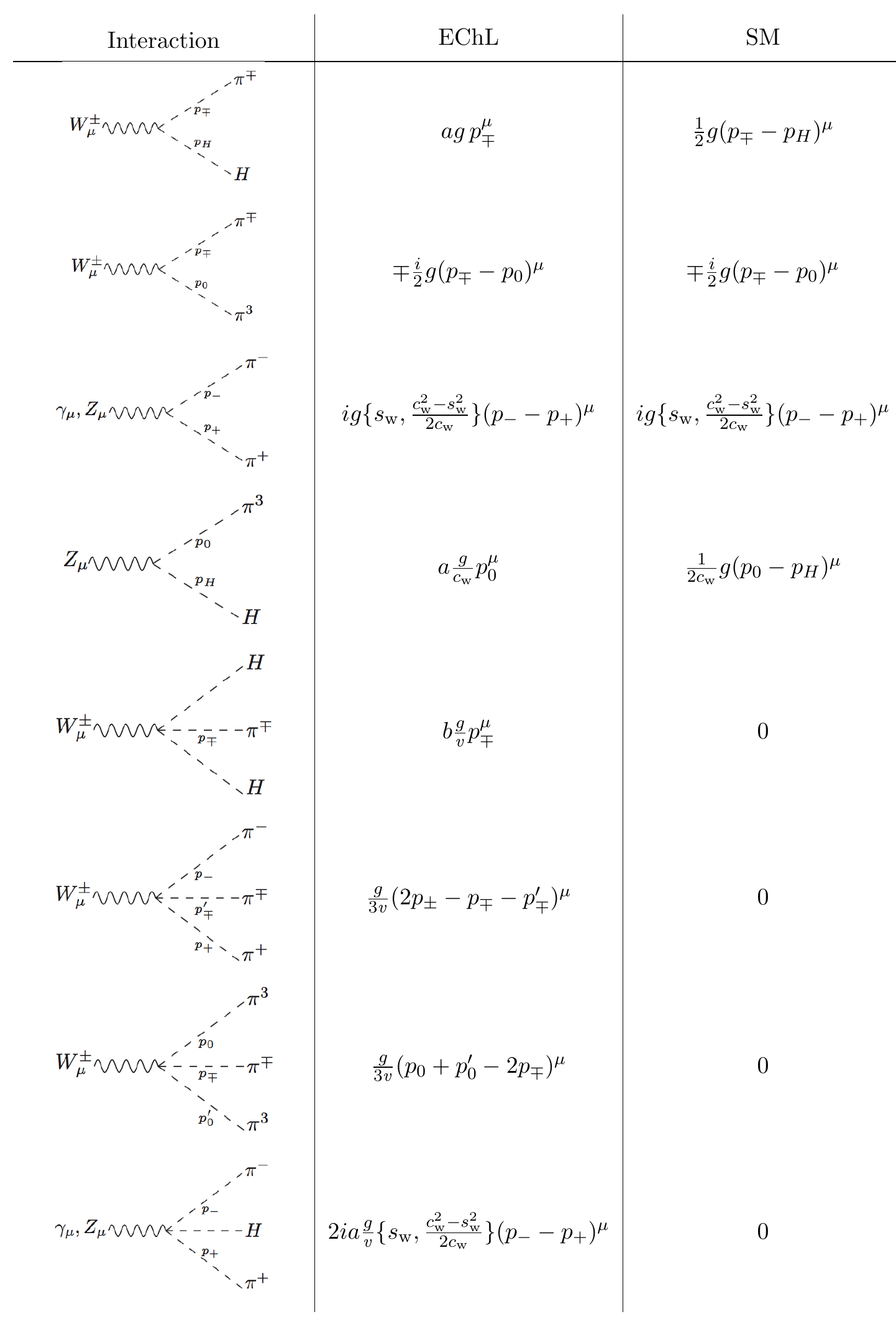} 
\caption{Additional relevant EChL Feynman rules from $\mL_2$, with one gauge boson, participating in the loops contributions to VBS processes.  In the rules with `$\{\,,\}$', the first and second arguments correspond to the photon and $Z$ boson, respectively. All momenta are incoming. The corresponding SM rules are included for comparison.}
\label{onegauge_FR}
\end{table}

\begin{table}[H]
\includegraphics[width=150mm]{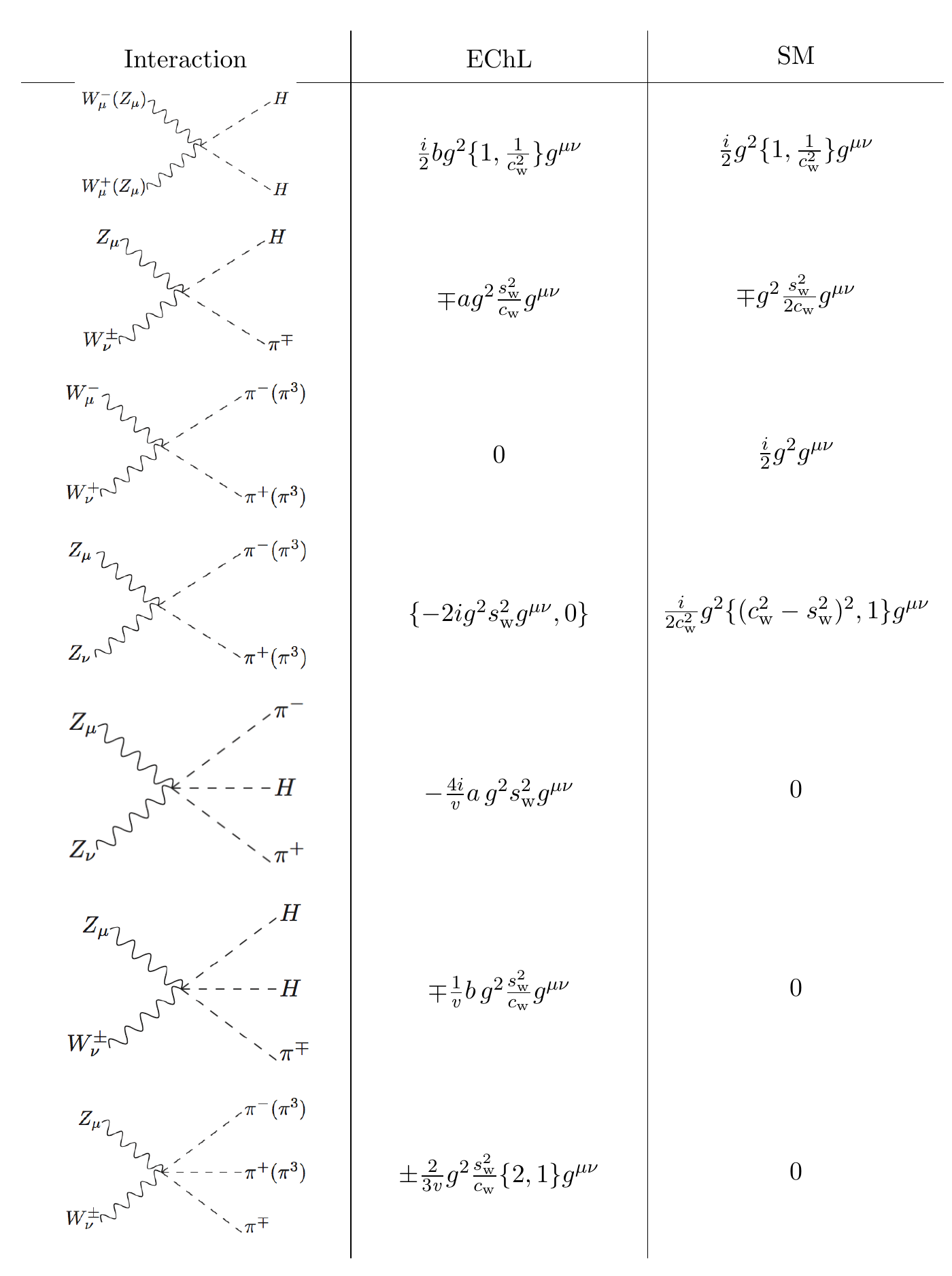} 
\caption{Additional relevant EChL Feynman rules from $\mL_2$, with two gauge bosons,  participating in the loops contributions to VBS processes.  All momenta are incoming.  The corresponding SM rules are included for comparison.}
\label{twogauge_FR}
\end{table}
\begin{table}[H]
\includegraphics[width=139mm]{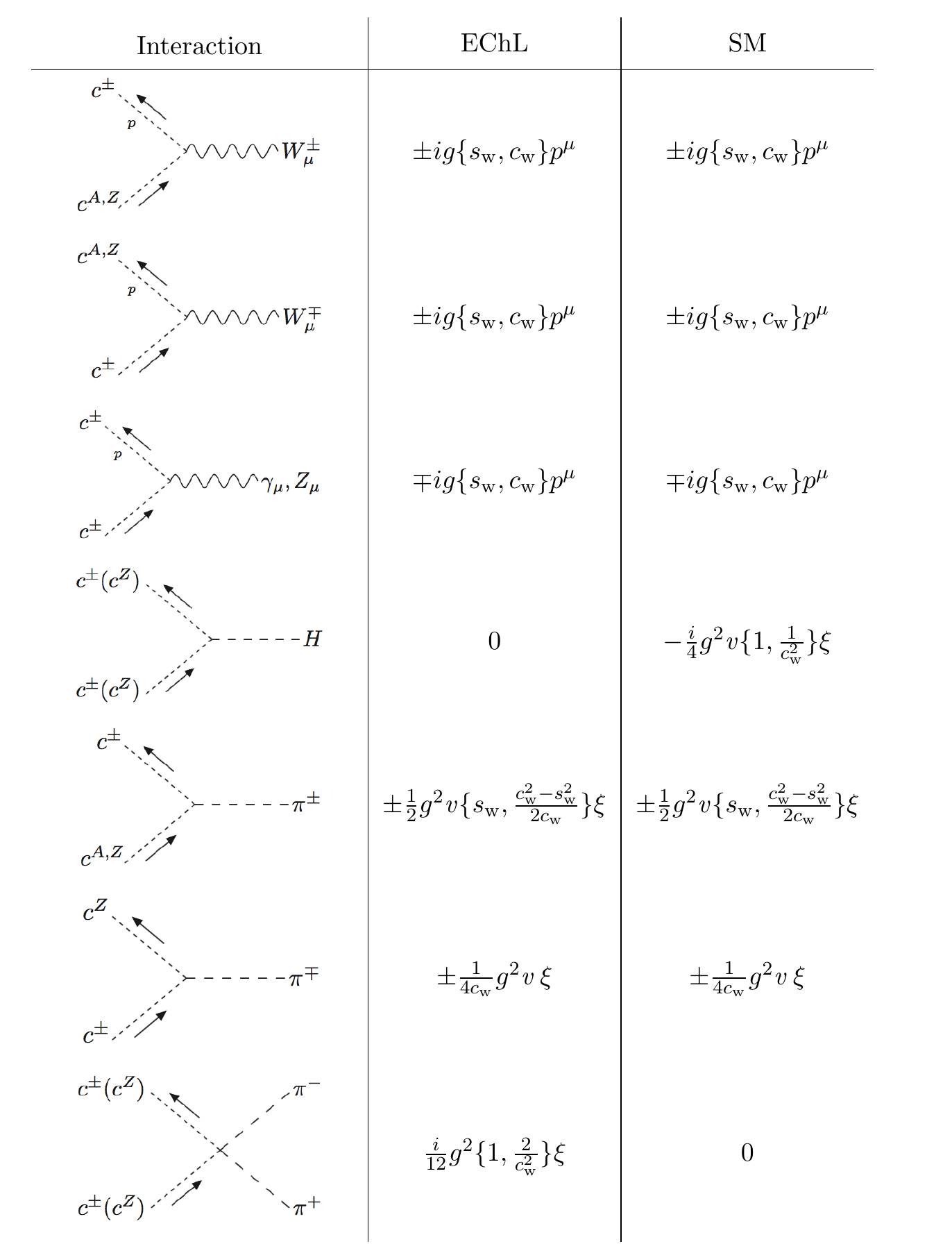} 
\caption{Additional relevant EChL Feynman rules from $\mL_2$, involving ghosts,  participating in the loops contributions to VBS processes.  All momenta are incoming (except for the $c$'s, given by the arrows).  The corresponding SM rules are included for comparison.}
\label{ghost-FR}
\end{table}
\section{Relevant \1loop diagrams}
\label{App-oneloopdiag}
Here we present the relevant \1loop diagrams entering in the computation of the 1PI functions within the 
EChL.   In order to shorten the number of diagrams in the full list that is generated with FeynArts~\cite{FeynArts},  we have grouped them by different topologies and used a generic notation for the internal propagators: dashed lines refer to all possible scalar particles $S$,  namely,  Higgs  boson and Goldstone bosons;  wavy lines refer to all possible EW gauge bosons; and dotted lines refer to the ghost fields.  

The generic loop diagrams entering in the specific 1PI functions that participate in the present computation of  the $WZ \to WZ$ scattering amplitude are then summarized in the following figures: \figref{SE-generic-loops} (self-energies), \figref{VVV-generic-loops} ($VVV$),  \figref{piWZ-generic-loops} ($\pi WZ$),  \figref{HVV-generic-loops} ($HVV$),  and \figref{WZWZ-generic-loops} ($WZWZ$).  

As we have remarked in \appref{App-FRules},  the Feynman rules in the EChL from ${\cal L}_2$ and the SM are different, and in consequence the loop-diagrams contributions are also different in both theories.  These differences can be summarized in 5 categories: {\it i)} the self interactions among GBs have momentum dependence in the EChL (this is a typical feature of the non-linear EFTs),  in contrast to the SM case,  {\it ii)} In the EChL there are new interactions of the Higgs with gauge bosons interactions given by the $a$ and $b$ coefficients,  and these coincide with the SM ones only for $a=b=1$,   {\it iii)} interactions of gauge bosons with multiple (more than 2) scalars are present in the EChL but not in the SM (these are also typical in the non-linear EFTs), {\it iv)} there are not Higgs-ghosts interactions in the EChL (since $H$ is a singlet),  in contrast to the SM, {\it v)} there are interactions of ghosts with multiple GBs in the EChL (due to the non-linear GBs  transformations under $SU(2)_L\times U(1)_Y$)  that are not present in the SM.  Other differences due to $\kappa_{3,4} \neq 1$ do not affect the present computation of the \1loop $WZ$ scattering amplitude.
The previous remarked differences give rise to different results for the loops computation in the EChL and the SM: {\it i)}-{\it ii)} derive in different results for the same topological diagrams in both theories;
 {\it iii)}-{\it v)} produce different (presence or absence) diagrams depending on the theory.  We comment these issues in more detail for each relevant 1PI in our computation, in the following.

We start with \figref{SE-generic-loops} corresponding to the self-energies. In the Higgs boson self-energy  (first column) the ghosts do not participate in the EChL case but they are present in the SM computation.  All the results in the EChL depend on $a$ or $b$ and the resulting momentum dependence in the self-energy is quite different due to the different behaviour of the scalar loop diagrams (chiral loops) in the EChL and the SM.
The second column corresponds to $\Sigma_{\pi\pi}$, in which the momentum dependence from the chiral loop diagrams is also quite different and there is an additional topology (the last one) coming from the two GB with two ghosts interactions proper of the EChL.
The main difference in the self-energy connecting a $W$ with its corresponding GB (third column) arises from the first topology since it is not present in the SM. 
The $\Sigma_{WW}$ topologies are the same in both theories, but the results involving scalars differ. Notice that the pure gauge and ghost loops give the same contributions in both theories.

The 1PI Green functions corresponding to three gauge bosons within the EChL are collected in \figref{VVV-generic-loops}. The generic topologies are the same as in the SM. However the corresponding results differ only for the loop diagrams with scalars.  The contributions from pure gauge and ghost loops  coincide in the EChL and the SM.

Regarding the $\pi WZ$ Green function of \figref{piWZ-generic-loops}, the first three topologies are genuine of the EChL and are not present in the SM since they involve interactions of three scalars with two or one gauge bosons. The remaining diagrams with scalars in the loop give different results in both theories. However, the pure gauge and ghost loops coincide again.

The \1loop diagrams for the  3-legs 1PI involving the Higgs and two gauge bosons ($WW$ and $ZZ$) are showed in \figref{HVV-generic-loops}. The first (first and second) topologies of $HWW$ ($HZZ$) are present in the EChL, but they are not present in the SM,  since they involve interactions of three scalars with one gauge boson.  As before,  the remaining diagrams with  scalars in the loops also differ, but the loops with only gauge bosons coincide in both theories.  However,  there are no diagrams with ghost in the loops within the EChL.

The observations for the $WZWZ$ 1PI function (in \figref{WZWZ-generic-loops}) are the same than in the three gauge bosons ones. In particular,  all the generic topologies coincide in the EChL and the SM (also the pure gauge and ghost loops predictions) while the corresponding results differ for the diagrams with scalars.

\begin{figure}[H]
\begin{center}
    \includegraphics[width=1.\textwidth]{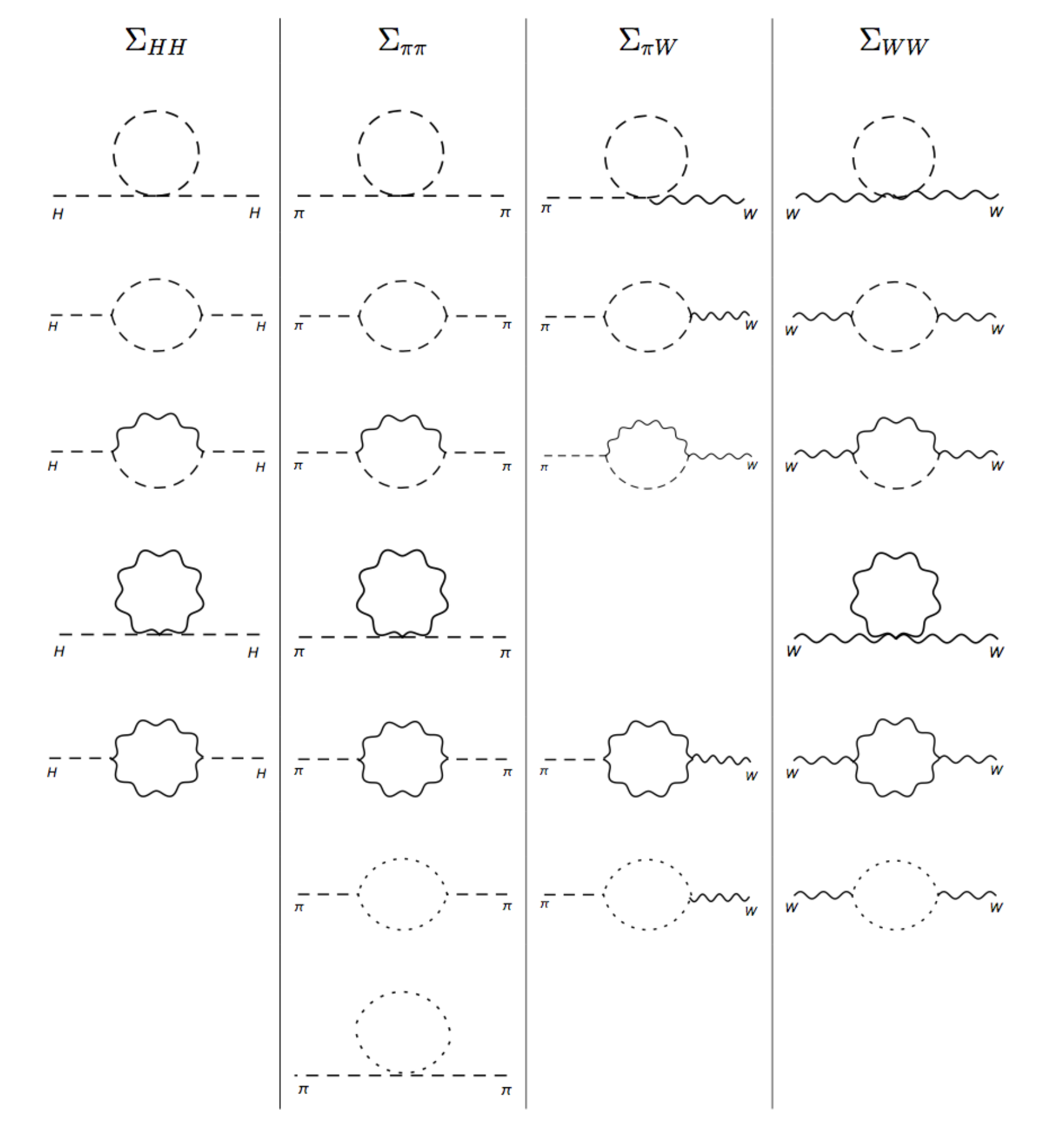}
\caption{Generic loop diagrams for the self-energies in the EChL.  Dashed lines represent the Higgs boson and the GB bosons,  wavy lines represent the EW gauge bosons,  and dotted lines are the ghost fields.  The same notation is used in the loop diagrams of all the figures. }
\label{SE-generic-loops}
\end{center}
\end{figure}

\begin{figure}[H]
\begin{center}
    \includegraphics[width=1.\textwidth]{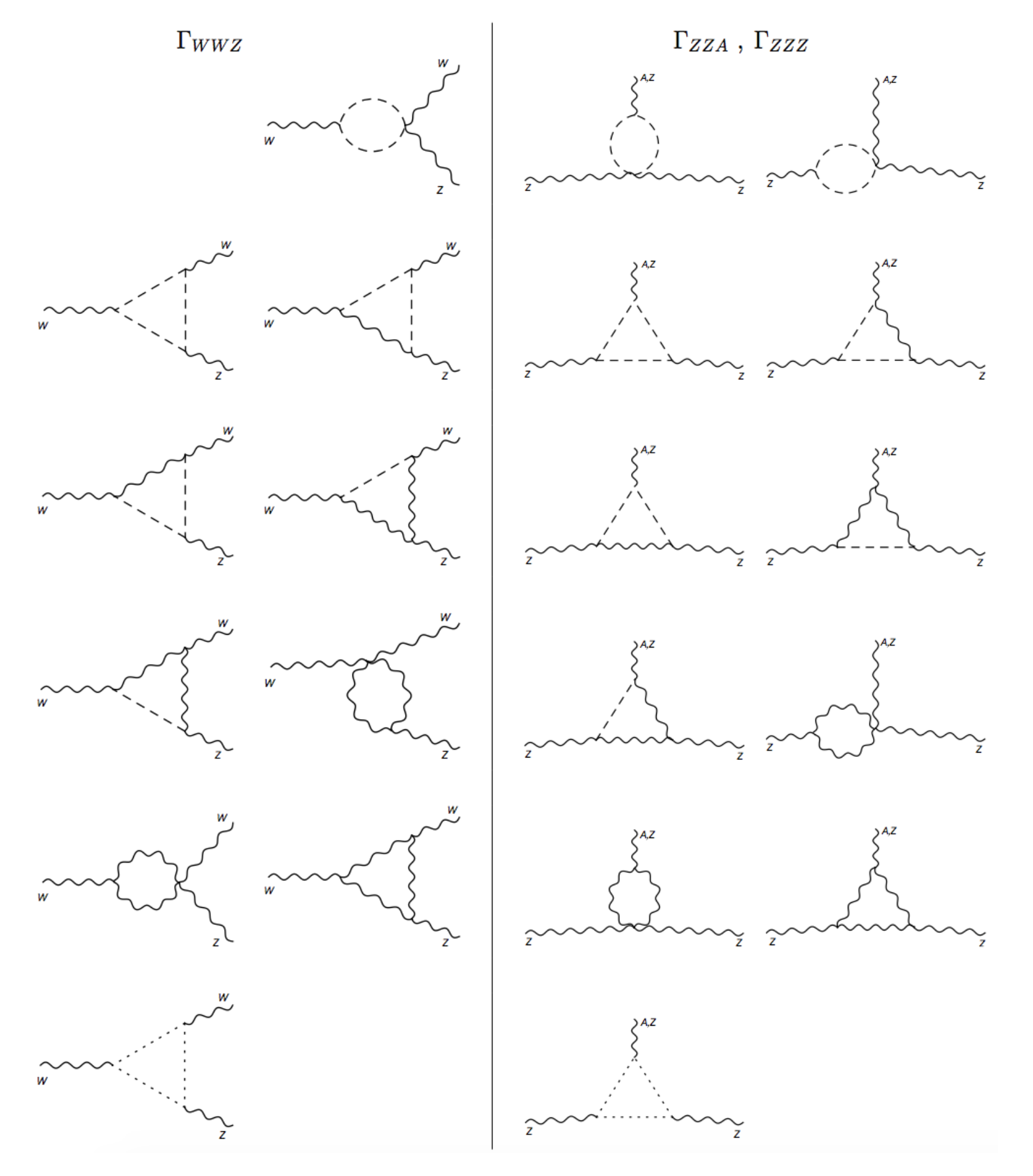}
\caption{Generic loop diagrams for the $WWZ$,  $ZZA$ and $ZZZ$ Green functions in the EChL.}
\label{VVV-generic-loops}
\end{center}
\end{figure}
%


\begin{figure}[H]
\begin{center}
    \includegraphics[width=1.\textwidth]{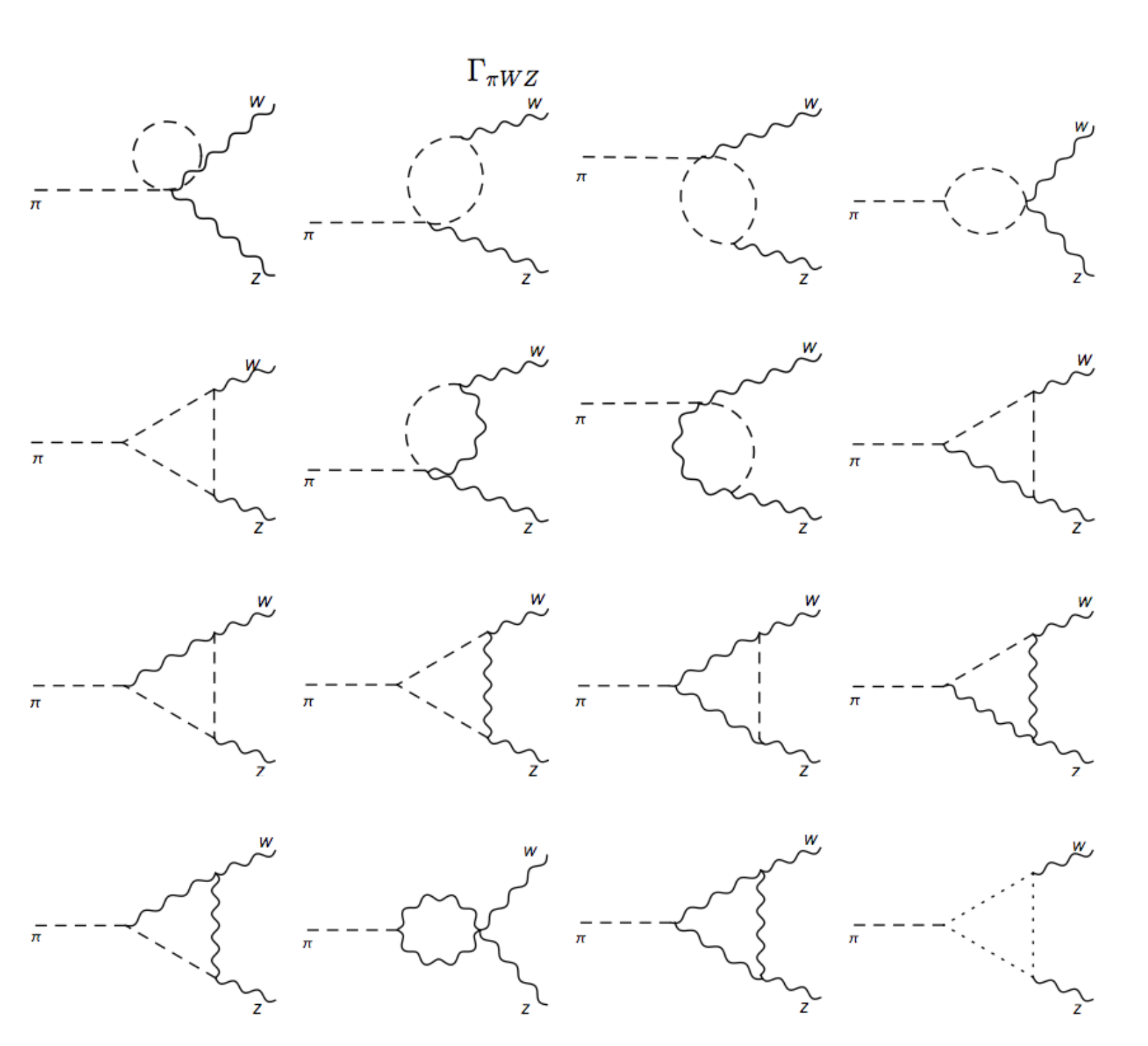}
\caption{Generic loop diagrams for the $\pi WZ$ Green functions in the EChL.}
\label{piWZ-generic-loops}
\end{center}
\end{figure}
\begin{figure}[H]
\begin{center}
    \includegraphics[width=1.1\textwidth]{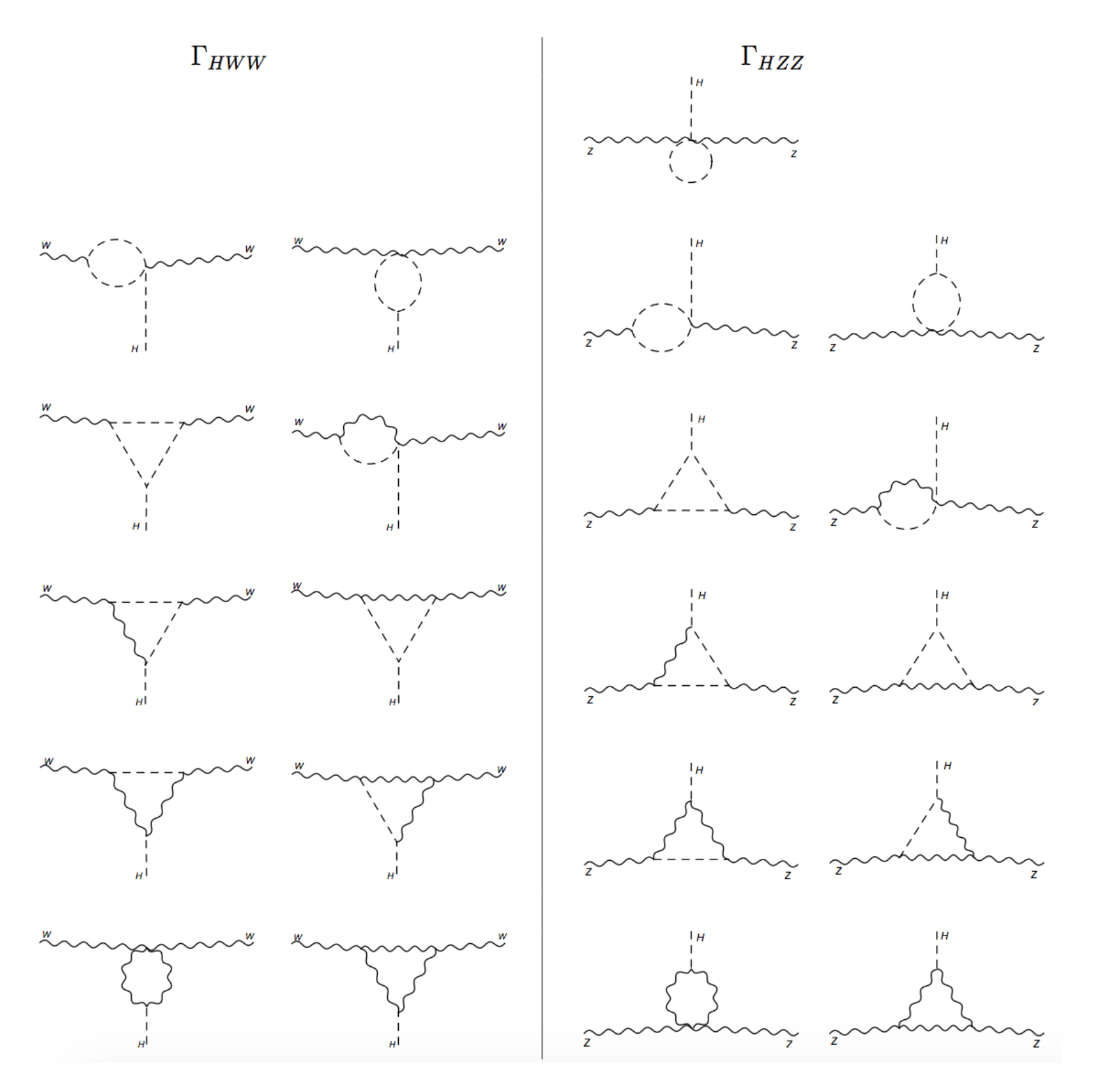}
\caption{Generic loop diagrams for the $HWW$ and $HZZ$ Green functions in the EChL.}
\label{HVV-generic-loops}
\end{center}
\end{figure}
%

\begin{figure}[H]
\begin{center}
    \includegraphics[width=.95\textwidth]{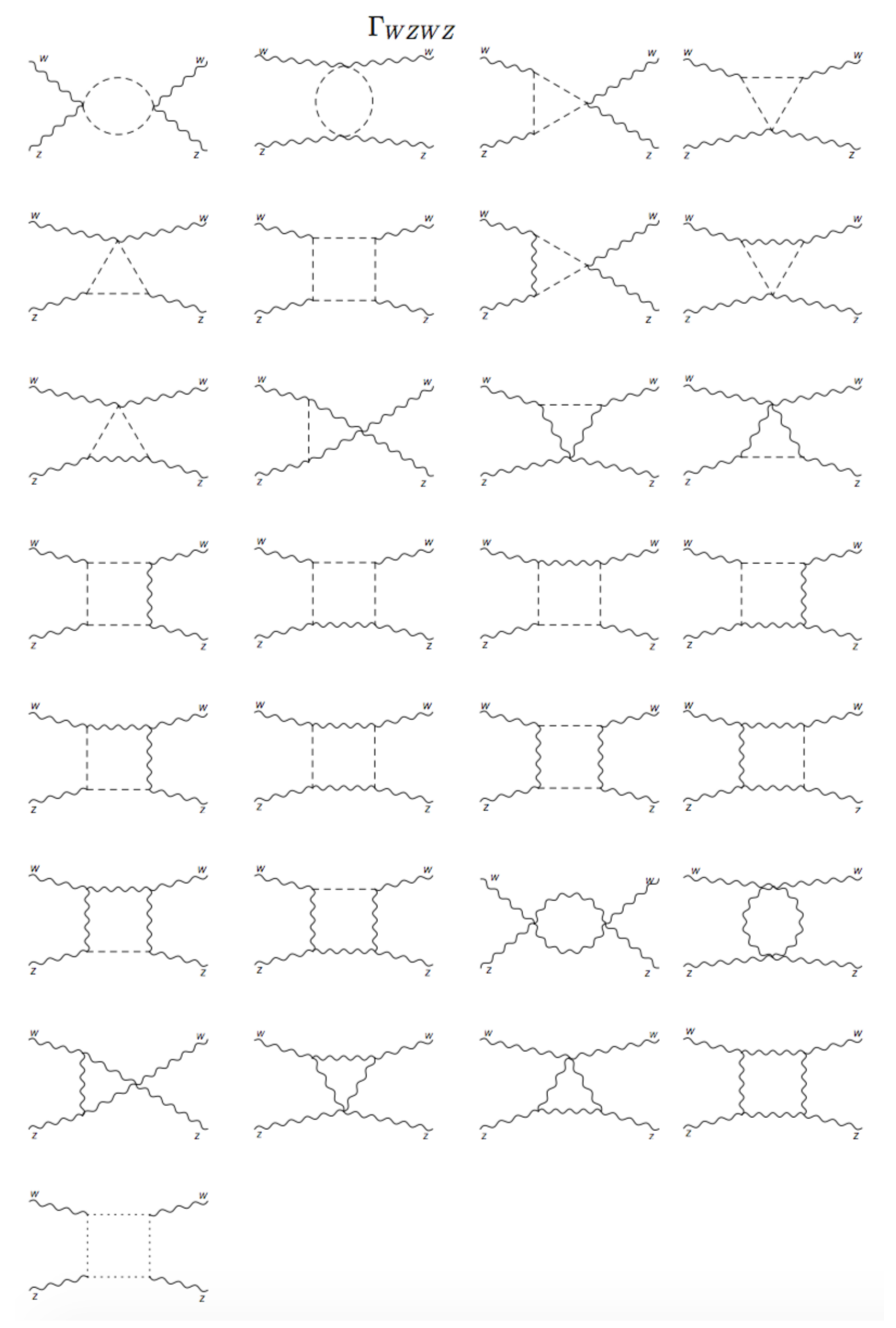}
\caption{Generic loop diagrams for the $WZWZ$  Green functions in the EChL.}
\label{WZWZ-generic-loops}
\end{center}
\end{figure}

\section{The SM case: One-loop divergences and CTs in the $R_\xi$ gauges}
\label{App-SMcompu}

The renormalized 1PI functions within the SM are obtained, as it is usual,  by adding the tree part,  the loop contributions and the CTs, like in \eqref{fgreen}.  
In this appendix, we collect the results for the \1loop divergences of the 1PI functions, as well as the corresponding CTs for the SM case.
To our knowledge,  these have not been provided in the literature in the $R_\xi$ gauges yet and, therefore, they are worth to be considered by themselves.  Besides,  we believe they are interesting to be compared  with the results within the EChL,  which is our main motivation in the present paper.
For illustrative purposes and to differentiate EChL and SM results clearly,  we use the `bar' notation for all the 1PI  functions in the SM, i.e., $\overline{T}$, $\overline{\Sigma}_i$ and $\overline{\Gamma}_i$ denote the SM tadpole, self-energies and $n$-legs functions, respectively.

After the computation of all the relevant \1loop diagrams in the SM,  already mentioned in the appendix \ref{App-oneloopdiag}, we find the following SM results  for the  divergencies in 1- and 2-legs 1PI functions that should be compared with the EChL results in \eqref{Loop-SE}: 
\bear
i\overline{T}^{\rm Loop}\vert_{div} &=& i\frac{\Delta_\epsilon}{16\pi^2}\frac{1}{2\vev} 
 (3\mh^4 +6\left(2 \mw^4+\mz^4\right)+\mh^2(2 \mw^2+\mz^2)\xi) \nn\\
-i\SMSigma_{HH}^{\rm Loop}(q^2)\vert_{div} &=& i\frac{\Delta_\epsilon}{16\pi^2}
\frac{1}{2\vev^2} 
( -2(3-\xi)(2\mw^2+\mz^2)q^2 +15\mh^4+18(2\mw^4+\mz^4)+\mh^2(2 \mw^2+\mz^2)\xi)  \nn\\
i\SMSigma_{WW}^{T\,{\rm Loop}} (q^2)\vert_{div} &=& i\frac{\Delta_\epsilon}{16\pi^2}\frac{g^2}{12}\left((50-12\xi)q^2 +3(3+\xi)(2\mw^2 -\mz^2)\right)  \nn\\
i\SMSigma_{AA}^{T\,{\rm Loop}} (q^2)\vert_{div} &=& i\frac{\Delta_\epsilon}{16\pi^2}e^2(4-\xi)q^2  \nn\\
i\SMSigma_{ZZ}^{T\,{\rm Loop}} (q^2)\vert_{div} &=& i\frac{\Delta_\epsilon}{16\pi^2}\frac{g^2}{12}\left((4-\frac{2}{\cw^2}+12\cw^2(4-\xi))q^2 \right.  \nn\\
&&\left.\hspace{16mm}+12\mw^2(3+\xi)-6(3+\xi)\mz^2-3(3+\xi)\frac{\mz^2}{\cw^2} \right)  \nn\\ 
i\SMSigma_{ZA}^{T\,{\rm Loop}} (q^2)\vert_{div} &=& i\frac{\Delta_\epsilon}{16\pi^2}\frac{e g}{2\cw}\left((1/3+2\cw^2(4-\xi))q^2 +\mw^2(3+\xi)\right)  \nn\\
i\SMSigma_{WW}^{L\,{\rm Loop}} (q^2)\vert_{div} &=& i\frac{\Delta_\epsilon}{16\pi^2}\frac{g^2}{4}(3+\xi)(2\mw^2-\mz^2) \nn\\
i\SMSigma_{W\pi}^{\rm Loop} (q^2)\vert_{div} &=& i\frac{\Delta_\epsilon}{16\pi^2}\frac{g^2}{4}\left(-2\xi\mw^2+3\mz^2\right)  \nn\\
-i\SMSigma_{\pi\pi}^{\rm Loop} (q^2)\vert_{div} &=& i\frac{\Delta_\epsilon}{16\pi^2}\frac{g^2}{2}\left(-2(3-\xi)(2\mw^2+\mz^2)q^2+3\mh^2+12\mw^4+6\mz^4+\mh^2(2\mw^2+\mz^2)\xi\right)  \nn\\
\label{Loop-SE-SM}
\eear

We find the following SM results for the loop divergencies in the $S(q)V_\mu(k_1)V'_\nu(k_2)$ 1PI functions  that should be compared with the EChL results in \eqref{Loop-SVV}:
\bear
\SMgreenfL_{HAA}\vert_{div} &=& 0  \nn\\
i\SMgreenfL_{HAZ}\vert_{div} &=& i\frac{\Delta_\epsilon}{16\pi^2}\frac{g^2\sw}{\vev}\mw\mz(\xi +3) \gmunu \nn\\
i\SMgreenfL_{HZZ}\vert_{div} &=& i\frac{\Delta_\epsilon}{16\pi^2} \frac{g^2}{2\vev\cw^2}  \left(4\cw^2\mw^2(3+\xi)-6\mw^2-3\mz^2 \right)\gmunu  \nn\\
i\SMgreenfL_{HWW}\vert_{div} &=& i\frac{\Delta_\epsilon}{16\pi^2} \frac{g^2}{2\vev} \left(2\mw^2(3+2\xi)-3\mz^2 \right)\gmunu  \nn\\
i\SMgreenfL_{\pi WA}\vert_{div} &=& -\frac{\Delta_\epsilon}{16\pi^2}\frac{ g^2\sw}{2\vev} \left( 2\mw^2\xi-3\mz^2 \right)\gmunu \nn\\
i\SMgreenfL_{\pi WZ}\vert_{div} &=& \frac{\Delta_\epsilon}{16\pi^2}\frac{ g^2\sw^2}{2\vev\cw} \left( 2\mw^2\xi-3\mz^2 \right)\gmunu
\label{Loop-SVV-SM}
\eear

We find the following SM results for the loop divergencies in the $W^-_\mu(k_1)W^+_\nu(k_2)V_\rho$ 1PI functions that should be compared with the EChL results in \eqref{Loop-WWV}:
\bear
i\SMgreenfL_{WWA}\vert_{div} &=& i\frac{\Delta_\epsilon}{16\pi^2}\frac{g^3\sw}{6}(16-9\xi) \left(\gmunu \left(k_1^\rho-k_2^\rho\right) +g^{\nu\rho}(k_1^\mu+2k_2^\mu) -g^{\rho\mu}(2k_1^\nu+k_2^\nu) \right)  \nn\\
i\SMgreenfL_{WWZ}\vert_{div} &=& i\frac{\Delta_\epsilon}{16\pi^2}\frac{g^3\cw}{6}(16-9\xi) \left(\gmunu \left(k_1^\rho-k_2^\rho\right) +g^{\nu\rho}(k_1^\mu+2k_2^\mu) -g^{\rho\mu}(2k_1^\nu+k_2^\nu) \right) 
\label{Loop-WWV-SM}
\eear

As in the EChL, the SM $Z_\mu(k_1)Z_\nu(k_2)V_\rho$ 1PI functions are finite.

Finally, we find the following SM results  for the loop  divergencies in the $V_\mu V'_\nu W^-_\rho W^+_\sigma$ 1PI  functions  that should be compared with the EChL results in \eqref{Loop-VVWW}:
\bear
i\SMgreenfL_{AAWW}\vert_{div} &=& -i\frac{\Delta_\epsilon}{16\pi^2}\frac{g^4\sw^2}{6}(12\xi-7)S^{\mu\nu,\rho\sigma}  \nn\\
i\SMgreenfL_{AZWW}\vert_{div} &=& i\frac{\Delta_\epsilon}{16\pi^2}\frac{g^4\sw\cw}{6}(12\xi-7)S^{\mu\nu,\rho\sigma}  \nn\\
i\SMgreenfL_{ZZWW}\vert_{div} &=& -i\frac{\Delta_\epsilon}{16\pi^2}\frac{g^4\cw^2}{6}(12\xi-7)S^{\mu\nu,\rho\sigma}  \nn\\
i\SMgreenfL_{WWWW}\vert_{div} &=& i\frac{\Delta_\epsilon}{16\pi^2}\frac{g^4}{6}(12\xi-7)S^{\mu\rho,\nu\sigma}
\label{Loop-VVWW-SM}
\eear

Overall,  the main difference of these SM results respect to our previous EChL results is that all the SM divergences found have the tree level Lorentz structure,  or in other words, the off-shell contributions in the SM are finite.  This fact can be understood since the SM is a fully renormalizable theory while the EChL is renormalizable perturbatively in the chiral expansion.  Namely,  as we have repeatedly said,  the new operators and coefficients in $\mL_4$ act as extra CTs and are the responsible for the cancelation of the extra divergencies arising in the loop diagrams that are generated by $\mL_2$.  This way,  one obtains finite 1PI Green functions in the EChL at all off-shell momenta.

Finally,  we summarize in the following  the resulting divergencies of the SM counterterms using the $R_\xi$ gauges that should be compared with the EChL results in \eqref{L2param-div}.  From the previous findings in this appendix we get the following SM results for the counterterms of the EW parameters:
\bear
&&\deltaCT\Zf_\phi=\frac{\div}{16\pi^2}\frac{3-\xi}{\vev^2}(2\mw^2+\mz^2)\,,\quad \deltaCT \overline{T}=\frac{\div}{16\pi^2} \frac{1}{2\vev}\left(3\mh^4 +6\left(2 \mw^4+\mz^4\right)+\mh^2(2\mw^2+\mz^2)\xi \right)\,,  \nn\\
&&\deltaCT\mh^2=\frac{\div}{16\pi^2}\frac{3}{2\vev^2}(5\mh^4+(-2+\xi)\mh^2(2\mw^2+\mz^2)+6(2\mw^4+\mz^4))\,,  \nn\\
\vspace{1mm}
&&\deltaCT\Zf_B=-\frac{\div}{16\pi^2}\frac{\gY^2}{6}\,, \qquad \deltaCT\Zf_W=\frac{\div}{16\pi^2}\frac{g^2}{6}(25-6\xi)\,,  \nn\\
&&\deltaCT\mw^2=-\frac{\div}{16\pi^2}\frac{g^2}{12}\left((68-6\xi)\mw^2 -(9+3\xi)\mz^2\right)\,,  \nn\\
    %
&&\deltaCT\mz^2=\frac{\div}{16\pi^2}\frac{g^2}{12\cw^2}\left((14+6\xi-84\cw^2)\mw^2 +(11+3\xi)\mz^2\right)\,,  \nn\\
&&\deltaCT\gY/\gY=0\,,  \qquad\deltaCT g/g=-\frac{\div}{16\pi^2}\frac{g^2}{2}(3+\xi)\,,  \nn\\
\vspace{1mm}
&&\deltaCT\xi_1=\frac{\div}{16\pi^2}\frac{g^2}{6}(25-6\xi)\,,  \nn\\
&&\deltaCT\xi_2=\frac{\div}{16\pi^2}\frac{2}{3\vev^2}(25\mw^2-9\mz^2)\,,  \nn\\
&&\deltaCT\vev/\vev=\frac{\div}{16\pi^2}\frac{2\mw^2+\mz^2}{\vev^2}\xi\,.
\label{SMparam-div}
\eear
Overall, we see that all the CTs, except $\deltaCT g'/g'$ and $\deltaCT g /g$,  are different in the SM and the EChL.  The results for $\deltaCT\Zf_B$, $\deltaCT\Zf_W$ and $\deltaCT \xi_1$ coincide in the EChL and the SM for $a=1$.  The tadpole and the mass CTs in the SM,   contain  an explicit dependence on the $\xi$ parameter, in contrast to the EChL result. Since there is just one  wave function renormalization in the SM case,  corresponding to the whole doublet,   the value for $\deltaCT Z_\phi$, is different than $\deltaCT Z_H$ and $\deltaCT Z_\pi$ of the EChL. 
The results for $\deltaCT \xi_2$ and $\deltaCT v /v$ are also different in the SM and the EChL (they only coincide for $a=b=1$ in the Landau gauge).  One of the most important differences among the SM and EChL results comes from the different role played by the Higgs tadpole in both theories. In fact, it implies that within the SM, only the proper contribution $\widetilde{\delta m}^2$ of both tadpole $\delta T$ and mass counterterms $\delta m^2$ to the pole of the propagators are $\xi$-independent.  In contrast to the EChL case, where due to the genuine character of $H$ being a singlet, the tadpole and the mass counterterms are $\xi$-independent,  separately, as we have already said.  We discuss these tadpole related issues in more detail in the next appendix.
\section{Tadpole related issues}
\label{App-tadpole}

Within the SM the Higgs tadpole appears due to the linear $H$ term in the potential once the doublet field  is
expanded in terms of their component fields:
\bear
{\it V_{SM}}(\Phi) &=& -\frac{1}{2}\mh^2\Phi^\dagger\Phi +\lambda(\Phi^\dagger\Phi)^2 \nn\\
&=& \overline{T} H +\frac{\mh^2}{2}H^2 + \overline{T}\left(\pi^+\pi^-+\pi_3^2/2\right)/\vev +\lambda\vev\left(H^3+H\pi^+\pi^-+H\pi_3^2\right)  \nn\\
&& +\lambda\left(\frac{H^4}{4}+\frac{\pi_3^4}{4}+(\pi^+\pi^-)^2+\frac{H^2\pi_3^2}{2}+H^2\pi^+\pi^-+\pi_3^2\pi^+\pi^-\right) \, , 
\label{SM_Higgs-pot}
\eear
where the $\Phi$ doublet is defined as:
\be
\Phi=\binom{i\pi^+}{\frac{\vev+H-i\pi_3}{\sqrt{2}}}.
\ee
Notice that the tadpole  $\overline{T}$ is also present in the quadratic terms of the GBs.
At the leading order,  the tadpole is absent since  $\overline{T}_{LO}=(-\mh^2/2+\lambda \vev^2)\vev=0$ (due to the relation among the tree level quantities setting $\mh^2=2 \lambda \vev^2$).  At the next to leading order,  the tadpole is present and it must be renormalized,  starting with the bare tadpole given by   
 $\overline{T}_0=(-\mh^{0\,2}/2+\lambda_0\vev_0^2)\vev_0$. 
 
Regarding the multiplicative renormalization prescription in the SM,  we use the usual choice for the scalar sector given by:
\bear
\Phi_0 &=& \sqrt{\Zf_\phi}\Phi \quad\Rightarrow\quad H_0 = \sqrt{\Zf_\phi}H \,,\,\, \pi_0^{1,2,3} = \sqrt{\Zf_\phi}\pi^{1,2,3} \,,  \nn\\
\vev_0 &=& \sqrt{\Zf_\phi}(\vev +\delta\vev)  \,,\quad \lambda_0 = \Zf_\phi^{-2}(\lambda +\delta\lambda)  \,,\quad \mh^{0\,2} = \mh^2 +\delta \mh^2 \,, 
\label{SM_Hsector-renorm}
\eear
where $\Zf_\phi=1+\delta\Zf_\phi$.  Notice the difference in the SM with respect to the EChL in \eqref{EChL-renorm-factors} where we had two renormalization constants $\Zf_H$ and $\Zf_\pi$.  The use of just one $Z_{\phi}$ in the scalar sector of the SM is convenient to preserve the $SU(2)$ invariance in the renormalization procedure.

In the OS scheme, where the renormalized Higgs boson mass is the physical mass, $\mh$,   and it is related to the renormalized coupling $\lambda$ and the renormalized $v$ by $\mh^2=2\lambda\vev^2$, the counterterm associated to the SM tadpole is: 
\be
\delta \overline{T} = (-\delta\mh^2/2 +\mh^2 \left(\frac{\delta\vev}{\vev}-\frac{\delta\Zf_\phi}{2}\right) +\delta\lambda\vev^2)\vev  \,,
\label{SM_tadpole-renorm}
\ee
and the resulting counterterms for the linear and quadratic contributions in the SM Higgs potential are:
\be
\delta \overline{T} \, H +\frac{1}{2}(\delta \mh^2 +\mh^2\delta\Zf_\phi) H^2 +\delta \overline{T}\left(\pi^+\pi^-+\pi_3^2/2\right)/\vev  \,.
\label{SM_count-potHiggs-OS}
\ee
Therefore, the self-energies of the Higgs and charged GBs in the SM are
\bear
-i\SERSM_{HH}(q^2) &=& -i\SMSigma_{HH}^{\rm Loop}(q^2) +i\left(\delta\Zf_\phi (q^2-\mh^2) -\delta\mh^2\right) \,,  \nn\\
-i\SERSM_{\pi\pi} (q^2) &=& -i\SMSigma_{\pi\pi}^{\rm Loop} (q^2) +i\left( \left( q^2 -\xi\mw^2 \right)\delta\Zf_\phi -\xi\delta\mw^2 -\xi\mw^2\delta\xi_2 -\delta \overline{T}/\vev\right) \,,
\eear
and the main difference with the EChL result is the presence of $\delta \overline{T}$ in the SM GB self-energy.
All the remaining 1PI Green functions in the SM have formally equal contributions from the CTs to the corresponding ones of the EChL by setting  $a=b=1$ and $\delta a=\delta a_i=0$ in \eqrefs{SE-phys-renorm}{VVWW-renorm}.

Since  we impose the same OS renormalization conditions of \eqrefs{condOS1}{condOS5}  in both the SM and the EChL, but  there are less counterterms to be fixed in the SM than in the EChL, then it is clear that a reduced number of 1PI functions need to be renormalized in the SM  respect to the EChL case.  Concretely,  for the present computation it is sufficient to renormalize the 1-leg and 2-legs functions in the SM case.   All the remaining 1PI will be finite in the SM as it happens in any fully renormalizable theory.  Consequently,  in the SM one can determine the renormalization constant $\delta\Zf_\phi$ from the residue of the Higgs boson propagator in \eqref{condOS2} and the counterterms of the unphysical charged sector (only $\delta\xi_1$ and $\delta\xi_2$) by the pole structure of the renormalized propagators in \eqref{condOS5}.  

Finally,  the tadpole  has also implications in the STI for the SM case,  since  it enters in all the reducible 2-legs Green functions of the unhysical sector and its contribution (contrary to the EChL case) do not cancel in the particular combination of self-energies defining the STI.  Let us see this feature on more detail next. 

Since we implement a (linear) covariant $R_\xi$ gauges as in \eqref{GF-lag} within both EChL and SM, the STI in terms of the undressed propagators in momentum-space is the same as in \eqref{STIundressed}.  
Once the undressed propagators are written in terms of the reducible 2-legs Green functions, the resulting STI at the \1loop level is:
\be
q^2\Gamma_{WW}^{L\,{\rm Loop}} +2q^2\Gamma_{W\pi}^{\rm Loop} -\mw^2\Gamma_{\pi\pi}^{\rm Loop} = 0 \,.
\label{STI-unrenormSE}
\ee
However these reducible 2-legs functions differ within the SM and EChL when they are written in terms of the 1PI contributions, i.e., in terms of the self-energies and tadpole. The difference comes from the non-linearity of the scalar sector interactions and the $a$-dependence:
\bear
&\Gamma_{WW}^{L\,{\rm Loop}} = \Sigma_{WW}^{L\,{\rm Loop}}+ag\mw\frac{T^{\rm Loop}}{\mh^2}  \,,\qquad &\overline{\Gamma}_{WW}^{L\,{\rm Loop}} = \SMSigma_{WW}^{L\,{\rm Loop}}+g\mw\frac{\overline{T}^{\rm Loop}}{\mh^2} \,;  \nn\\
&\Gamma_{W\pi}^{\rm Loop} = \Sigma_{W\pi}^{\rm Loop}-ag\mw\frac{T^{\rm Loop}}{\mh^2}  \,,\qquad &\overline{\Gamma}_{W\pi}^{\rm Loop} = \SMSigma_{W\pi}^{\rm Loop}-\frac{g\mw}{2}\frac{\overline{T}^{\rm Loop}}{\mh^2} \,;  \nn\\
&\Gamma_{\pi\pi}^{\rm Loop} = \Sigma_{\pi\pi}^{\rm Loop}-a\frac{g\,q^2}{\mw}\frac{T^{\rm Loop}}{\mh^2}  \,,\qquad &\overline{\Gamma}_{\pi\pi}^{\rm Loop} = \SMSigma_{\pi\pi}^{\rm Loop}+\frac{g\mh^2}{2\mw}\frac{\overline{T}^{\rm Loop}}{\mh^2} \,.
\label{reducible-SE}
\eear
With these definitions in \eqref{STI-unrenormSE} we arrive to \eqref{charged-ST-unrenorm} in the EChL, in which the tadpole contribution cancel out. The analog expression in the SM is, in contrast: 
\be
q^2\SMSigma_{WW}^{L\,{\rm Loop}} +2q^2\SMSigma_{W\pi}^{\rm Loop} -\mw^2\SMSigma_{\pi\pi}^{\rm Loop}-\frac{g\mw}{2}\overline{T}^{\rm Loop} = 0 \,,
\label{SM_charged-ST-unrenorm}
\ee
where the presence of the tadpole is manifest.
Adding the corresponding counterterms and using the tadpole renormalization OS condition, we have the same formal expression, corresponding to  \eqref{charged-ST-renorm},  at renormalized level in the SM~\cite{Bohm:1986rj}:
\be
q^2\SERSM_{WW}^{L}(q^2) +2q^2\SERSM_{W\pi}(q^2) -\mw^2\SERSM_{\pi\pi}(q^2) = \frac{q^2 -\xi\mw^2}{\xi} \overline{f}_{ST}(q^2) \,,
\label{charged-ST-renormSM}
\ee
with the finite function given by $\overline{f}_{ST}(q^2)=-\left( \delta\Zf_W -\delta\xi_1 \right)q^2 +\xi\delta\mw^2 +\xi\mw^2\left( \delta\Zf_\phi +\delta\xi_2 \right)$. Notice that all the involved counterterms are already determined by the conditions in \eqrefs{condOS1}{condOS5}.

The last difference on the tadpole role in both theories resides in its contribution to the poles of the physical propagators.
If we consider the counterterm contribution independent of the momentum to the reducible 2-legs Green functions ($\Gamma_{HH}$, $\Gamma_{WW}^T$ and $\Gamma_{ZZ}^T$), we arrive to the following combination ($\widetilde{\delta m}^2$) of both tadpole and mass' counterterms to these reducible functions in both EChL and SM
\bear
&\widetilde{\delta m}^2_{\rm H} \vert_{EChL} = \delta\mh^2-\frac{3\kappa_3\mh^2}{\vev}\frac{\delta T}{\mh^2} \,,\qquad &\widetilde{\delta m}^2_{\rm H} \vert_{SM} = \delta\mh^2-\frac{3\mh^2}{\vev}\frac{\delta \overline{T}}{\mh^2} \,;  \nn\\
&\widetilde{\delta m}^2_{\rm W} \vert_{EChL} = \delta\mw^2-a\,g\mw\frac{\delta T}{\mh^2} \,,\qquad &\widetilde{\delta m}^2_{\rm W} \vert_{SM} = \delta\mw^2-g\mw\frac{\delta \overline{T}}{\mh^2} \,;  \nn\\
&\widetilde{\delta m}^2_{\rm Z} \vert_{EChL} = \delta\mw^2-a\,g\frac{\mz}{\cw}\frac{\delta T}{\mh^2} \,,\qquad &\widetilde{\delta m}^2_{\rm Z} \vert_{SM} = \delta\mz^2-g\frac{\mz}{\cw}\frac{\delta \overline{T}}{\mh^2} \,.
\label{deltam2bar}
\eear

As we already anticipated in \secref{divEWparams} and \appref{App-SMcompu}, only these $\widetilde{\delta m}^2_{\rm H}$, $\widetilde{\delta m}^2_{\rm W}$ and $\widetilde{\delta m}^2_{\rm Z}$ combinations of mass and tadpole counterterms are $\xi$-independent in the SM~\cite{Degrassi:1992ff}.  Whereas in the EChL, each contribution is $\xi$-independent separately.  This is one of the implications of the Higgs boson being a singlet in the EChL where the role of the tadpole is different than in the SM,  and ends up in the $\xi$-independence of some quantities in the EChL respect to the SM.

\bibliography{HMrenorm-arXiv-v2}
\end{document}